# Noninertial Relativistic Symmetry

## Stephen G. Low*

The definition of invariant time is fundamental to relativistic symmetry. Invariant time may be formulated as a degenerate orthogonal metric on a flat phase space with time $t$, position $q$, energy $\varepsilon$ and momentum $p$ degrees of freedom that is also endowed with a symplectic metric $\omega = -dt \wedge d\varepsilon + \delta_{i,j}\, dq^i \wedge dp^j$. For Einstein proper time, the degenerate orthogonal metric is $d\tau^{\circ 2} = dt^2 - \frac{1}{c^2} dq^2$ and, in the limit $c \to \infty$, becomes Newtonian absolute time, $dt^2$. We show that the the resulting symmetry group leaving $\omega$ and $dt^2$ invariant is the Jacobi group that gives the expected transformations between noninertial states defined by Hamilton's equations. The symmetry group for $\omega$ and $d\tau^{\circ 2}$ is the semidirect product of the Lorentz group and an abelian group parameterized by the time derivative of the energy-momentum tensor that characterizes noninertial states in special relativity. This leads to the consideration of invariant time based on a nondegenerate Born orthogonal metric, $d\tau^2 = dt^2 - \frac{1}{c^2} dq^2 - \frac{1}{b^2} dp^2 + \frac{1}{b^2 c^2} d\varepsilon^2$. $b$ is a universal constant with dimensions of force that, together with $c$, $\hbar$ define the dimensional scales of phase space. We determine the resulting symmetry group for transformations between noninertial states that is essentially a noncompact unitary group. It reduces to the noninertial symmetry group for Einstein proper time in the $b \to \infty$ limit and to the noninertial symmetry group for Hamiltonian mechanics in the $b, c \to \infty$ limit. The causal cones in phase space defined by the null surfaces $d\tau^2 = 0$ bound the rate of change of momentum as well as position; that is, force and velocity. Furthermore, spacetime is no longer an invariant subspace of phase space but depends on the noninertial state; there is neither an absolute rest state nor an absolute inertial state that all observers agree on.

\*     *Email: Stephen.Low@utexas.edu*
       *arxiv.org/abs/2104.05392*
       *Version 2, May 10, 2021*

# Table of Contents

# 1     Introduction

Special relativity, and its limiting form, Galilean relativity, addresses the transformation of clocks and measuring rods between inertial states on a flat manifold where gravity is negligible. Fundamental to special relativity is that Einstein proper time defined in terms of the Minkowski metric is invariant. It contracts to absolute Newton time in the limit of relative rates of change of position (velocities) that are small compared to the speed of light $c$.

We consider in this paper the transformation between noninertial states that have a nonzero relative rates of change of position, momentum and energy (velocity, force, power) on a flat phase space manifold. As the manifold is assumed to be flat, this noninertial motion is assumed to be caused by forces other than gravity such as, by way of example, electromagnetism.

The noninertial symmetry is considered for various invariant time assumptions: Newton absolute time, Einstein proper time and reciprocally invariant proper time. The concepts introduced in this introduction are elaborated in detail in the sections that follow.

## 1.1     Special relativity inertial symmetry with Einstein invariant proper time

The Lorentz group symmetry for observers in inertial states, that is fundamental to special relativity, has been verified to a high degree of precision. Key to this symmetry is invariant Einstein proper time $\tau^\circ$ that is defined by an orthogonal metric on flat spacetime $\mathbb{M} \simeq \mathbb{R}^{n+1}$,

$$d\tau^{\circ 2} = dt^2 - \frac{1}{c^2} dq^2 = dt^2 \left(1 - \frac{v^2}{c^2}\right), \tag{1.1}$$

with $(t, q) \in \mathbb{M}$, $\tau^\circ, t \in \mathbb{R}$, $q, v \in \mathbb{R}^n$ where $n = 3$ is the physical case. $c$ is the universal dimensional constant that is the speed of light (or any zero rest mass particle state) in a vacuum. It also is the upper bound on velocity of massive particle states and defines the causal cones that bounds the rate that information can propagate in spacetime.

$\tau^\circ$ is proper time measured at inertial rest to the observer's physical state. $t$ is the time measured by the observer for a physical state with constant relative velocity $v$ to the

$$\tau^\circ \qquad\qquad t$$
$$v$$

τ°                                                                                                                                          t

observer's state. Proper time $\tau°$ is invariant but local time $t$ is not as it depends on the relative velocity $v$ of the state.

Invariant mass for inertial states is given on energy-momentum space $\tilde{\mathbb{M}} \simeq \mathbb{R}^{n+1}$ by $m = \frac{\mu}{c}$ where

$$\mu^2 = \frac{1}{c^2}\varepsilon^2 - p^2, \tag{1.2}$$

with $(\varepsilon, p) \in \tilde{\mathbb{M}}$, $\mu, \varepsilon \in \mathbb{R}$ and $p \in \mathbb{R}^n$.

The group that leaves the Minkowski metric, that defines Einstein proper time, invariant is the extended Lorentz group,

$$O(1, n) \simeq \mathbb{Z}_{2,2} \otimes_s \mathcal{L}(1, n), \tag{1.3}$$

where $\mathbb{Z}_{2,2} \simeq \mathbb{P} \otimes \mathbb{T} \simeq \mathbb{Z}_2 \otimes \mathbb{Z}_2$ is the finite parity - time-reversal symmetry and $\mathcal{L}(1, n)$ is the orthochronous Lorentz group (that we refer to simply as the Lorentz group) that is the normal subgroup that is the connected component of $O(1, n)$. This group also leaves invariant the invariant mass $m$.

The connected Lorentz group $\mathcal{L}(1, n)$ is the symmetry of nature that has been experimentally verified to a high degree of precision. The discrete symmetry $\mathbb{Z}_{2,2}$ is an approximate symmetry as there are physical particle states that violate the parity $\mathbb{P}$ and the time reversal $\mathbb{T}$ symmetries.

Special relativity may be formulated on phase space $\mathbb{P} \simeq \mathbb{M} \otimes \tilde{\mathbb{M}}$ by defining the symplectic metric,

$$\omega = -dt \wedge d\varepsilon + \delta_{i,j}\, dq^i \wedge dp^j, \tag{1.4}$$

that is a nondegenerate closed 2-form on $\mathbb{P}$. Einstein proper time $d\tau°^2$ is a degenerate orthogonal metric on the phase space $\mathbb{P}$ and we can also define the degenerate orthogonal metric for invariant mass on $\mathbb{P}$,

$$d\mu^2 = \frac{1}{c^2} d\varepsilon^2 - dp^2, \tag{1.5}$$

that, when restricted to the subspace $\tilde{\mathbb{M}}$, is an orthogonal metric on energy-momentum space. We will show that requiring invariance of $\omega$, $d\tau°^2$, and $d\mu^2$ results precisely in the extended Lorentz group $O(1, n)$ symmetry where, furthermore, $\mathbb{M}$ and $\tilde{\mathbb{M}}$ are invariant subspaces. When we restrict to the spacetime subspace $\mathbb{M}$, the symmetry is the $O(1, n)$ group discussed above and likewise for the energy momentum space $\tilde{\mathbb{M}}$.

By way of example, we can consider the one dimensional case $n = 1$. The transformations for the connected component $\mathcal{L}(1, 1)$ of a basis of the cotangent space $ds = (dt, dq, d\varepsilon, dp) \in T^*\mathbb{P}$ with $t, q, \varepsilon, p \in \mathbb{R}$, is

$$\begin{aligned} d\tilde{t} &= \gamma°(v)\left(dt + \frac{v}{c^2} dq\right), & d\tilde{q} &= \gamma°(v)\,(dq + v\, dt), \\ d\tilde{\varepsilon} &= \gamma°(v)\,(d\varepsilon + v\, dp), & d\tilde{p} &= \gamma°(v)\left(dp + \frac{v}{c^2} d\varepsilon\right), \end{aligned} \tag{1.6}$$

where $\gamma°(v) = \sqrt{1 - \frac{v^2}{c^2}}$. This may be written in matrix notation as $d\tilde{s} = \Gamma(v)\, ds$ where

$$\Gamma(v) = \gamma^\circ(v) \begin{pmatrix} 1 & \frac{v}{c^2} & 0 & 0 \\ v & 1 & 0 & 0 \\ 0 & 0 & 1 & v \\ 0 & 0 & \frac{v}{c^2} & 1 \end{pmatrix} \in \mathcal{L}(1,1). \tag{1.7}$$

## 1.2 'Nonrelativistic' inertial symmetry with Newton absolute time

'Nonrelativistic' physics[2] of inertial states is the effective theory for special relativity for small velocities relative to $c$, $\frac{v}{c} \ll 1$. Mathematically this is this is equivalent to the limit of $c \to \infty$. As reviewed in Chapter 7 in [24], Newtonian time is the limit,

$$dt^2 = \lim_{c \to \infty} d\tau^{\circ 2} = \lim_{c \to \infty} \left( dt^2 - \frac{1}{c^2} dq^2 \right). \tag{1.8}$$

This defines a degenerate orthogonal metric on the spacetime. Thus, in the nonrelativistic theory, the local time of the physical state is invariant and independent of its relative velocity to the observer's state. Thus, we say that Newtonian time is absolute.

The 'nonrelativistic' inertial relativistic symmetry group is the extended Euclidean group that is the group contraction of the extended Lorentz group in the limit of $c \to \infty$,

$$\lim_{c \to \infty} O(1,n) = \mathbb{Z}_{2,2} \otimes_s \mathcal{E}(n), \quad \mathcal{E}(n) \simeq SO(n) \otimes_s \mathcal{A}(n). \tag{1.9}$$

This is the homogeneous subgroup of the Galilean group.

The abelian group is parameterized by velocity, $v \in \mathcal{A}(n)$ and is consequently simply additive and unbounded. The real subspace $\mathbb{R}$ of spacetime $\mathbb{M}$ parameterized by time, $t \in \mathbb{R} \subset \mathbb{M}$ is invariant under the action of the Euclidean group $\mathcal{E}(n)$.

This may be formulated on phase space $\mathbb{P} \simeq \mathbb{M} \otimes \tilde{\mathbb{M}}$. The symplectic metric $\omega$ does not scale with $c$ and so continues to have the form given in (1.4) in the limit $c \to \infty$. $dt^2$ is also a degenerate orthogonal metric on the phase space $\mathbb{P}$. We can also define the degenerate orthogonal metric $dp^2$ that is defined by the limit,

$$dp^2 = -\lim_{c \to \infty} d\mu^2 = -\lim_{c \to \infty} \left( \frac{1}{c^2} d\varepsilon^2 - dp^2 \right). \tag{1.10}$$

We will show that requiring invariance of $\omega$, $dt^2$ and $dp^2$ results precisely in the extended Euclidean group $\mathbb{Z}_{2,2} \otimes_s \mathcal{E}(n)$ symmetry on phase space. Furthermore, $\mathbb{M}$ and $\tilde{\mathbb{M}}$ are invariant subspaces, and when the symmetry is restricted to these subspaces, it is the $\mathbb{Z}_{2,2} \otimes_s \mathcal{E}(n)$ symmetries discussed above. In addition the time subspace, $t \in \mathbb{R} \subset \mathbb{M}$ is invariant for Newton absolute time.

By way of example, the $n=1$ transformations for $\mathcal{E}(1)$ of a basis of the cotangent space $ds = (dt, dq, d\varepsilon, dp) \in T^*\mathbb{P}$ is

$$\begin{aligned} d\tilde{t} &= dt, & d\tilde{q} &= dq + v\, dt, \\ d\tilde{\varepsilon} &= d\varepsilon + v\, dp, & d\tilde{p} &= dp. \end{aligned} \tag{1.11}$$

In the energy transformation, '$v\,dp$' is the kinetic energy term. This may be written in matrix notation as $d\tilde{s} = \Gamma(v)\, ds$ where

$$\Gamma(v) = \begin{pmatrix} 1 & 0 & 0 & 0 \\ v & 1 & 0 & 0 \\ 0 & 0 & 1 & v \\ 0 & 0 & 0 & 1 \end{pmatrix} \in \mathcal{E}(1). \tag{1.12}$$

As the states are assumed to be inertial, there is no work term, '$-f\, dq$' or power term '$r\, dt$' where $f$ is the force and $r$ the power between the states.

## 1.3    Noninertial symmetry with Newton absolute time

The above symmetries are for the case where the physical state of the observer and for the physical state being observed are inertial states. Clearly, particle states are generally noninertial with nonzero rates of change of momentum. This paper focusses on the noninertial symmetries of these states.

We start with Hamilton's mechanics with Newton absolute time. Newton time is invariant regardless of the observer and observed states; the observer may be in a noninertial frame of reference and the observed state may be in a noninertial trajectory with nonzero rates of change of momentum and energy with time. Furthermore, with the appropriate choice of the Hamiltonian, Hamilton's equations are valid for noninertial states, both for the observer and observed states. The inertial states are a special case.

For noninertial states, in our 1 dimensional example, we expect the energy and momentum to transform as

$$d\tilde{\varepsilon} = d\varepsilon + v\, dp - f\, dp + r\, dt, \quad d\tilde{p} = dp + f\, dt. \tag{1.13}$$

As the states are noninertial, both the work term, '$-f\, dq$' and power term '$r\, dt$' where $f$ is the force and $r$ the power between the states, are required in addition to the kinetic term '$v\, dp$'. Of course, $v$, $f$ and $r$ are not constant in these expressions.

Time is invariant Newton absolute time and position continue to transform as given in (1.11),

$$d\tilde{t} = dt, \quad d\tilde{q} = dq + v\, dt. \tag{1.14}$$

Momentum, and hence $dp^2$, is no longer invariant but change with time due to the relative force. This leads us to remove the requirement above that $dp^2$ be invariant and to consider the symmetry on $\mathbb{P}$ that leaves invariant only $\omega$ and $dt^2$. We will show that the resulting symmetry group acting on $\mathbb{P}$ is the Jacobi group[3],

$$\mathcal{HS}p(2n) \simeq \mathcal{S}p(2n) \otimes_s \mathcal{H}(n). \tag{1.15}$$

In this expression, $\mathcal{S}p(2n)$ is the symmetry of the symplectic submanifold $(p,q) \in \mathbb{P}^\circ \subset \mathbb{P}$ that has the symplectic metric

$$\omega^\circ = \delta_{i,j}\, dq^i \wedge dp^j \tag{1.16}$$

The symplectomorphisms $\varphi : \mathbb{P}^\circ \to \mathbb{P}^\circ$ such that $\varphi^* \omega^\circ = \omega^\circ$ are the standard canonical transformations of Hamilton's mechanics.

$\mathcal{H}(n)$ is the Weyl-Heisenberg group that is the semidirect product of two abelian groups. Some readers may be surprised to see it showing up in a classical (non-quantum) 'nonrelativistic' symmetry of Hamilton's mechanics as it is usually associated with quantum mechanics. However, just as Lie groups such as the orthogonal, unitary, abelian and so forth, appear in many different physical contexts, so does the Weyl-Heisenberg group. (The properties of the Weyl-Heisenberg group are reviewed in

Chapter 8 in [24].)

To illustrate the Weyl-Heisenberg group symmetry in the current context, we can consider the $n = 1$ dimensional case for which the symmetry is $\mathcal{H}Sp(2) \simeq Sp(2) \otimes_s \mathcal{H}(1)$. We have given the transformation equations in (1.13-14) and these provide a realization of $\mathcal{H}(1)$ by writing these equations in matrix form. That is, with $ds = (dt, dq, d\varepsilon, dp) \in T^*\mathbb{P}$ and $d\tilde{s} = \Gamma \, ds$, then (1.13-14) may be written in matrix notation as

$$d\tilde{s} = \begin{pmatrix} d\tilde{t} \\ d\tilde{q} \\ d\tilde{\varepsilon} \\ d\tilde{p} \end{pmatrix} = \Gamma(v, f, r) \, ds = \begin{pmatrix} 1 & 0 & 0 & 0 \\ v & 1 & 0 & 0 \\ r & -f & 1 & v \\ f & 0 & 0 & 1 \end{pmatrix} \begin{pmatrix} dt \\ dq \\ d\varepsilon \\ dp \end{pmatrix}. \tag{1.17}$$

We will show that this is a matrix realization of $\mathcal{H}(1)$,

$$\Gamma(v, f, r) = \begin{pmatrix} 1 & 0 & 0 & 0 \\ v & 1 & 0 & 0 \\ r & -f & 1 & v \\ f & 0 & 0 & 1 \end{pmatrix} \in \mathcal{H}(1). \tag{1.18}$$

Furthermore, in $n$ dimensions, we will show that the the diffeomorphisms $\rho: \mathbb{P} \to \mathbb{P}$ that have a Jacobian $\frac{\partial \rho}{\partial s} = \Gamma(v, f, r)$ are Hamilton's equations. As a consequence, we will show that the symplectomorphisms $\varrho$ such that $\varrho^* \omega = \omega$ that also preserve the Newton absolute time degenerate orthogonal metric $\varrho^* dt^2 = dt^2$ (and therefore have a Jacobian matrix $\frac{\partial \varrho}{\partial s} \in \mathcal{H}Sp(2n)$) are Hamilton's equations up to a canonical transformation on $\mathbb{P}^\circ \subset \mathbb{P}$, $\varrho = \rho \circ \varphi$. Therefore, the Jacobi group $\mathcal{H}Sp(2n)$ is the noninertial symmetry group for Hamilton's mechanics with absolute Newton time.

If we restrict the symplectic group $Sp(2n)$ to its subgroup of physical rotations, $O(n) \subset Sp(2n)$, then the symmetry group is the extended Hamilton group $\mathbb{Z}_{2,2} \otimes_s \mathcal{H}a(n)$ where the Hamilton group is defined as

$$\mathcal{H}a(n) \simeq SO(n) \otimes_s \mathcal{H}(n). \tag{1.19}$$

## 1.4   Noninertial symmetry with Einstein proper time

Special relativity assumes that the spacetime manifold $\mathbb{M}$ is flat. The observer state is assumed to be inertial but the non-inertial motion of, as an example, a physical particle state (e.g electron) under the influence of an electromagnetic force is admissible. The velocity of these particle states is not constant. We can ask the question as to how to transform between a cotangent frame $ds = (dt, dp, d\varepsilon, dp) \in T^*\mathbb{P}$ of the inertial observer state and the cotangent frame $d\tilde{s} = (d\tilde{t}, d\tilde{p}, d\tilde{\varepsilon}, d\tilde{p}) \in T^*\mathbb{P}$ of the particle undergoing noninertial motion. This is studied in the paper [73]. The first assertion is that Einstein proper time $\tau^\circ$ defined by (4.1) continues to be valid although the velocity $v$ is no longer constant but varies with time. The relation between proper time $\tau^\circ$ and local time $t$ is given by the instantaneous value of $v(\tau^\circ)$,

$$dt = \frac{d\tau^\circ}{\sqrt{1 - \frac{1}{c^2} v(\tau^\circ)^2}} \ . \tag{1.20}$$

At this point we digress and briefly discuss gravity and general relativity as it is frequently assumed to generalize special relativity to noninertial states. General relativity brings gravity into conformance with special relativity locally but does not actually address noninertial states. The equivalence principle of general relativity states that a particle state accelerating under the influence of an *apparent* force of gravity is equivalent to a locally inertial particle state on a certain curved spacetime manifold defined by the Einstein gravitational field equations. In a world were only gravity is present (and there are no other forces), the particle states, both massive and massless, follow geodesics that are the inertial trajectories on the curved manifold. The Riemann connection transforms a locally inertial basis for $T^*_x \mathbb{M}$ at one point to a locally inertial basis at a neighboring point $T^*_{x+dx} \mathbb{M}$. Consequently, in pure general relativity there are only locally inertial physical states and no noninertial states. However, if a force such as electromagnetism is present, then there is noninertial motion relative to these locally inertial states in the curved manifold.

In our discussion in this paper, we are interested in noninertial states and so for simplicity in this initial development, assume gravity is not present so that the manifold is flat.

Einstein invariant proper time $d\tau^{\circ 2}$ defines a degenerate orthogonal metric on the phase space $\mathbb{P}$ that is also endowed with a symplectic metric $\omega$ defined in (1.4). We will show that the symmetry group that leaves $\omega$ and $d\tau^{\circ 2}$ invariant is the group,

$$\mathcal{O}a(1, n) \simeq \mathcal{O}(1, n) \otimes_s \mathcal{A}(m) \simeq \mathbb{Z}_{2,2} \otimes_s \mathcal{L}a(1, n), \quad m = \frac{1}{2} n(n+3), \tag{1.21}$$

where $\mathcal{L}a(1, n)$ is the connected component defined in terms of the orthochronous Lorentz group $\mathcal{L}(1, n)$ and an abelian group $\mathcal{A}(m)$,

$$\mathcal{L}a(1, n) \simeq \mathcal{L}(1, n) \otimes_s \mathcal{A}(m). \tag{1.22}$$

To illustrate essential properties, the $\mathcal{L}a(1, 1)$ transformations of the cotangent basis for the $n = 1$ case are

$$d\tilde{t} = \gamma^\circ(v)\left(dt + \frac{v}{c^2} dq\right), \qquad d\tilde{q} = \gamma^\circ(v)(dq + v\, dt),$$
$$d\tilde{\varepsilon} = \gamma^\circ(v)(d\varepsilon + v\, dp - f\, dq + r\, dt), \quad d\tilde{p} = \gamma^\circ(v)\left(dp + \frac{v}{c^2} d\varepsilon + f\, dt - \frac{r}{c^2} dq\right). \tag{1.23}$$

Note that $d\mu^2$ defined in (1.5) is not invariant under these transformations. If $f = r = 0$, these equations reduce to the inertial transformations (1.6) and $d\mu^2$ is invariant. These equations differ from the inertial transformations (1.6) through the noninertial term

$$\begin{pmatrix} d\tilde{\varepsilon} \\ d\tilde{p} \end{pmatrix} = \mathrm{M} \begin{pmatrix} dt \\ dq \end{pmatrix}, \quad \mathrm{M} = \gamma^\circ(v) \begin{pmatrix} r & -f \\ f & -\frac{1}{c^2} r \end{pmatrix}. \tag{1.24}$$

We will show that M is a $(1, 1)$ tensor under the Lorentz transformations that is a stress-power-force tensor that is the proper time derivative of the stress-energy-momentum tensor in the $n$ dimensional case. (The $(0, 2)$ form of the tensor is symmetric and traceless.)

The above transformations clearly contract to the 'nonrelativistic' transformations
$$c \to \infty \qquad\qquad n$$

(1.13-14) in the limit $c \to \infty$ and we will show that this is true for the general $n$ case. Furthermore, we will show that

$$\lim_{c \to \infty} \mathcal{O}a(1, n) \simeq \lim_{c \to \infty} \mathbb{Z}_{2,2} \otimes_s \mathcal{L}a(1, n) = \mathbb{Z}_{2,2} \otimes_s \mathcal{H}a(n), \tag{1.25}$$

where $\mathcal{H}a(n)$ is the Hamilton group defined in (1.19).

## 1.5     Noninertial symmetry with reciprocally invariant proper time

The relationship between Einstein invariant proper time $\tau°$ and local time $t$ depends only on the velocity $v(\tau°)$ as given in (1.20). It does not depend on the relative change of momentum or energy, that is, force $f$ and power $r$, that characterize the noninertial motion. This is due to the fact that $d\tau^{°2}$ is given by a degenerate orthogonal metric on phase space $\mathbb{P}$ that is only nondegenerate on the spacetime subspace, $\mathbb{M} \subset \mathbb{P}$.

While this relationship (1.20) has been verified to a high degree of precision for the case that all states are inertial in which case velocity is constant, $v(\tau°) = v$, it has not been verified to as high degree of precision for extreme noninertial motion.

The Einstein time line element defines causal cones in spacetime that bounds the rate of change of position for any massive or massless physical state to be less than or equal to $c$. It therefore bounds the rate at which information can propagate in spacetime. In the nonrelativistic limit, this bound becomes infinite. However, there is not a bound on the rate of change of momentum for physical states in momentum space. For inertial states, momentum is constant and this rate is zero and so it is not an issue. For noninertial states, this becomes particularly problematic when we consider the quantum theory. A state may be equally well measured and consequently diagonalized in either position or momentum; from a quantum perspective these are on the 'same footing'. Why should the position measurement have to conform to the bound (up to the quantum uncertainty) in its rate of propagation by a causal cone and not the momentum measurement?

This leads us to consider a more general definition of invariant time on phase space. In the inertial limit this must reduce to special relativity with its invariants $d\tau^{°2}$ on spacetime $\mathbb{M}$ and $d\mu^2$ on energy-momentum space $\tilde{\mathbb{M}}$. In the well defined limit of small rates of change of momentum and energy, it must reduce to the noninertial symmetry with Einstein invariant time described in the previous Section 1.1.

Furthermore, the more general definition of time should also bound the rates of change of momentum just as the rate of change of position is bounded by $c$.

We will show that a generalized time $d\tau$ defined by non-degenerate orthogonal metric on phase space $\mathbb{P}$ given by

$$d\tau^2 = dt^2 - \frac{1}{c^2} dq^2 + \frac{1}{b^2}\left(\frac{1}{c^2} d\varepsilon^2 - dp^2\right) = dt^2\left(1 - \frac{v^2}{c^2} - \frac{f^2}{b^2} + \frac{r^2}{c^2 b^2}\right), \tag{1.26}$$

has all of these properties [14,15,16,22]. In this expression $v$, $f$ and $r$ are the rates of change of position, momentum and energy with time. $b$ is a fundamental universal constant with the dimensions of force that, together with $c$ and $\hbar$ defines the dimensional scales of the phase space.

Born recognized the problem of the 'asymmetry' between momentum and position in the context of reconciling quantum mechanics with special relativity. He proposed a *reciprocity principle* that states physical laws should be invariant under the discrete symmetries [4, 5],

$$(q, p) \to (p, -q), \quad (t, \varepsilon) \to (\varepsilon, -t). \tag{1.27}$$

He applied this principle by proposing that the Klein-Gordon equation should be generalized to a reciprocally invariant wave equation through the introduction of a minimum length scale that defined a nondegenerate orthogonal metric on phase space. This generalized wave equation did not prove to be fruitful, but the reciprocity principle has continued to appear in many contexts[5].

The above metric (1.26) is reciprocally invariant under this discrete symmetry. For this reason, as a tribute to Born's pioneering work, we call $d\tau$ *reciprocally invariant time* and the nondegenerate orthogonal metric on phase space, the *Born metric* [15, 16].

We will show that the Born metric is invariant under the following symmetry group,

$$\mathcal{U}u(1, n) \simeq \mathcal{U}_2(1) \otimes_s \mathcal{SU}c(1, n), \tag{1.28}$$

where $\mathcal{U}_2(1) = \mathcal{U}(1) \otimes \mathcal{U}(1)$ and $\mathcal{SU}c(1, n)$ is the connected subgroup of $\mathcal{SU}(1, n)$. Heuristically the group $\mathcal{U}u(1, n)$ may be thought of as the unitary group with a overall phase, '$\mathcal{U}u(1, n) \simeq \mathcal{U}(1) \otimes \mathcal{U}(1, n)$'. However we show that the phase is topologically nontrivially intertwined with $\mathcal{U}(1, n)$ as $\mathcal{U}(1) \cap \mathcal{U}(1, n) \simeq \mathbb{Z}_2$ and so it cannot be written as the trivial direct product.

The inertial subgroup with $f = r = 0$ is the extended Lorentz group $O(1, n)$. Furthermore, we will show that in the limit $b \to \infty$,

$$\lim_{b \to \infty} \mathcal{U}u(1, n) \simeq Oa(1, n) = O(1, n) \otimes_s \mathcal{A}\left(\frac{1}{2}n(n+1)\right). \tag{1.29}$$

To illustrate essential properties, the transformations of the cotangent basis for the $n = 1$ case, $\mathcal{SU}c(1, n)$, are

$$\begin{aligned}
d\tilde{t} &= \gamma(v, f, r)\left(dt + \frac{v}{c^2}dq + \frac{f}{b^2}dp - \frac{r}{c^2 b^2}d\varepsilon\right),\\
d\tilde{q} &= \gamma(v, f, r)\left(dq + v\,dt + \frac{r}{b^2}dp - \frac{f}{b^2}d\varepsilon\right),\\
d\tilde{\varepsilon} &= \gamma(v, f, r)(d\varepsilon + v\,dp - f\,dq + r\,dt),\\
d\tilde{p} &= \gamma(v, f, r)\left(dp + f\,dt - \frac{r}{c^2}dq + \frac{v}{c^2}d\varepsilon\right),
\end{aligned} \tag{1.30}$$

where

$$\gamma(v, f, r) = \left(1 - \frac{v^2}{c^2} - \frac{f^2}{b^2} + \frac{r^2}{c^2 b^2}\right)^{-\frac{1}{2}}. \tag{1.31}$$

These transformation equations reduce to the equations (1.6) in the inertial case and to (1.23) in the limit $b \to \infty$. Furthermore, the null surfaces are given by $\gamma(v, f, r) = 0$,

$$\frac{v^2}{c^2} + \frac{f^2}{b^2} = 1 + \frac{r^2}{c^2 b^2}. \tag{1.32}$$

Consequently, both the rate of change of momentum and position, $f$ and $v$ are bounded, but this bound depends on the rate of change of energy $r$. We will show that this is also true in the $n$ dimensional case and defines the causal cones for reciprocally invariant time. Clearly the effects of a reciprocally invariant time as compared to Einstein invariant time become most apparent at scales where $f \to b$ and $r \to bc$. If $b$ experimentally turns out to be 'large', then these effects are not readily apparent in the near inertial realm. It is an experimental question as to whether these effects that distinguish recipro-cally invariant proper time from Einstein proper time manifest physically.

## 1.6 Paper outline

We will start the more detailed analysis by briefly reviewing the phase space symplectic manifold for the flat manifold case ($\mathbb{P} \simeq \mathbb{R}^{2n+2}$) and its dimensional scales based on the universal dimensional constants $c$, $b$ and $\hbar$ in Section 2.2. Then, in the subsequent Sections 3.3-5, we develop in more detail noninertial relativity based on Newton absolute time, Einstein proper time and reciprocally invariant proper time. These three relativistic symmetries for the different definitions of invariant time have the following basic properties:

### *Newton absolute time*

Time is invariant for all observers and defines an invariant subspace of phase space that all observers in any physical state agree on. There is an absolute rest state and an absolute inertial state that all observers agree on. Position-time (aka spacetime) and momentum-time are invariant subspaces of phase space that all observers in any physical state agree on. Velocity and force are unbounded; instantaneous action at a distance is possible in position space and momentum space.

### *Einstein proper time (on a flat manifold with no gravity)*

Time depends on the observer relative velocity. Time is not an invariant subspace of spacetime nor phase space. There is not an absolute rest state but there continues to be an absolute inertial state that all observers agree on. Spacetime is an invariant subspaces of phase space that all observers in any physical state agree on. Velocity is bounded by the causal cone defined by $c$ in spacetime but force is unbounded and so instantaneous action at a distance is not possible in spacetime but is possible in momentum-time space.

### *Reciprocal invariant proper time (on a flat manifold with no gravity)*

Time depends on the observer physical state. Time is not an invariant subspace of spacetime nor phase space. There is neither an absolute rest state nor an absolute inertial state that all observers agree on. Spacetime is an not an invariant subspaces of phase space but depends on the observers physical state. Velocity and force are bounded by the causal cone defined by $c$ and $b$ in phase space.

### Quantum symmetry

In the papers [16-23], we study the quantum symmetry defined by the projective representations of the corresponding inhomogeneous symmetry groups. All of the above symmetry groups of interest are subgroups of the symplectic group $Sp(2n+2)$ and so the corresponding inhomogeneous groups admit a noncommutative Weyl-Heisenberg subgroup in their central extension.

## 2    Phase space symplectic manifold

The symplectic structure of phase space is foundational to our understanding of Hamiltonian mechanics. A symplectic manifold is a $2m$ dimensional differential manifold $\mathbb{P}$ with a nondegenerate 2-form $\omega$ that is closed, $d\omega = 0$. Darboux's theorem states that locally coordinates $\tilde{s}$ exists where this 2-form is of the form

$$\omega = \tilde{\zeta}_{\alpha,\beta}\, d\tilde{s}^\alpha\, d\tilde{s}^\beta, \quad \alpha, \beta, \ldots = 0, 1, \ldots n, \overline{0}, \overline{1}, \ldots \overline{n}, \quad m = n+1, \tag{2.1}$$

where the matrix of components is

$$\tilde{\zeta} = [\![\tilde{\zeta}_{\alpha,\beta}]\!] = \begin{pmatrix} 0 & 1_{n+1} \\ -1_{n+1} & 0 \end{pmatrix}. \tag{2.2}$$

We will refer to this nondegenerate closed symplectic 2-form as a symplectic metric where we use the term *metric* in a generalized sense. The canonical form of the matrix of components is not unique. The order of the coordinates may be permuted, for example, by exchanging $0 \leftrightarrow \overline{0}$,

$$\tilde{s} = \left(\tilde{s}^0, \tilde{s}^i, \tilde{s}^{\overline{0}}, \tilde{s}^{\overline{i}}\right) \mapsto s = \left(s^0, s^i, s^{\overline{0}}, s^{\overline{i}}\right) = \left(\tilde{s}^{\overline{0}}\, \tilde{s}^i, \tilde{s}^0, \tilde{s}^{\overline{i}}\right), \quad i, j, \ldots = 1, \ldots, n.$$

The symplectic metric is unchanged,

$$\omega = \tilde{\zeta}_{\alpha,\beta}\, d\tilde{s}^\alpha\, d\tilde{s}^\beta = \zeta_{\alpha,\beta}\, ds^\alpha\, ds^\beta, \tag{2.3}$$

whereas the corresponding matrix of components becomes

$$\zeta = [\![\zeta_{\alpha,\beta}]\!] = \begin{pmatrix} 0 & \eta \\ -\eta & 0 \end{pmatrix}, \quad \eta = \begin{pmatrix} -1 & 0 \\ 0 & 1_n \end{pmatrix}. \tag{2.4}$$

Of course, while $\eta's$ appear in the matrix of components in these coordinates, the symplectic metric does not have a topologically invariant signature in the sense of the orthogonal metrics. We adopt matrix notation and suppress indices going forward. In matrix notation, the symplectic metric is

$$\omega = ds^t\, \zeta\, ds, \quad s \in \mathbb{R}^{2n+2}. \tag{2.5}$$

A diffeomorphism $\varrho : \mathbb{P} \to \mathbb{P} : s \mapsto s' = \varrho(s)$ induces

$$\varrho^* : T^*\mathbb{P} \to T^*\mathbb{P} : ds \mapsto ds' = \Sigma(s)\, ds, \tag{2.6}$$

where

$$\Sigma(s) = \frac{\partial \varrho(s)}{\partial s} \in \mathcal{GL}(2n+2, \mathbb{R}). \tag{2.7}$$

A symplectomorphism on $\mathbb{P}$ is a diffeomorphism that preserves the symplectic metric, $\omega = \varrho^* \omega$. In coordinates, this is

$$\omega = ds'^t\, \zeta\, ds' = (\Sigma(s)\, ds)^t\, \zeta\, \Sigma(s)\, ds, \tag{2.8}$$

and therefore the symplectic condition on the matrix of components is

$$\Sigma(s)^t \zeta \Sigma(s) = \zeta. \tag{2.9}$$

As we have studied previously in Chapter 9 in [24], this defines the real symplectic metric, $\Sigma(s) \in Sp(2n+2, \mathbb{R})$. Define,

$$\Sigma = \begin{pmatrix} \Sigma_1 & \Sigma_2 \\ \Sigma_3 & \Sigma_4 \end{pmatrix}, \tag{2.10}$$

where the submatrices must satisfy

$$\Sigma_1^t \eta \Sigma_4 - \Sigma_3^t \eta \Sigma_2 = \eta, \; \Sigma_1^t \eta \Sigma_3 = \left(\Sigma_1^t \eta \Sigma_3\right)^t, \; \Sigma_2^t \eta \Sigma_4 = \left(\Sigma_2^t \eta \Sigma_4\right)^t. \tag{2.11}$$

Again, the appearance of the $\eta$'s is simply due to our choice of ordering of the coordinates; the symplectic group does not have a signature.

Phase space in Hamiltonian mechanics is a symplectic manifold for which the symplectomorphisms are generally referred to as canonical transformations. The symplectic metric defines the concept of canonical conjugacy between vectors in the tangent space and 1-forms in the cotangent space. Furthermore, as we studied in Chapter 9 in [24], the symplectic structure is key to the origin of Heisenberg commutation relations when we consider the quantum symmetry implications of the theory.

For our considerations, the symplectic manifold $(\mathbb{P}, \omega)$ is the phase space with the physical time, position, energy and momentum degrees of freedom. This will hold throughout this paper. For our purposes, it is sufficient to consider the case where $\mathbb{P}$ is a flat manifold diffeomorphic to $\mathbb{R}^{2n+2}$. In this case global coordinates exist and Darboux's theorem holds globally and we can write $\mathbb{P} \simeq \mathbb{M} \otimes \tilde{\mathbb{M}}$ with coordinates $s = (x, y)$ where $\mathbb{M}$ is the spacetime submanifold with coordinates $x \in \mathbb{M} \simeq \mathbb{R}^{n+1}$ and $\tilde{\mathbb{M}}$ is the energy-momentum submanifold with coordinates $y \in \tilde{\mathbb{M}} \simeq \mathbb{R}^{n+1}$. The symplectic metric is

$$\omega = \eta_{a,b} \, dx^a \wedge dy^b, \quad a, b, \ldots = 0, 1, \ldots, n. \tag{2.12}$$

We can also write it more explicitly with the coordinates of time $t \in \mathbb{R}$, energy $\varepsilon \in \mathbb{R}$, position $q \in \mathbb{R}^n$ and momentum $p \in \mathbb{R}^n$ with $s = (t, q, \varepsilon, p)$ and the symplectic metric is then

$$\omega = -dt \wedge d\varepsilon + \delta_{i,j} \, dq^i \wedge dp^j. \tag{2.13}$$

## 2.1 Dimensional scales of phase space

The physical degrees of freedom including time, energy, position and momentum are dimensioned quantities that are measured by dimensional scales. We reviewed these dimensional scales in our discussion of the Galilean relativity as the small velocity (relative to the speed of light $c$) limit of special relativity Chapter 7 of [24].

The dimensional scales of length $\lambda_q$, time $\lambda_t$, momentum $\lambda_p$ and energy $\lambda_\varepsilon$ may be defined in terms of three universal invariant dimensional scale constants $\{c, b, \hbar\}$ that satisfy the relations

$$\frac{\lambda_q}{\lambda_t} = c = \frac{\lambda_\varepsilon}{\lambda_p}, \; \frac{\lambda_p}{\lambda_t} = b = \frac{\lambda_\varepsilon}{\lambda_q}, \; \lambda_q \lambda_p = \hbar = \lambda_\varepsilon \lambda_t. \tag{2.14}$$

Solving these equation results in the dimensional scales,

$$\lambda_t = \sqrt{\frac{\hbar}{bc}}, \quad \lambda_q = \sqrt{\frac{\hbar c}{b}}, \quad \lambda_p = \sqrt{\frac{\hbar b}{c}}, \quad \lambda_\varepsilon = \sqrt{\hbar b c}. \tag{2.15}$$

$c$ is the universal constant that has the dimensions of velocity that is the speed of light in a vacuum. It has fundamental importance in special relativity as the upper bound of velocity for any massive particle and the velocity of any massless particles, including photons. It defines the causal cones that bounds the rate at which information may be transmitted in spacetime.

$\hbar$ is the universal constant that has the dimensions of action that is Planck's constant fundamental to quantum mechanics.

$b$ is the universal constant that has dimensions of force. The usual Planck scales take the third universal constant to be the Newton gravitational constant $G_N$. Then $b$ may be defined in terms of the Newton gravitational universal constant and the universal constant $c$ as

$$b = \frac{c^4}{G_N}. \tag{2.16}$$

In this case, the dimensional scales $\{\lambda_t, \lambda_q, \lambda_\varepsilon, \lambda_p\}$ are the familiar Planck scales[6] defined in terms of the three universal constants $\{c, G_N, \hbar\}$,

$$\lambda_t = \sqrt{\frac{\hbar G_N}{c^5}}, \quad \lambda_q = \sqrt{\frac{\hbar G_N}{c^3}}, \quad \lambda_p = \sqrt{\frac{\hbar c^3}{G_N}}, \quad \lambda_\varepsilon = \sqrt{\frac{\hbar c^5}{G_N}}. \tag{2.17}$$

The constants $c$ and $\hbar$ are universally applicable to all physical theories. $G_N$ on the other hand, is the gravitational coupling constant specific to gravity, just as the electromagnetic-weak and strong coupling constants are specific to their corresponding forces. Consequently, in this paper we will take the perspective that $b$ is the third fundamental constant that is experimentally determined along with $c$ and $\hbar$. The dimensional Newton gravitational constant is is defined in terms of the universal constants $b$ and $c$ and the dimensionless constant $\alpha_N$,

$$G_N = \alpha_N \frac{c^4}{b}. \tag{2.18}$$

$\alpha_N$ is the dimensionless gravitational coupling constant. This will play a pivotal role in our discussion. Determining $b$ fixes $\alpha_N$ and vice-versa. If $\alpha_N$ turns out to be unity, $\alpha_N = 1$, then the dimensional scales are the usual Planck scales (2.17).

Our convention will always be that the coordinates $s \in \mathbb{P}$ and $x \in \mathbb{M}$, $y \in \tilde{\mathbb{M}}$ and are dimensionless. These may be expanded out in terms of the dimensioned coordinates $\{t, q, \varepsilon, p\}$ that are scaled by the $\lambda_\alpha$ defined in (2.15) to form dimensionless quantities,

$$s = (x, y) = \left(\frac{t}{\lambda_t}, \frac{q}{\lambda_q}, \frac{\varepsilon}{\lambda_\varepsilon}, \frac{p}{\lambda_p}\right), \quad s \in \mathbb{P}, \ x, y \in \mathbb{R}^{n+1}, \ t, \varepsilon \in \mathbb{R}, \ q, p \in \mathbb{R}^n. \tag{2.19}$$

If natural units are chosen so that $c = b = \hbar = 1$, then the dimensional scales reduce to unity, $\lambda_t = \lambda_q = \lambda_\varepsilon = \lambda_p = 1$.

The symplectic metric using these coordinates is

$$\omega = -\frac{dt}{\lambda_t} \wedge \frac{d\varepsilon}{\lambda_\varepsilon} + \delta_{i,j} \frac{dq^i}{\lambda_q} \wedge \frac{dp^j}{\lambda_p} = \frac{1}{\hbar}\left(-dt \wedge d\varepsilon + \delta_{i,j} dq^i \wedge dp^j\right). \quad (2.20)$$

Of course, in natural units $c = b = \hbar = 1$, this is the same as (2.13).

We have already seen that the understanding of Galilei relativity as a limiting form of special relativity $\frac{v}{c} \to 0$ or $c \to \infty$ required this explicit handling of dimensional scales. In that case, the calculations worked out so that only $c$ was relevant. As we consider noninertial relativity based on the Born metric in a non-quantum mechanical setting, both $c, b$ play a key role. In the quantum theory that we develop in [16], all three of these scales manifest.

# 3     Noninertial relativity with Newton absolute time

In this section we formulate standard nonrelativistic Hamilton's mechanics on the phase space that is a symplectic manifold $\mathbb{P}$ with time, position, energy and momentum degrees of freedom. The key assumption in this formulation is that time is Newtonian absolute invariant time. We show that the full invariance group is the Jacobi group and that this leads directly to Hamilton's equations [17,18,19,21]. The formulation on $\mathbb{P}$ allows us to study generalizations in the following sections that assume Einstein proper time and reciprocally invariant proper time instead of Newton absolute time and to show that the nonrelativistic case is the small velocity and force limit.

## 3.1     Invariant Newton time symmetry group

A most basic property of nonrelativistic mechanics is that time is invariant Newtonian time. Observers in any state, inertial or noninertial, measure the same passage of time on clocks. In our study of inertial Galilean relativity in Chapter 7 in [24], invariant Newtonian time is realized as the degenerate orthogonal metric $dt^2$. This applies also in Hamiltonian mechanics with $dt^2$ regarded as a degenerate orthogonal metric on the phase space $\mathbb{P}$.

It is convenient to start by choosing natural units $c = b = \hbar = 1$ so that the dimensional scales $\lambda_t = \lambda_q = \lambda_\varepsilon = \lambda_p = 1$. We will make these scales explicit later in this section and show that they just cancel out. It is also convenient for us to start with a choice of ordering of the coordinates of $\mathbb{P}$ to be $s = (p, q, \varepsilon, t)$ with $p, q \in \mathbb{R}^n$ and $t, \varepsilon \in \mathbb{R}$.

The Newtonian invariant time line element is the degenerate orthogonal metric,

$$dt^2 = ds^t \eta^\circ ds, \quad (3.1)$$

where the matrix of components with the above coordinate ordering is

$$\eta^\circ = \begin{pmatrix} 0_{2n+1} & 0 \\ 0 & 1 \end{pmatrix}. \quad (3.2)$$

$0_{2n+1}$ is a $2n+1$ square matrix of zeros and the other $0's$ are the corresponding $2n+1$ dimensional row and column zero matrices.

Consider $\Gamma : T^*\mathbb{P} \to T^*\mathbb{P} : ds \mapsto d\tilde{s} = \Gamma \, ds$ that leaves invariant the Newtonian time line element,

$$dt^2 = d\tilde{s}^t\, \eta^\circ\, d\tilde{s} = (\Gamma\, ds)^t\, \eta^\circ\, \Gamma(s)\, ds = ds^t\, \eta^\circ\, ds, \tag{3.3}$$

and therefore $\Gamma^t\, \eta^\circ\, \Gamma = \eta^\circ$.

It is straightforward matrix algebra to show that the $\Gamma$ matrices leaving invariant $dt^2$ have the form

$$\Gamma = \begin{pmatrix} \Sigma^\circ & b & w \\ c^t & a & r \\ 0 & 0 & 1 \end{pmatrix} \begin{pmatrix} 1_{2n} & 0 & 0 \\ 0 & 1 & 0 \\ 0 & 0 & \epsilon \end{pmatrix} = \begin{pmatrix} \Sigma^\circ & b & \epsilon w \\ c^t & a & \epsilon r \\ 0 & 0 & \epsilon \end{pmatrix}, \tag{3.4}$$

where $\Sigma^\circ$ is a $2n \times 2n$ real matrix, $w, b, c \in \mathbb{R}^n$ and $a, r \in \mathbb{R}$ and $\epsilon = \pm 1$. Consequently, $\Gamma(s) \in \mathbb{Z}_2 \otimes_s \mathcal{IGL}(2n+1, \mathbb{R})$. The $\epsilon = \pm 1$ characterize the finite $\mathbb{Z}_2$ subgroup. This $\mathbb{Z}_2$ subgroup is time-reversal symmetry.

$$dt \mapsto \epsilon\, dt = \pm\, dt. \tag{3.5}$$

The $\Gamma$ with $\epsilon = 1$ is the $\mathcal{IGL}(2n+1, \mathbb{R})$ normal subgroup for which $dt$ is invariant.

The phase space $\mathbb{P}$ also is a symplectic manifold as we reviewed in the previous section. In the ordering of coordinates $s = (p, q, \varepsilon, t)$, the matrix of components of the symplectic metric $\omega = ds^t\, \zeta\, ds$ (2.5) are

$$\zeta = \begin{pmatrix} \zeta^\circ & 0 & 0 \\ 0 & 0 & -1 \\ 0 & 1 & 0 \end{pmatrix}, \quad \zeta^\circ = \begin{pmatrix} 0 & 1_n \\ -1_n & 0 \end{pmatrix}. \tag{3.6}$$

In nonrelativistic Hamiltonian mechanics, we are interested in the canonical transformations that leave invariant the symplectic metric, $\omega = \varrho^* \omega$ and the degenerate orthogonal line element for invariant Newtonian time, $dt^2 = \varrho^* dt^2$. That is, $\varrho: \mathbb{P} \to \mathbb{P}$ such that the Jacobian matrix $\Gamma(s) = \frac{\partial \varrho(s)}{\partial s}$, satisfies

$$dt^2 = (\Gamma(s)\, ds)^t\, \eta^\circ\, \Gamma(s)\, ds = ds^t\, \eta^\circ\, ds, \quad \omega = (\Gamma(s)\, ds)^t\, \zeta\, \Gamma(s)\, ds = ds^t\, \zeta\, ds. \tag{3.7}$$

Therefore, the elements $\Gamma(s)$ are in the intersection of the symplectic and inhomogeneous groups,

$$\Gamma(s) \simeq Sp(2n+2) \cap \mathbb{Z}_2 \otimes_s \mathcal{IGL}(2n+1), \tag{3.8}$$

with

$$\Gamma(s)^t\, \zeta\, \Gamma(s) = \zeta, \quad \Gamma(s)^t\, \eta^\circ\, \Gamma(s) = \eta^\circ. \tag{3.9}$$

The group that this intersection defines is summarized in the following proposition.

**Proposition 3.1:** *Let $Sp(2n+2)$ be the real symplectic group that leaves invariant $\omega$, and let $\mathbb{Z}_2 \otimes_s \mathcal{IGL}(2n+1)$ be the extended inhomogeneous general linear group that leaves invariant the degenerate orthogonal metric $dt^2$. Let $\mathcal{HSp}(2n) \simeq Sp(2n) \otimes_s \mathcal{H}(n)$ be the Jacobi group, where $\mathcal{H}(n)$ is the Weyl-Heisenberg group. Then the extended Jacobi group $\mathbb{Z}_2 \otimes_s \mathcal{HSp}(2n)$ is the intersection,*

$$Sp(2n+2) \cap \mathbb{Z}_2 \otimes_s \mathcal{IGL}(2n+1) \simeq \mathbb{Z}_2 \otimes_s \mathcal{HSp}(2n). \tag{3.10}$$

*and it leaves invariant $\omega$ and $dt^2$. Furthermore, the Jacobi group $\mathcal{HSp}(2n)$ that leaves invariant $\omega$ and $dt$ is the intersection*

$$Sp(2n+2) \cap \mathcal{IGL}(2n+1) \simeq \mathcal{HSp}(2n). \tag{3.11}$$

The proof is through basic matrix algebra. Start with $\Gamma \in \mathbb{Z}_2 \otimes_s \mathcal{IGL}(2n+1)$ defined in (3.3) and require that it also leaves invariant $\zeta = \Gamma^t \zeta \Gamma$ with $\zeta$ defined in (3.6). This results in the condition,

$$\begin{pmatrix} \Sigma^{\circ t} \zeta^\circ \Sigma^\circ & \Sigma^{\circ t} \zeta^\circ b & \epsilon(\Sigma^{\circ t} \zeta^\circ w - c) \\ -(\Sigma^{\circ t} \zeta^\circ b)^t & b^t \zeta^\circ b & \epsilon(b^t \zeta^\circ w - a) \\ -\epsilon(\Sigma^{\circ t} \zeta^\circ w - c)^t & -(b^t \zeta^\circ w - \epsilon a) & w^t \zeta^\circ w \end{pmatrix} = \begin{pmatrix} \zeta^\circ & 0 & 0 \\ 0 & 0 & -1 \\ 0 & 1 & 0 \end{pmatrix}. \tag{3.12}$$

First, $\Sigma^{\circ t} \zeta^\circ \Sigma^\circ = \zeta^\circ$ implies that $\Sigma^\circ \in Sp(2n)$. Note that $w^t \zeta^\circ w \equiv 0$ and $b^t \zeta^\circ b \equiv 0$ as $\zeta^\circ$ is antisymmetric and $\Sigma^{\circ t} \zeta^\circ b \equiv 0$ implies that $b = 0$. The remaining terms imply that

$$a = \epsilon, \quad c = \Sigma^{\circ t} \zeta^\circ w. \tag{3.13}$$

Thus, $\Gamma$ has the form

$$\Gamma(\epsilon, \Sigma^\circ, w, r) = \begin{pmatrix} \Sigma^\circ & 0 & \epsilon w \\ -w^t \zeta^\circ \Sigma^\circ & \epsilon & \epsilon r \\ 0 & 0 & \epsilon \end{pmatrix}, \quad w \in \mathbb{R}^{2n}, \; r \in \mathbb{R}, \; \epsilon = \pm 1. \tag{3.14}$$

The group product and inverse are

$$\begin{aligned} \Gamma(\epsilon'', \Sigma^{\circ\prime\prime}, w'', r'') &= \Gamma(\epsilon', \Sigma^{\circ\prime}, w', r') \Gamma(\epsilon, \Sigma^\circ, w, r), \\ \Gamma^{-1}(\epsilon, \Sigma^\circ, w, r) &= \Gamma(\epsilon, \Sigma^{\circ -1}, -\epsilon \Sigma^{\circ -1} w, -r), \end{aligned} \tag{3.15}$$

where

$$\begin{aligned} \epsilon'' &= \epsilon' \epsilon, \quad \Sigma^{\circ\prime\prime} = \Sigma^{\circ\prime} \Sigma^\circ, \\ w'' &= \epsilon(\epsilon' w' + \Sigma^{\circ\prime} w), \\ r'' &= \epsilon(\epsilon' r + \epsilon' r' - w'^t \zeta^\circ \Sigma^{\circ\prime} w). \end{aligned} \tag{3.16}$$

The inner automorphism are

$$\Gamma(\epsilon'', \Sigma^{\circ\prime\prime}, w'', r'') = \zeta_{\Gamma(\epsilon', \Sigma^{\circ\prime}, w', r')} \Gamma(\epsilon, \Sigma^\circ, w, r), \tag{3.17}$$

where

$$\begin{aligned} \epsilon'' &= \epsilon, \quad \Sigma^{\circ\prime\prime} = \Sigma^{\circ\prime} \Sigma^\circ \Sigma^{\circ\prime -1}, \\ w'' &= \epsilon \left( w' + \epsilon' \Sigma^{\circ\prime} w - \Sigma^{\circ\prime} \Sigma^\circ \Sigma^{\circ\prime -1} w' \right), \\ r'' &= \epsilon r + \epsilon' w^t \zeta^\circ \Sigma^\circ \Sigma^{\circ\prime -1} w' + w'^t \zeta^\circ \Sigma^{\circ\prime} \Sigma^\circ \Sigma^{\circ\prime -1} w' - \epsilon \epsilon' w'^t \zeta^\circ \Sigma^{\circ\prime} w. \end{aligned} \tag{3.18}$$

The subgroup $\Delta_T(\epsilon) = \Gamma(\epsilon, 1_{2n}, 0, 0) \in \mathbb{Z}_2$ is the time reversal symmetry $dt \mapsto \pm dt$. The $\Gamma(\Sigma^\circ, w, r) = \Gamma(1, \Sigma^\circ, w, r)$ are elements of $\mathcal{HSp}(2n)$. It is a normal subgroup as

$$\zeta_{\Delta_T(\epsilon)} \Gamma(\Sigma^\circ, w, r) = \Delta_T(\epsilon) \Gamma(\Sigma^\circ, w, r) \Delta_T^{-1}(\epsilon) = \Gamma(\Sigma^\circ, \epsilon w, r). \tag{3.19}$$

Therefore the elements $\Gamma(\Sigma^\circ, w, r) \in \mathcal{HSp}(2n)$ leave Newtonian time invariant, $dt \mapsto dt$. The matrix realization of these elements is

$$\Gamma(\Sigma^\circ, w, r) = \begin{pmatrix} \Sigma^\circ & 0 & w \\ -w^t \zeta^\circ \Sigma^\circ & 1 & r \\ 0 & 0 & 1 \end{pmatrix}, \quad w \in \mathbb{R}^{2n}, \; r \in \mathbb{R}. \tag{3.20}$$

We can establish that $\mathcal{HSp}(2n)$ is the semidirect product of the symplectic and Weyl-Heisenberg group. Define

$$\Sigma^\circ \simeq \Gamma(\Sigma^\circ, 0, 0) \in Sp(2n), \quad \Upsilon(w, r) = \Gamma(1_{2n}, w, r) \in \mathcal{H}(n). \tag{3.21}$$

Using (3.18) the automorphisms of the Weyl-Heisenberg subgroup establish that it is a normal subgroup,

$$\zeta_{\Sigma^\circ} \Upsilon(w, r) = \Gamma(\Sigma^\circ, 0, 0)\, \Gamma(1_{2n}, w, r)\Gamma^{-1}(\Sigma^\circ, 0, 0) = \Upsilon(\Sigma^\circ w, r). \tag{3.22}$$

Finally, as $Sp(2n) \cap \mathcal{H}(n) \simeq e$ and $\Upsilon(w, r)\, \Sigma^\circ \simeq \Gamma(\Sigma^\circ, w, r)$, it is the semidirect product as stated. This completes the proof of the proposition.

As noted above, the elements of the semidirect product may be factored directly as

$$\Gamma(\Sigma^\circ, w, r) = \Gamma(1_{2n}, w, r)\, \Gamma(\Sigma^\circ, 0, 0) \simeq \Upsilon(w, r)\, \Sigma^\circ. \tag{3.23}$$

The matrix realization of these elements is

$$\Gamma(\Sigma^\circ, w, r) = \begin{pmatrix} \Sigma^\circ & 0 & w \\ -w^t\, \zeta^\circ\, \Sigma^\circ & 1 & r \\ 0 & 0 & 1 \end{pmatrix} = \begin{pmatrix} 1_{2n} & 0 & w \\ -w^t\, \zeta^\circ & 1 & r \\ 0 & 0 & 1 \end{pmatrix} \begin{pmatrix} \Sigma^\circ & 0 & 0 \\ 0 & 1 & 0 \\ 0 & 0 & 1 \end{pmatrix}. \tag{3.24}$$

We have established that the Jacobi group $\mathcal{H}Sp(2n)$ is the subgroup of the symplectic group $Sp(2n+2)$ that leaves Newtonian time invariant. The Jacobi group is in turn, the semidirect product of a symplectic group $Sp(2n)$ and a normal Weyl-Heisenberg subgroup $\mathcal{H}(n)$. Many readers are surprised to see a Weyl-Heisenberg group appearing under such basic assumptions in classical Hamiltonian mechanics; it is normally associated with quantum mechanics and the Heisenberg commutation relations. The Weyl-Heisenberg group is just a real, solvable, simply connected matrix group that is given by semidirect product of two abelian groups of the form $\mathcal{H}(n) \simeq \mathcal{A}(n) \otimes_s \mathcal{A}(n+1)$. Just as the abelian, orthogonal, general linear and other groups appear in many physical contexts, so also does the Weyl-Heisenberg group.

We can now state the following key result.

**Proposition 3.2:** *Consider a map $\varrho: \mathbb{P} \to \mathbb{P}$ that is a canonical transformation (i.e. symplectomorphism) that leaves invariant the symplectic metric, $\omega = \varrho^* \omega$, It also leaves invariant Newtonian time, $dt = \varrho^* dt$, if and only if its Jacobian matrix is an element of the Jacobi group,*

$$\left[\!\!\left[ \frac{\partial \varrho(s)}{\partial s} \right]\!\!\right] = \Gamma(\Sigma^\circ, w, r) \in \mathcal{H}Sp(2n) \subset Sp(2n+2). \tag{3.25}$$

The proof follows immediately from Proposition 3.1.

## 3.2  Jacobi group symmetry of Hamilton's equations

Nonrelativistic Hamiltonian mechanics assumes that time is absolute invariant Newtonian time. Another basic result is that Hamilton's equations are invariant under canonical transformations of momentum-position phase space. Furthermore, particle trajectories satisfying Hamilton's equations may be thought of as canonical transformations continuously evolving with invariant Newtonian time.

We show in this section that Proposition 3.2 embodies all of these results. That is, if we assume invariance of a symplectic metric $\omega$ on the phase space $\mathbb{P}$ and the invariance of Newtonian time, this leads directly to Hamilton's equations and canonical transforma-
$\mathbb{P}^\circ$

tions on $\mathbb{P}^\circ$ that leave Hamilton's equations invariant.

Let $\varrho : \mathbb{P} \to \mathbb{P}$ be a canonical transformation leaving Newtonian time invariant and hence satisfies the conditions of Proposition 3.2. The Jacobian matrix is given by the matrix realization of $\mathcal{H}Sp(2n)$ in (3.20) that may be written as a product of a matrix realization of the Weyl-Heisenberg group $\mathcal{H}(n)$ and the symplectic group $Sp(2n)$ using (3.23),

$$\left[\!\left[\frac{\partial \varrho(s)}{\partial s}\right]\!\right] = \Gamma(\Sigma^\circ, w, r) = \Gamma(1_{2n}, w, r)\,\Gamma(\Sigma^\circ, 0, 0). \tag{3.26}$$

Consequently, we can write $\varrho = \rho \circ \varphi$ where

$$\begin{aligned}
\varphi : \mathbb{P} \to \mathbb{P} : s \mapsto s' = \varphi(s) &\quad \left[\!\left[\frac{\partial \varphi(s)}{\partial s}\right]\!\right] = \Gamma(\Sigma^\circ(s), 0, 0), \\
\rho : \mathbb{P} \to \mathbb{P} : s' \mapsto s'' = \rho(s') &\quad \left[\!\left[\frac{\partial \rho(s')}{\partial s'}\right]\!\right] = \Gamma(1_{2n}, w(s'), r(s')).
\end{aligned} \tag{3.27}$$

and consistent with (3.26),

$$\left[\!\left[\frac{\partial \varrho(s)}{\partial s}\right]\!\right] = \left[\!\left[\frac{\partial \rho(\varphi(s))}{\partial s}\right]\!\right] = \left[\!\left[\frac{\partial \rho(s')}{\partial s'}\right]\!\right]\left[\!\left[\frac{\partial \varphi(s)}{\partial s}\right]\!\right]. \tag{3.28}$$

Consider first the symplectic subgroup with elements $\Gamma(\Sigma^\circ, 0, 0)$. We can expand out $s = (s^\circ, \varepsilon, t)$ with $s^\circ = (p, q) \in \mathbb{R}^{2n}$ and likewise label the component of $\varphi = (\varphi^\circ, \varphi^\varepsilon, \varphi^t)$. The Jacobian matrix is

$$\begin{pmatrix} \frac{\partial \varphi^\circ(s)}{\partial s^\circ} & \frac{\partial \varphi^\circ(s)}{\partial \varepsilon} & \frac{\partial \varphi^\circ(s)}{\partial t} \\ \frac{\partial \varphi^\varepsilon(s)}{\partial s^\circ} & \frac{\partial \varphi^\varepsilon(s)}{\partial \varepsilon} & \frac{\partial \varphi^\varepsilon(s)}{\partial t} \\ \frac{\partial \varphi^t(s)}{\partial s^\circ} & \frac{\partial \varphi^t(s)}{\partial \varepsilon} & \frac{\partial \varphi^t(s)}{\partial t} \end{pmatrix} = \Gamma(\Sigma^\circ, 0, 0) = \begin{pmatrix} \Sigma^\circ & 0 & 0 \\ 0 & 1 & 0 \\ 0 & 0 & 1 \end{pmatrix}. \tag{3.29}$$

This is straightforwardly solved to give the functional dependency,

$$\varphi^\circ(s) = \varphi^\circ(s^\circ), \quad \frac{\partial \varphi^\circ(s)}{\partial s^\circ} = \Sigma(s^\circ) \in Sp(2n), \quad \varphi^\varepsilon(s) = \varphi^\varepsilon(\varepsilon) = \varepsilon, \quad \varphi^t(s) = \varphi^t(t) = t. \tag{3.30}$$

$\varphi^\circ$ are canonical transformations on the position-momentum phases space given by the symplectic manifold $(\mathbb{P}^\circ, \omega^\circ)$ where $\mathbb{P}^\circ \simeq \mathbb{R}^{2n} \subset \mathbb{P}$ and

$$\omega^\circ \simeq ds^{\circ t} \zeta^\circ ds^\circ = \delta_{i,j}\,dq^i \wedge dp^j. \tag{3.31}$$

$\varphi^\varepsilon, \varphi^t$ are the identity transforms on $\mathbb{P}\setminus\mathbb{P}^\circ$.

Next, consider the Weyl-Heisenberg subgroup $\Gamma(1_{2n}, w, r)$ where we again label the components $\rho = (\rho^\circ, \rho^\varepsilon, \rho^t)$,

$$\begin{pmatrix} \frac{\partial \rho^\circ(s)}{\partial s^\circ} & \frac{\partial \rho^\circ(s)}{\partial \varepsilon} & \frac{\partial \rho^\circ(s)}{\partial t} \\ \frac{\partial \rho^\varepsilon(s)}{\partial s^\circ} & \frac{\partial \rho^\varepsilon(s)}{\partial \varepsilon} & \frac{\partial \rho^\varepsilon(s)}{\partial t} \\ \frac{\partial \rho^t(s)}{\partial s^\circ} & \frac{\partial \rho^t(s)}{\partial \varepsilon} & \frac{\partial \rho^t(s)}{\partial t} \end{pmatrix} = \Gamma(1_{2n}, w, r) = \begin{pmatrix} 1_{2n} & 0 & w \\ -w^t \zeta^\circ & 1 & r \\ 0 & 0 & 1 \end{pmatrix}. \tag{3.32}$$

The zero and unit terms reduce the functional dependency to

$$\rho^\circ(s) = \rho^\circ(s^\circ, t) = s^\circ + \gamma(t),$$
$$\rho^\varepsilon(s) = \varepsilon + H(s^\circ, t), \qquad (3.33)$$
$$\rho^t(s) = \rho^t(t) = t.$$

In this expression, $\gamma : \mathbb{R} \to \mathbb{P}^\circ$ and $H : \mathbb{P}^\circ \otimes \mathbb{R} \to \mathbb{R}$ are functions that must satisfy the remaining expressions from the (3.32) relations,

$$\frac{\partial \rho^\circ(s)}{\partial t} = \frac{d\gamma(t)}{dt} = w, \quad \frac{\partial \rho^\varepsilon(s)}{\partial s^\circ} = \frac{\partial H(s^\circ, t)}{\partial s^\circ} = -w^t \zeta^\circ, \quad \frac{\partial \rho^\varepsilon(s)}{\partial t} = \frac{\partial H(s^\circ, t)}{\partial t} = r. \quad (3.34)$$

Solving for $w$ and equating,

$$\frac{d\gamma(t)}{dt} = w = \zeta^\circ \left(\frac{\partial H(s^\circ, t)}{\partial s^\circ}\right)^t, \quad r = \frac{\partial H(s^\circ, t)}{\partial t}. \qquad (3.35)$$

These are Hamilton's equations in matrix form!

The diffeomorphism $\varrho$ that is given in (3.27) in terms of an element of the full Jacobi group is $\varrho = \rho \circ \varphi$. That is, it is the composition of a canonical transformation $\varphi$ on $\mathbb{P}^\circ$ with the $\rho$ diffeomorphism that is the solution of Hamilton's equations,

$$\varrho^\circ(s) = \rho^\circ(\varphi^\circ(s^\circ), t) = \varphi^\circ(s^\circ) + \gamma(t),$$
$$\varrho^\varepsilon(s) = \varepsilon + H(\varphi^\circ(s^\circ), t), \qquad (3.36)$$
$$\varrho^t(s) = \rho^t(t) = t.$$

The Jacobian $\left[\!\left[\frac{\partial \varrho(s)}{\partial s}\right]\!\right] = \Gamma(\Sigma^\circ, w, r)$ with $\tilde{s}^\circ = \varphi^\circ(s^\circ)$ results in the relations

$$\frac{\partial \varrho^\circ(s)}{\partial t} = \frac{d\gamma(t)}{dt} = w, \quad \frac{\partial \varrho^\varepsilon(s)}{\partial s^\circ} = \frac{\partial H(\tilde{s}^\circ, t)}{\partial \tilde{s}^\circ} \frac{\partial \varphi^\circ(s^\circ)}{\partial s^\circ} = \frac{\partial H(\tilde{s}^\circ, t)}{\partial \tilde{s}^\circ} \Sigma^\circ = -w^t \zeta^\circ \Sigma^\circ. \quad (3.37)$$

The $\Sigma^\circ$ cancel and consequently, Hamilton's equations continue to hold.

Alternatively, we can transform with the $\rho$ diffeomorphism first, $\tilde{s}^\circ = \rho(s^\circ)$ and then compose with the canonical transformation, $\tilde{\varrho} = \varphi \circ \rho$. This corresponds to the matrix group

$$\left[\!\left[\frac{\partial \tilde{\varrho}(s)}{\partial s}\right]\!\right] = \tilde{\Gamma}(\Sigma^\circ, w, r) = \Gamma(\Sigma^\circ, 0, 0)\, \Gamma(1_{2n}, w, r) = \begin{pmatrix} \Sigma^\circ & 0 & \Sigma^\circ w \\ -w^t \zeta^\circ & 1 & r \\ 0 & 0 & 1 \end{pmatrix}. \qquad (3.38)$$

This results in the relations

$$\frac{\partial \rho^\circ(s)}{\partial t} = \frac{d\gamma(t)}{dt} = \Sigma^\circ w, \quad \frac{\partial \varrho^\varepsilon(s)}{\partial s^\circ} = \frac{\partial H(\tilde{s}^\circ, t)}{\partial \tilde{s}^\circ} \frac{\partial \varphi^\circ(s^\circ)}{\partial s^\circ} = \frac{\partial H(\tilde{s}^\circ, t)}{\partial \tilde{s}^\circ} \Sigma^\circ = -w^t \zeta^\circ. \quad (3.39)$$

Solving for $w$ gives

$$\Sigma^{\circ\,-1}\frac{d\gamma(t)}{dt} = w = \zeta^\circ \Sigma^{\circ t}\left(\frac{\partial H(\tilde{s}^\circ, t)}{\partial \tilde{s}^\circ}\right)^t, \quad r = \frac{\partial H(\tilde{s}^\circ, t)}{\partial t}. \qquad (3.40)$$

and as $\Sigma^\circ \zeta^\circ \Sigma^{\circ t} = \zeta^\circ$, Hamilton's equations result in the transformed coordinates,

$$\frac{d\gamma(t)}{dt} = w = \zeta^\circ \left(\frac{\partial H(\tilde{s}^\circ, t)}{\partial \tilde{s}^\circ}\right)^t, \quad r = \frac{\partial H(\tilde{s}^\circ, t)}{\partial t}. \qquad (3.41)$$

The above results are summarized in the proposition

**Proposition 3.3:** *Consider a map $\varrho : \mathbb{P} \to \mathbb{P}$ from Proposition 3.2 that is a canonical transformation (i.e. symplectomorphism) that leaves invariant the symplectic metric, $\omega = \varrho^* \omega$ and Newtonian time, $dt = \varrho^* dt$. These maps can be written as $\varrho = \rho \circ \varphi$ where $\rho$ is a solution of Hamilton's equations and $\varphi$ is a canonical transformation leaving invariant $\omega^\circ$ on position-momentum phase space $\mathbb{P}^\circ \subset \mathbb{P}$, $\mathbb{P}^\circ \simeq \mathbb{R}^{2n}$, $\varphi^* \omega^\circ = \omega^\circ$. In local coordinates, these maps have Jacobian matrices satisfying*

$$\left[\frac{\partial \rho(s)}{\partial s}\right] = \Gamma(1_{2n}, w, r) \in \mathcal{H}(n) \subset \mathcal{H}Sp(2n),$$
$$\left[\frac{\partial \varphi(s)}{\partial s}\right] = \Gamma(\Sigma^\circ, 0, 0) \in Sp(2n) \subset \mathcal{H}Sp(2n). \tag{3.42}$$

The proof follows immediately from Propositions 3.1-2 and the above discussion.

## Position and momentum coordinates

The above results can be further expanded by setting $s^\circ = (p, q)$, $p, q \in \mathbb{R}^n$, $\gamma = (\gamma_p, \gamma_q)$ and $w = (f, v)$, $f, v \in \mathbb{R}^n$. $v$ is velocity, the rate of change of position with time, $f$ is force, the rate of change of momentum with time, and $r \in \mathbb{R}$ is power, the rate of change of energy with time.

Hamilton's equations (3.35) then take the familiar form

$$\frac{d\gamma_q(t)}{dt} = v = \left(\frac{\partial H(p, q, t)}{\partial p}\right)^t, \quad \frac{d\gamma_p(t)}{dt} = f = -\left(\frac{\partial H(p, q, t)}{\partial q}\right)^t, \quad r = \frac{\partial H(p, q, t)}{\partial t} \tag{3.43}$$

Setting $p(t) = \gamma_p(t)$, $q(t) = \gamma_q(t)$ and making indices explicit, this is Hamilton's equations in their most basic form,

$$\frac{dq^i(t)}{dt} = v^i = \frac{\partial H(p, q, t)}{\partial p^i}, \quad \frac{dp^i(t)}{dt} = f^i = -\frac{\partial H(p, q, t)}{\partial q^i}, \quad r = \frac{\partial H(p, q, t)}{\partial t}. \tag{3.44}$$

The transformation equations (3.33) are

$$\rho^t(s) = \rho^t(t) = t,$$
$$\rho^q(s) = \rho^q(q, t) = q + \gamma_q(t),$$
$$\rho^\varepsilon(s) = \rho^\varepsilon(q, t, \varepsilon, p) = \varepsilon + H(q, p, t),$$
$$\rho^p(s) = \rho^p(p, t) = p + \gamma_p(t). \tag{3.45}$$

The transformation of a basis of the cotangent space $T_s^* \mathbb{P}$ is

$$d\tilde{t} = dt,$$
$$d\tilde{q} = dq + \frac{d\gamma_q(t)}{dt} dt,$$
$$d\tilde{\varepsilon} = d\varepsilon + \frac{\partial H(p,q,t)}{\partial p} \cdot dp + \frac{\partial H(p,q,t)}{\partial q^i} \cdot dq + \frac{\partial H(p,q,t)}{\partial t} dt,$$
$$d\tilde{p} = dp^i + \frac{d\gamma_p(t)}{dt} dt. \tag{3.46}$$

Applying Hamilton's equations, this is

$$\begin{aligned} d\tilde{t} &= dt, \\ d\tilde{q} &= dq + v\,dt, \\ d\tilde{\varepsilon} &= d\varepsilon + v\cdot dp - f\cdot dq + r\,dt, \\ d\tilde{p} &= dp + f\,dt. \end{aligned} \qquad (3.47)$$

In the energy term, $v\cdot dp$ is the 'kinetic' term, $-f\cdot dp$ is the 'work' term and $r\,dt$ is the power term that we expect from basic nonrelativistic mechanics.

It is convenient at this point to change the basis ordering to $s = (t, q, \varepsilon, p)$, where $t, \varepsilon \in \mathbb{R}$ and $q, p \in \mathbb{R}^n$ as originally defined in (2.19). We will use this ordering for the remainder of the paper.

In this ordering of the basis, the transformation equations (3.47) may be written in matrix form as

$$\begin{pmatrix} d\tilde{t} \\ d\tilde{q} \\ d\tilde{\varepsilon} \\ d\tilde{p} \end{pmatrix} = \begin{pmatrix} 1 & 0 & 0 & 0 \\ v & 1 & 0 & 0 \\ r & -f & 1 & v \\ f & 0 & 0 & 1 \end{pmatrix} \begin{pmatrix} dt \\ dq \\ d\varepsilon \\ dp \end{pmatrix}. \qquad (3.48)$$

This may be compactly written as $d\tilde{s} = \Upsilon(v, f, r)\,ds$ where the matrix realization of $\Upsilon(v, f, r) \in \mathcal{H}(n)$, where $\mathcal{H}(n)$ is the Weyl-Heisenberg group, is

$$\Upsilon(v, f, r) = \begin{pmatrix} 1 & 0 & 0 & 0 \\ v & 1 & 0 & 0 \\ r & -f & 1 & v \\ f & 0 & 0 & 1 \end{pmatrix}. \qquad (3.49)$$

This is the same as (3.32) with $w$ expanded out as $(v, f)$ and the basis ordering permuted to $s = (t, q, \varepsilon, p)$. In this notation, the group composition and inverse for the Weyl Heisenberg group take the form,

$$\Upsilon(v'', f'', r'') = \Upsilon(v', f', r')\,\Upsilon(v, f, r), \qquad (3.50)$$

where

$$v'' = v' + v,\; f'' = f' + f,\; r'' = r' + r + f\cdot v' - v\cdot f'. \qquad (3.51)$$

The inverse and identity are

$$\Upsilon^{-1}(v, f, r) = \Upsilon(-v, -f, -r),\; e \simeq \Upsilon(0, 0, 0). \qquad (3.52)$$

### Dimensioned phase space coordinates

The phase space coordinates are dimensioned physical quantities as we reviewed in Section 2.2. We have chosen natural scales $c = b = \hbar = 1$ in our definition $s = (t, q, \varepsilon, p)$ up to this point in this section. We now make these scales explicit by setting $s = \left(\frac{t}{\lambda_t}, \frac{q}{\lambda_q}, \frac{\varepsilon}{\lambda_\varepsilon}, \frac{p}{\lambda_p}\right)$ as given in (2.19) where $\{\lambda_t, \lambda_q, \lambda_\varepsilon, \lambda_p\}$ are defined in (2.15). $s$ is dimensionless whereas $\{t, q, \varepsilon, p\}$ are dimensioned. Likewise, velocity force and power scale as $c$, $b$ and $c\,b$ so that $\left\{\frac{v}{c}, \frac{f}{b}, \frac{r}{c\,b}\right\}$ are dimensionless quantities.

The mathematical formulation is expressed in terms of dimensionless quantities that includes the coordinates of the phase space and the parameters of the groups.

The group elements of $\mathcal{H}(n)$ in terms of $\{v, f, r\}$ that are dimensional quantities is

$$\Upsilon\left(\frac{v}{c}, \frac{f}{b}, \frac{r}{cb}\right) = \begin{pmatrix} 1 & 0 & 0 & 0 \\ \frac{v}{c} & 1 & 0 & 0 \\ \frac{r}{cb} & -\frac{f}{b} & 1 & \frac{v}{c} \\ \frac{f}{b} & 0 & 0 & 1 \end{pmatrix} \in \mathcal{H}(n). \qquad (3.53)$$

Furthermore, a basis of the cotangent space of $\mathbb{P}$ is $ds = \left(\frac{dt}{\lambda_t}, \frac{dq}{\lambda_q}, \frac{d\varepsilon}{\lambda_\varepsilon}, \frac{dp}{\lambda_p}\right)$. Consequently the transformation equations $d\tilde{s} = \Upsilon\, ds$ are

$$\begin{pmatrix} \frac{d\tilde{t}}{\lambda_t} \\ \frac{d\tilde{q}}{\lambda_q} \\ \frac{d\tilde{\varepsilon}}{\lambda_\varepsilon} \\ \frac{d\tilde{p}}{\lambda_p} \end{pmatrix} = \begin{pmatrix} 1 & 0 & 0 & 0 \\ \frac{v}{c} & 1 & 0 & 0 \\ \frac{r}{cb} & -\frac{f}{b} & 1 & \frac{v}{c} \\ \frac{f}{b} & 0 & 0 & 1 \end{pmatrix} \begin{pmatrix} \frac{dt}{\lambda_t} \\ \frac{dq}{\lambda_q} \\ \frac{d\varepsilon}{\lambda_\varepsilon} \\ \frac{dp}{\lambda_p} \end{pmatrix}. \qquad (3.54)$$

Collecting the scaling terms $\lambda_\alpha$ and using the relations (2.14) that $\frac{\lambda_q}{\lambda_t} = c = \frac{\lambda_\varepsilon}{\lambda_p}$ and $\frac{\lambda_p}{\lambda_t} = b = \frac{\lambda_\varepsilon}{\lambda_q}$ this simplifies to

$$\begin{pmatrix} d\tilde{t} \\ d\tilde{q} \\ d\tilde{\varepsilon} \\ d\tilde{p} \end{pmatrix} = \begin{pmatrix} \frac{\lambda_t}{\lambda_t} & 0 & 0 & 0 \\ \frac{\lambda_q}{\lambda_t}\frac{v}{c} & \frac{\lambda_q}{\lambda_q} & 0 & 0 \\ \frac{\lambda_\varepsilon}{\lambda_t}\frac{r}{cb} & -\frac{\lambda_\varepsilon}{\lambda_q}\frac{f}{b} & \frac{\lambda_\varepsilon}{\lambda_\varepsilon} & \frac{\lambda_\varepsilon}{\lambda_p}\frac{v}{c} \\ \frac{\lambda_p}{\lambda_t}\frac{f}{b} & 0 & 0 & \frac{\lambda_p}{\lambda_p} \end{pmatrix} \begin{pmatrix} dt \\ dq \\ d\varepsilon \\ dp \end{pmatrix} = \begin{pmatrix} 1 & 0 & 0 & 0 \\ v & 1 & 0 & 0 \\ r & -f & 1 & v \\ f & 0 & 0 & 1 \end{pmatrix} \begin{pmatrix} dt \\ dq \\ d\varepsilon \\ dp \end{pmatrix}, \qquad (3.55)$$

that is precisely (3.48).

The point is that in Hamilton's mechanics, the scaling always works out so that the scales $c, b$ do not appear explicitly. Of course, this is not true in special relativity where $c$ has an important explicit role. Nonrelativistic symmetry is the limit $c \to \infty$ and so any terms explicitly depending on $c$ are scaled out as we reviewed in Chapter 7 in [24]. We will show later in this paper that this is also true for the scale constant $b$.

## 3.3  Inertial Galilean relativity transformations

The case $f \equiv f' \equiv 0$, $r \equiv r' \equiv 0$ is the inertial case where the velocities $v, v'$ are constants. The group composition (3.50-52) shows that velocities are elements of the abelian group $\mathcal{A}(n) \subset \mathcal{H}(n)$,

$$\begin{aligned}\Upsilon(v'', 0, 0) &= \Upsilon(v', 0, 0)\,\Upsilon(v, 0, 0) = \Upsilon(v' + v, 0, 0), \\ \Upsilon^{-1}(v, 0, 0) &= \Upsilon(-v, 0, 0).\end{aligned} \qquad (3.56)$$

The cotangent basis transformation equations for the inertial case reduce to

$$d\tilde{t} = dt, \qquad d\tilde{\varepsilon} = d\varepsilon + v \cdot dp,$$
$$d\tilde{q} = dq + v\,dt, \quad d\tilde{p} = dp. \qquad (3.57)$$

Spacetime $\mathbb{M}$, with coordinates $(t, q)$, and energy-momentum space $\tilde{\mathbb{M}}$, with coordinates $(\varepsilon, p)$, are both invariant submanifolds under these transformations and consequently $\mathbb{P} = \mathbb{M} \otimes \tilde{\mathbb{M}}$. These equations, of course, are the expected velocity transformations on $T^*\mathbb{M}$ and on $T^*\tilde{\mathbb{M}}$ for Galilean relativity.

Galilean relativity also has a rotational symmetry due to the isotropy of spacetime and energy-momentum space. The transformations including rotations are

$$d\tilde{t} = dt, \qquad d\tilde{\varepsilon} = d\varepsilon + v \cdot dp,$$
$$d\tilde{q} = R\,dq + v\,dt, \quad d\tilde{p} = R\,dp, \qquad (3.58)$$

where $R \in SO(n)$. This is expressed in terms of the coordinates $s$ of the phase space $\mathbb{P}$ as $d\tilde{s} = \Gamma^\circ(R, v)\,ds$ where

$$\Gamma^\circ(R, v) = \begin{pmatrix} \Lambda^\circ(R, v) & 0 \\ 0 & \tilde{\Lambda}^\circ(R, v) \end{pmatrix}. \qquad (3.59)$$

We can set $s = (x, y)$ with $x = \left(\frac{t}{\lambda_t}, \frac{q}{\lambda_q}\right) \in \mathbb{R}^{n+1} \simeq \mathbb{M}$ and $y = \left(\frac{\varepsilon}{\lambda_\varepsilon}, \frac{p}{\lambda_p}\right) \in \mathbb{R}^{n+1} \simeq \tilde{\mathbb{M}}$ so that

$$d\tilde{s} = \Gamma^\circ(R, v)\,ds, \quad d\tilde{x} = \Lambda^\circ(R, v)\,dx, \quad d\tilde{y} = \tilde{\Lambda}^\circ(R, v)\,dy, \qquad (3.60)$$

where

$$\Lambda^\circ(R, v) = \begin{pmatrix} 1 & 0 \\ v & R \end{pmatrix}, \quad \tilde{\Lambda}^\circ(R, v) = \begin{pmatrix} 1 & v^t \\ 0 & R \end{pmatrix}. \qquad (3.61)$$

The elements $\Gamma^\circ(R, v)$, $\Lambda^\circ(R, v)$ and $\tilde{\Lambda}^\circ(R, v)$ are matrix realizations of elements of the Euclidean group $\mathcal{E}(n)$. These inertial transformations are the small velocity limit $\frac{v}{c} \mapsto 0$ (or $c \to \infty$) of the Lorentz transformations of special relativity for inertial states on spacetime and energy-momentum space that we reviewed in Chapter 7 in [24]. The group transformation and inverse are

$$\Gamma^\circ(R', v')\,\Gamma^\circ(R, v) = \Gamma^\circ(R'R, v' + R'v), \quad \Gamma^\circ(R, v)^{-1} = \Gamma^\circ(R^{-1}, -R^{-1}v), \qquad (3.62)$$

and likewise for $\Lambda^\circ(R, v)$ and $\tilde{\Lambda}^\circ(R, v)$.

## 3.4  The Hamilton group

The symplectic group $Sp(2n, \mathbb{R})$ is a symmetry of the symplectic manifold $(\mathbb{P}^\circ, \omega^\circ)$ as defined in (3.31). That is, an element $\Sigma^\circ \in Sp(2n, \mathbb{R})$ acts on the basis $ds^\circ = (dq, dp)$ of the cotangent space $T^*\mathbb{P}^\circ$, $d\tilde{s}^\circ = \Sigma^\circ\,ds^\circ$. The orthogonal group $O(n)$ is the subgroup of the symplectic group $Sp(2n, \mathbb{R})$ that also leaves invariant the two degenerate orthogonal metrics on $\mathbb{P}^\circ$,

$$dq^2 = ds^{\circ t}\,\eta^\circ_q\,ds^\circ, \quad dp^2 = ds^{\circ t}\,\eta^\circ_p\,ds^\circ,$$

where $\eta^\circ_p$, $\eta^\circ_q$ are the $2n$ dimensional square matrices

$$\eta^\circ_q = \begin{pmatrix} 1_n & 0 \\ 0 & 0 \end{pmatrix}, \quad \eta^\circ_p = \begin{pmatrix} 0 & 0 \\ 0 & 1_n \end{pmatrix}. \qquad (3.63)$$

That is, symplectic transformations $d\tilde{s}^\circ = \Sigma^\circ \, ds^\circ$ that leave these metrics invariant

$$\Sigma^{\circ t} \eta_p^\circ \Sigma^\circ = \eta_p^\circ, \quad \Sigma^{\circ t} \eta_q^\circ \Sigma^\circ = \eta_q^\circ, \tag{3.64}$$

in addition to the symplectic metric $\omega^\circ$ have the form

$$\Sigma^\circ(\tilde{R}) = \begin{pmatrix} \tilde{R} & 0 \\ 0 & \tilde{R} \end{pmatrix}, \quad \tilde{R} \in O(n). \tag{3.65}$$

We further restrict the group to the connected component that is the special orthogonal group $SO(n)$. This is the normal subgroup defined by the kernel of the determinant,

$$\text{Det} : O(n) \to \mathbb{Z}_2, \quad \ker(\text{Det}) \simeq SO(n). \tag{3.66}$$

(We will return to full orthogonal group and its discrete symmetries in the following section.) The special orthogonal subgroup of the symplectic group is

$$\Sigma^\circ(R) = \begin{pmatrix} R & 0 \\ 0 & R \end{pmatrix}, \quad \text{Det } R = 1, \quad R \in SO(n). \tag{3.67}$$

The Hamilton group $\mathcal{H}a(n)$ is the name we give to the subgroup of the Jacobi group $\mathcal{H}Sp(2n)$ where the symplectic group is restricted to the special orthogonal group. It follows immediately that it is a semidirect product of the form

$$\mathcal{H}a(n) = SO(n) \otimes_s \mathcal{H}(n). \tag{3.68}$$

The semidirect product structure of the Hamilton group enables its elements, that we write as $\Gamma(R, v, f, r) \in \mathcal{H}a(n)$, to be factored as

$$\Gamma(R, v, f, r) = \Gamma(R, v, f, r) \, \Gamma(R, 0, 0, 0) = \Upsilon(v, f, r) \, \Sigma^\circ(R), \tag{3.69}$$

where $\Upsilon(v, f, r) \in \mathcal{H}(n)$ and $\Sigma^\circ(R) \in SO(n)$. Using (3.49) and (3.67), this expand out as the matrices

$$\Gamma(R, v, f, r) = \begin{pmatrix} 1 & 0 & 0 & 0 \\ v & 1_n & 0 & 0 \\ r & -f^t & 1 & v^t \\ f & 0 & 0 & 1_n \end{pmatrix} \begin{pmatrix} 1 & 0 & 0 & 0 \\ 0 & R & 0 & 0 \\ 0 & 0 & 1 & 0 \\ 0 & 0 & 0 & R \end{pmatrix} = \begin{pmatrix} 1 & 0 & 0 & 0 \\ v & R & 0 & 0 \\ r & -f^t R & 1 & v^t R \\ f & 0 & 0 & R \end{pmatrix}. \tag{3.70}$$

The group product and inverse may be computed to be

$$\Gamma(R'', v'', f'', r'') = \Gamma(R', v', f', r') \, \Gamma(R, v, f, r) = \Gamma(R'R, v' + Rv, f' + Rf,$$
$$\Gamma(R, v, f, r)^{-1} = \Gamma(R^{-1}, -Rv, -Rf, -r). \tag{3.71}$$

and the inner automorphisms are

$$\varsigma_{\Gamma(R', v', f', r')} \Gamma(R, v, f, r) = \Gamma\big(\varsigma_{R'} R, \, R'v + \tilde{R} v',$$
$$R' f + \tilde{R} f', \, r - f' \tilde{R} v + v' \tilde{R} f + f^t RR^{-1} v' - v^t RR^{-1} f'\big), \tag{3.72}$$

where $\tilde{R} = 1_n - \varsigma_{R'} R$.

Elements of the Hamilton group act on the cotangent space $T^*\mathbb{P}$, $d\tilde{s} = \Gamma(R, v, f, r) \, ds$ and this expands out as the noninertial transformations (3.58).

The Euclidean group $\mathcal{E}(n)$ defined in (3.62) is the subgroup of the Hamilton group $\mathcal{H}a(n)$ where force and power are zero, $\Gamma^\circ(R, v) = \Gamma(R, v, 0, 0)$. This is the homogeneous group of Galilean symmetry for the inertial case (3.57).

### Discrete symmetries

Proposition 3.1 states that the extended Jacobi group $\mathbb{Z}_2 \otimes_s \mathcal{H}Sp(2n)$ leaves invariant the symplectic metric $\omega$ and the degenerate orthogonal metric that is the Newtonian line element $dt^2$. The discrete $\mathbb{Z}_2$ group is time reversal and energy sign change,

$$\Delta_T(dt) = -dt, \quad \Delta_T(d\varepsilon) = -d\varepsilon, \quad \Delta_T(dq) = dq, \quad \Delta_T(dp) = dp. \tag{3.73}$$

The extended Hamilton group leaves invariant the symplectic 2 form $\omega$, the Newtonian time line element $dt^2$ as well as the position length $dq^2$ and momentum length $dp^2$. The extended Hamilton group also has time reversal and energy sign change symmetry. In the basis ordering $ds = (dt, dq, d\varepsilon, dp)$, this $\mathbb{Z}_2$ group is realized by the matrices

$$\Delta_e = 1_{2n+2}, \quad \Delta_T = \begin{pmatrix} \eta & 0 \\ 0 & \eta \end{pmatrix}, \quad \eta = \begin{pmatrix} -1 & 0 \\ 0 & 1_n \end{pmatrix}. \tag{3.74}$$

In addition, the orthogonal group (3.64) has a discrete $\mathbb{Z}_2$ symmetry. For $n$ odd, this is the direct product $O(n) = \mathbb{Z}_2 \otimes SO(n)$ where the $\mathbb{Z}_2$ symmetry is parity,

$$\Delta_P(dt) = dt, \quad \Delta_P(d\varepsilon) = d\varepsilon, \quad \Delta_P(dq) = -dq, \quad \Delta_P(dp) = -dp. \tag{3.75}$$

In this basis, this parity $\mathbb{Z}_2$ group is realized by the matrices,

$$\Delta_e = 1_{2n+2}, \quad \Delta_P = \begin{pmatrix} -\eta & 0 \\ 0 & -\eta \end{pmatrix}. \tag{3.76}$$

The product of these two $\mathbb{Z}_2$ symmetries is the $\mathbb{Z}_{2,2}$ discrete symmetries of time-energy reversal, parity and the combination, parity-time-energy reversal $\{\Delta_e, \Delta_T, \Delta_P, \Delta_{PT}\}$

$$\Delta_{PT}(dt) = -dt, \quad \Delta_{PT}(d\varepsilon) = -d\varepsilon, \quad \Delta_{PT}(dq) = -dq, \quad \Delta_{PT}(dp) = -dp, \tag{3.77}$$

where $\Delta_{PT} = \Delta_P \Delta_T = \Delta_T \Delta_P$ and its matrix representation is $\Delta_{PT} = -1_{2n+2}$. These discrete symmetries are automorphisms of the Hamilton group. A matrix computation gives

$$\varsigma_{\Delta_T} \Gamma(R, v, f, r) = \Gamma(R, -v, -f, -r), \quad \varsigma_{\Delta_P} \Gamma(R, v, f, r) = \Gamma(R, -v, -f, r),$$
$$\varsigma_{\Delta_{PT}} \Gamma(R, v, f, r) = \Gamma(R, v, f, -r). \tag{3.78}$$

### The extended Hamilton group

The extended Hamilton group $\mathcal{H}O(n) \simeq \mathbb{Z}_{2,2} \otimes_s \mathcal{H}a(n)$ is the subgroup of the Jacobi group $\mathcal{H}Sp(2n)$. We can use the notation $\Delta(\epsilon, \tilde{\epsilon})$ with

$$\Delta_e = \Delta(1, 1), \quad \Delta_T = \Delta(-1, 1), \quad \Delta_P = \Delta(1, -1), \quad \Delta_{PT} = \Delta(-1, -1). \tag{3.79}$$

The extended Hamilton group has elements

$$\Gamma(\epsilon, \tilde{\epsilon}, R, v, f, r) = \Gamma(R, v, f, r) \Delta(\epsilon, \tilde{\epsilon}). \tag{3.80}$$

The matrix realization is

$$\Gamma(\epsilon, \tilde{\epsilon}, R, v, f, r) = \begin{pmatrix} \epsilon & 0 & 0 & 0 \\ \epsilon v & \tilde{\epsilon} R & 0 & 0 \\ \epsilon r & -\tilde{\epsilon} f^t R & \epsilon & \tilde{\epsilon} v^t R \\ \epsilon f & 0 & 0 & \tilde{\epsilon} R \end{pmatrix}. \tag{3.81}$$

The group product and inverse for the extended group is straightforwardly computed.

## 3.5   Noninertial relativity

Galilean relativity addresses how clocks and lengths are measured by observers in inertial physical states with uniform relative velocity. On the phase space $\mathbb{P}$ it is described by the elements $\Gamma^\circ(R, v) = \Gamma(R, v, 0, 0)$ of Euclidean subgroup $\mathcal{E}(n)$ of the Hamilton group $\mathcal{H}a(n)$ for which $f = r = 0$. This Euclidean group is the small velocity $\frac{v}{c} \mapsto 0$ or $c \to \infty$ limit of the Lorentz group of special relativity. The Euclidean inertial transformations of a basis of the cotangent space $T^*\mathbb{P}$, $d\tilde{s} = \Gamma^\circ(R, v)\, ds$, is given in (3.47) and we repeat them here as they are key to this discussion,

$$d\tilde{t} = dt, \qquad d\tilde{\varepsilon} = d\varepsilon + v \cdot dp, \qquad (3.82)$$
$$d\tilde{q} = R\, dq + v\, dt, \quad d\tilde{p} = R\, dp.$$

The Galilean inertial relativity has the following properties:

- Time is Newtonian time that is invariant (i.e. absolute) in the sense that observers in any inertial physical state observe the same rate of passage of time $dt$. Consequently the time submanifold $t \in \mathbb{R} \subset \mathbb{P}$ is an invariant submanifold under the transformations (3.82).
- Simultaneity is absolute in the sense that observers in any inertial physical state agree on the ordering of events.
- Rates of change of position with respect to Newtonian time (i.e velocity) are simply additive and consequently unbounded.
- There is an absolute rest state that all observers in inertial states agree on.
- There is an absolute inertial state that all observers in inertial states agrees on.
- Information can propagate instantaneously in position space. That is, there are no causal cone (i.e. light cones) constraining the rate of propagation of information in position space. 'Instantaneous action at a distance' in position space is possible.
- Spacetime with coordinates $(t, q)$ is an invariant subspace of the phase space $\mathbb{P}$ that all observers in inertial physical states agree on as described in (3.82).

Particle states in Hamilton mechanics are generally not inertial. Rather, they satisfy Hamilton's equations for which the rate of change of position, momentum and energy are generally nonzero. (i.e. velocity $v \neq 0$, force $f \neq 0$ and $r \neq 0$.) Hamilton's equations have the symmetry of the Jacobi group in which position and momentum appear on completely equal footing.

The Hamilton group is the subgroup of the Jacobi group for which position and momentum lengths are invariant in inertial the rest frame and it is the symmetry group for Hamiltonian noninertial relativity. As with Galilean inertial relativity, Hamiltonian relativity is valid only in the regime of small rates of change of position relative to the universal scale $c$ (i.e. small velocity $\frac{v}{c} \mapsto 0$ or $c \to \infty$.) We shall see in the sections that follow that it is also valid only in the regime of small rates of change of momentum relative to a universal scale $b$ (i.e. small force $\frac{f}{b} \mapsto 0$ or $b \to \infty$.) As was the case above for the Euclidean group, we will also see that the Hamiltonian group is the contraction limit of a more general noninertial relativity group valid in regimes where $v$ is not small relative to $c$ and $f$ is not small relative to $b$. (These more general relativity symmetries that are valid outside of this limiting regime will be the topic of the remaining sections

of this paper.)

The Hamiltonian noninertial transformations of the basis of the cotangent space $T^*\mathbb{P}$, $d\tilde{s} = \Gamma^\circ(R, v, f, r)\, ds$, are given in (3.47) and we repeat them here as they are key to this discussion,

$$d\tilde{t} = dt, \qquad d\tilde{\varepsilon} = d\varepsilon + v \cdot dp - f \cdot dq + r\, dt, \qquad (3.83)$$
$$d\tilde{q} = R\, dq + v\, dt, \quad d\tilde{p} = R\, dp + f\, dt.$$

The Hamiltonian noninertial relativity has the following properties:

- Time is Newtonian time that is invariant (i.e. absolute) such that observers in *any* (inertial or noninertial) physical state observe the same rate of passage of time $dt$. This is a defining property of the Jacobi group, of which the Hamilton group is a subgroup, and consequently the time submanifold $t \in \mathbb{R} \subset \mathbb{P}$ is an invariant submanifold under the transformations (3.83).

- Simultaneity is absolute in the sense that observers in *any* physical state agree on the ordering of events.

- Rates of change of position with respect to Newtonian time (i.e. velocity) are simply additive and consequently unbounded.

- Rates of change of momentum with respect to Newtonian time (i.e. force) are simply additive and consequently unbounded.

- There is an absolute inertial state that all observers in *any* states agree on.

- There is an absolute rest state that all observers in *any* states agree on.

- Information can propagate instantaneously in position space. That is, there are no causal (i.e. light cones) constraining the rate of propagation of information in position space. 'Instantaneous action at a distance' in position space is possible.

- Information can propagate instantaneously in momentum space. That is, there are no 'causal cones' constraining the propagation of information in momentum space. 'Instantaneous action at a distance' in momentum space is possible.

- Position-time space (i.e spacetime) with coordinates $(t, q)$ is an invariant subspace of the phase space $\mathbb{P}$ that all observers in *any* physical states agree on as described in (3.83).

- Momentum-time space with coordinates $(t, p)$ is an invariant subspace of the phase space $\mathbb{P}$ that all observers in *any* physical states agree on as described in (3.83).

# 4 Noninertial relativity with Einstein proper time

## 4.1 Introduction

In the previous section, we reviewed that for the Hamilton formulation of nonrelativistic mechanics, Newton time may be defined by an invariant degenerate orthogonal metric $dt^2$ on the symplectic phase space $\mathbb{P}$. This lead directly to the Jacobi group and Hamilton's equations. The time subspace $t \in \mathbb{R} \subset \mathbb{P}$ is an invariant subspace. Physically, this invariant Newton time means that observers in any physical state, inertial or noninertial, measure the same rate of passage of time. That is, all clocks tick at the same rate.

Einstein's special and general relativity fundamentally altered our concept of time. In special relativity, the rate of passage of time depends on the relative velocity of the state and the inertial observer state. In special relativity, the spacetime is flat, $\mathbb{M} \simeq \mathbb{R}^{n+1}$ with coordinates $(t, q)$, $t \in \mathbb{R}$, $q \in \mathbb{R}^n$ and the invariant time is Einstein *proper* time $\tau$ that is defined by

$$d\tau^{\circ 2} = dt^2 - \frac{1}{c^2} dq^2 = dt^2 \left(1 - \frac{v^2}{c^2}\right). \tag{4.1}$$

The $d\tau^\circ$ is the rate clocks tick for an observer at inertial rest relative to the clock of interest. In this case, $v = 0$ and $dt = d\tau^\circ$. However, an observer in an inertial state sees the clock in another inertial state with a relative velocity $v$, ticking at the rate

$$dt = \frac{d\tau^\circ}{\sqrt{1 - \frac{v^2}{c^2}}}. \tag{4.2}$$

This, of course is the time dilatation formula that is fundamental to special relativity. While $dt$ is specific to the relative velocity of the inertial states, all inertial observers agree on the rate that the clock ticks for observers at inertial rest relative to it given by $d\tau^\circ$. In the limit of small velocities relative to $c$, $\frac{v}{c} \mapsto 0$, or equivalently, $c \to \infty$, Einstein proper time $d\tau^\circ$ contracts to Newton absolute time $dt$.

The proper time (4.1) defines an orthogonal metric with signature $(1, n)$ with $n = 3$ the usual physical case. Using natural scales with $c = 1$, and coordinates $x \in \mathbb{R}^{n+1}$ for $\mathbb{M}$ this is the Minkowski metric

$$d\tau^{\circ 2} = -\eta_{\mu,\nu} dx^\mu dx^\nu, \quad \eta = \begin{pmatrix} -1 & 0 \\ 0 & 1_n \end{pmatrix}, \quad \mu, \nu = 0, ..., n,$$

that is globally defined on $\mathbb{M}$.

This metric is invariant under the extended Lorentz group $O(1, n)$ for which the proper, orthochronous group $\mathcal{L}(1, n)$ is the connected normal subgroup. This was reviewed in Chapter 6 in [24].

Special relativity assumes that gravity is negligible or absent and so $\mathbb{M}$ is flat, $\mathbb{M} \simeq \mathbb{R}^{n+1}$. Mathematically, this is accomplished by setting the gravitational coupling constant to zero, $\alpha_N = 0$, in (2.16) and consequently $G_N = 0$.

General relativity incorporates gravity based on the equivalence principle. The equivalence principle states that a particle moving in an *apparent* noninertial state (i.e. accelerating) under the effect of gravity is equivalent to an inertial state on a certain curved spacetime manifold. That is, particle states (massive and massless) follow geodesics that are the inertial trajectories in the curved spacetime manifold that is defined by the Einstein gravitational field equations. These field equations may be solved to determine the metric $g(x)_{\mu,\nu}$ that defines Einstein invariant time,

$$d\tau^{\circ 2} = -g(x)_{\mu,\nu}\, dx^\mu\, dx^\nu,$$

At any point in the manifold $\mathbb{M}$, we can choose locally inertial coordinates $\tilde{x}^\mu$ so that this has the form,

$$d\tau^{\circ 2}\,|_p = -\eta_{\mu,\nu}\, d\tilde{x}^\mu\, d\tilde{x}^\nu\,|_p, \tag{4.3}$$

at the point $p$. This is the locally inertial frame for $T^*_p \mathbb{M}$ at the given point $p$. The Riemann connection enables us to parallel transport this to the locally inertial frame at a neighboring point. This is what the covariant derivative does in the geodesic equation; ensuring that the particle state remains locally inertial on the curved manifold. The rate at which Einstein proper time passes depends on the specific trajectory the particle state follows on the curved manifold. The Einstein field equations work out so that this is always the case; in a world where only gravity is acting, there are only locally inertial states - no particle state undergoes noninertial motion.

The curvature of the manifold may result in many interesting phenomena, one of the most celebrated being that the curvature may be so great that the null geodesics, describing the motion of massless particles such as photons, become closed loops defining the event horizons of black holes.

The world that we live in is not purely gravitational. Other forces exist such as the electrodynamic, weak and strong interactions. These forces cause particle states to which they couple to undergo noninertial motion. There has been almost a century of effort to describe these purely geometrically, but these efforts are yet to be successful.

So we can ask the question, what is the invariant proper time for a particle undergoing noninertial motion due to one of these non-gravitational forces. To be concrete, consider an electron undergoing corkscrew motion in a magnetic field; how does its clock tick?

In special relativity, the answer is straightforward, it continues to be given by the Minkowski line element (4.2) but now the velocity is not constant but rather determined by the electrodynamic equations. The time dilation (4.2) is based on the instantaneous velocity. Furthermore, the reverse is true. The observer in the electron's noninertial state sees the the corresponding time dilation for the inertial observer. Thus, in special relativity, we implicitly assume that the Einstein proper time is invariant for any state, inertial or noninertial.

If gravity is present, the same argument holds relative to the locally inertial gravitational states. For example, the electrodynamic field equations are now defined using the covariant derivative. The Riemann connection inherent in the covariant derivative defines the noninertial states due to the electromagnetic forces relative to the the locally inertial states in the curved manifold. But relative to these locally inertial states, the above argument holds and (4.3) defines the proper time for these noninertial states.

Consequently, in Einstein mechanics, the invariant of time is defined by the orthogonal line element $d\tau^{\circ 2}$ for states that are inertial or noninertial. This is the direct generalization of the invariance of Newton time $dt^2$. As we have noted above, locally in the $\mathbb{P}$

$\mathbb{M} \simeq \mathbb{R}^{n+1}$

$$d\tau^{\circ 2}$$

neighbourhood of some point in spacetime $\mathbb{P}$ coordinates always exist so that the line element has the form (4.3). If gravity is absent and so the spacetime manifold is flat, $\mathbb{M} \simeq \mathbb{R}^{n+1}$, the Minkowski metric (4.2) is global. Our focus is on the symmetry of noninertial states for which gravity is not central as its states are always locally inertial. Consequently we assume gravity is absent ($\alpha_N = G_N = 0$) in what follows. We will return to curved manifolds in a later paper.

## 4.2  Invariant Einstein proper time symmetry group

In Section 3.3, we determined the symmetry group for 'nonrelativistic' Hamiltonian mechanics acting on $T^*\mathbb{P}$ that leaves invariant the symplectic metric $\omega$ and the degenerate orthogonal Newton time line element $dt^2$ that are invariant for observers in any state, inertial or noninertial [20].

In this section, we determine the symmetry group acting on $T^*\mathbb{P}$ with $\mathbb{P} \simeq \mathbb{R}^{2n+2}$ with coordinates $s = (x, y)$, $x, y \in \mathbb{R}^{n+1}$, that leaves invariant the symplectic metric $\omega$ given in (2.2) and the degenerate orthogonal Einstein time line element $d\tau^{\circ 2}$ on $\mathbb{P}$ that is defined by

$$d\tau^{\circ 2} = -ds^{\mathrm{t}}\, \eta^x\, ds = -dx^{\mathrm{t}}\, \eta\, dx, \quad \eta^x = \begin{pmatrix} \eta & 0 \\ 0 & 0_{n+1} \end{pmatrix}. \tag{4.4}$$

(We have adopted matrix notation as in the previous section and have added the $\circ$ to the proper time, $\tau^{\circ}$. In the next section we will introduce a generalization of proper time $\tau$ that reduces in a limit to Einstein proper time $\tau^{\circ}$ just as Einstein proper time reduces to Newtonian absolute time in a limit.)

When we restrict to the spacetime $\mathbb{M} \subset \mathbb{P}$, $\mathbb{M} \simeq \mathbb{R}^{n+1}$, this is just the usual proper time orthogonal line element given in (4.2).

Let $\Sigma : T_s^*\mathbb{P} \to T_s^*\mathbb{P} : d\tilde{s} = \Sigma\, ds$. Using the block submatrices

$$\Sigma = \begin{pmatrix} \Sigma_1 & \Sigma_2 \\ \Sigma_3 & \Sigma_4 \end{pmatrix}.$$

Invariance of $d\tau^{\circ 2}$ requires $\Sigma^{\mathrm{t}}\, \eta^x\, \Sigma = \eta^x$ results in the conditions

$$\Sigma_1^{\mathrm{t}}\, \eta\, \Sigma_1 = \eta, \quad \Sigma_2^{\mathrm{t}}\, \eta\, \Sigma_1 = 0. \tag{4.5}$$

This has the solution $\Sigma_1 = \Lambda$, $\Lambda \in O(1, n)$ with $\Lambda^{\mathrm{t}}\, \eta\, \Lambda = \eta$ and $\Sigma_2 \equiv 0$. In addition, the invariance of the symplectic metric requires $\Sigma^{\mathrm{t}}\, \zeta\, \Sigma = \zeta$ and this results in the additional conditions (2.11),

$$\Sigma_1^{\mathrm{t}}\, \eta\, \Sigma_4 - \Sigma_3^{\mathrm{t}}\, \eta\, \Sigma_2 = \eta, \quad \Sigma_1^{\mathrm{t}}\, \eta\, \Sigma_3 = \left(\Sigma_1^{\mathrm{t}}\, \eta\, \Sigma_3\right)^{\mathrm{t}}, \quad \Sigma_2^{\mathrm{t}}\, \eta\, \Sigma_4 = \left(\Sigma_2^{\mathrm{t}}\, \eta\, \Sigma_4\right)^{\mathrm{t}}.$$

As $\Sigma_2 = 0$, this requires $\Sigma_4 = \Lambda$. Next set $\Sigma_3 = \mathrm{M}$ and it follows immediately that

$$\mathrm{M}^{\mathrm{t}} = \eta\, \Lambda^{-1}\, \mathrm{M}\, \Lambda^{-1}\, \eta. \tag{4.6}$$

Note that $\mathrm{tr}(\mathrm{M}^{\mathrm{t}}) = \mathrm{tr}(\mathrm{M}) = \mathrm{tr}\!\left(\eta\, \Lambda^{-1}\, \mathrm{M}\, \Lambda^{-1}\, \eta\right)$ is identically satisfied for all $\Lambda \in O(1, n)$ only if $\mathrm{tr}(\mathrm{M}) \equiv 0$.

Therefore, the elements $\Gamma(\Lambda, \mathrm{M})$ satisfying these conditions have the form,

$$\Gamma(\Lambda, M) = \begin{pmatrix} \Lambda & 0 \\ M & \Lambda \end{pmatrix} = \begin{pmatrix} 1_{n+1} & 0 \\ M\Lambda^{-1} & 1_{n+1} \end{pmatrix} \begin{pmatrix} \Lambda & 0 \\ 0 & \Lambda \end{pmatrix}. \tag{4.7}$$

The determinant is $\text{Det}\,\Gamma(\Lambda, M) = \text{Det}\,\Lambda^2 \equiv 1$. The elements have the group product and inverse,

$$\begin{aligned}\Gamma(\Lambda'', M'') &= \Gamma(\Lambda', M')\,\Gamma(\Lambda, M) = \Gamma(\Lambda'\Lambda,\, \Lambda' M + M'\Lambda), \\ \Gamma^{-1}(\Lambda, M) &= \Gamma(\eta\,\Lambda^t\,\eta,\, -\eta\,M^t\,\eta) = \Gamma(\Lambda^{-1},\, -\Lambda^{-1} M \Lambda^{-1}).\end{aligned} \tag{4.8}$$

Therefore, the elements $\Gamma(\Lambda, M)$ define a subgroup, that we denote as $Oa(1, n)$ of the symplectic group,

$$\Gamma(\Lambda, M) \in Oa(1, n) \subset Sp(2n+2). \tag{4.9}$$

Note that

$$\begin{aligned}\Gamma(\Lambda', 0)\,\Gamma(\Lambda, 0) &= \Gamma(\Lambda'\Lambda, 0), \qquad & \Gamma(\Lambda, 0)^{-1} &= \Gamma(\Lambda^{-1}, 0), \\ \Gamma(1_{n+1}, M')\,\Gamma(1, M) &= \Gamma(1_{n+1}, M+M'), \qquad & \Gamma(1, M)^{-1} &= \Gamma(1_{n+1}, -M)\,,\; M^t = \eta\,M\,\eta\,.\end{aligned} \tag{4.10}$$

The subgroup, $\Gamma(\Lambda, 0) \in O(1, n) \subset Sp(2n+2)$ is the symmetry group on $\mathbb{P}$ for inertial special relativity that was reviewed in Section 1.1. Furthermore, $\Gamma(1, M) \in \mathcal{A}(m) \subset Sp(2n+2)$ where $\mathcal{A}(m)$ is an abelian subgroup. The dimension of the abelian group is $m = (n+1)(n+2)/2 - 1 = n(n+3)/2$ due to the transpose condition (4.6) that also required $\text{tr}(M) = 0$.

The automorphisms are

$$\begin{aligned}\varsigma_{\Gamma(\Lambda', M')} \Gamma(\Lambda, M) &= \Gamma(\Lambda', M')\,\Gamma(\Lambda, M)\,\Gamma(\Lambda', M')^{-1} \\ &= \Gamma\!\left(\varsigma_{\Lambda'}\Lambda,\, (\Lambda' M + M'\Lambda - \varsigma_{\Lambda'}\Lambda\, M')\,\Lambda'^{-1}\right),\end{aligned} \tag{4.11}$$

and therefore $\mathcal{A}(m)$ is a normal subgroup,

$$\varsigma_{\Gamma(\Lambda', M')} \Gamma(1_{n+1}, M) = \Gamma(1_{n+1}, \varsigma_{\Lambda'} M). \tag{4.12}$$

The group factors as $\Gamma(\Lambda, M) = \Gamma(1_{n+1}, \tilde{M})\,\Gamma(\Lambda, 0)$ where $\tilde{M} = M\Lambda^{-1}$ and as $\Lambda$ is nonsingular, this defines the equivalence relation $Oa(1, n) \simeq \mathcal{A}(m)\,O(1, n)$. Finally as $O(1, n) \cap \mathcal{A}(m) = 1_{2n+2}$, it follows that $Oa(1, n)$ is a semidirect product of the form,

$$Oa(1, n) \simeq O(1, n) \otimes_s \mathcal{A}(m). \tag{4.13}$$

The orthogonal group may be written as $O(1, n) \simeq \mathbb{Z}_{2,2} \otimes_s \mathcal{L}(1, n)$ where $\mathcal{L}(1, n)$ is the orthochronous Lorentz group that is the connected component. Then,

$$Oa(1, n) \simeq \mathbb{Z}_{2,2} \otimes_s \mathcal{L}a(1, n), \qquad \mathcal{L}a(1, n) \simeq \mathcal{L}(1, n) \otimes_s \mathcal{A}(m), \tag{4.14}$$

where $\mathcal{L}a(1, n)$ is the connected component of $Oa(1, n)$.

Setting $M = \tilde{M}\Lambda$ simply defines another coordinate system for the group. It has corresponding matrix realization

$$\Gamma(\Lambda, \tilde{M}) = \begin{pmatrix} \Lambda & 0 \\ \tilde{M}\Lambda & \Lambda \end{pmatrix} = \begin{pmatrix} 1_{n+1} & 0 \\ \tilde{M} & 1_{n+1} \end{pmatrix} \begin{pmatrix} \Lambda & 0 \\ 0 & \Lambda \end{pmatrix}, \quad \tilde{M}^t = \eta\,\tilde{M}\,\eta. \tag{4.15}$$

As it simply factors, we refer to it as the 'factoring' coordinate system. The group product and inverse are straightforwardly computed in this coordinate system. The advantage of this coordinate system is that M transforms as a (1,1) tensor under Lorentz transformations on spacetime and obeys the usual 'raising and lowering' of indices as one would expect

$$\tilde{M} = [\![m^a{}_b]\!], \quad \tilde{M}^t = [\![m_a{}^b]\!], \quad m_a{}^b = \eta_{a,c}\, \eta^{b,d}\, m^c{}_d. \tag{4.16}$$

**Transformation equations**

The transformation equations on $T^*\mathbb{P}$ are

$$d\tilde{s} = \Gamma(\Lambda, M)\, ds. \tag{4.17}$$

Setting, $s = (x, y)$, these are simply

$$\begin{aligned} d\tilde{x} &= \Lambda\, dx, \\ d\tilde{y} &= \Lambda\, dy + M\, dx. \end{aligned} \tag{4.18}$$

In terms of the 'factoring' coordinate system, this is

$$\begin{aligned} d\tilde{x} &= \Lambda\, dx = d\, x', \\ d\tilde{y} &= \Lambda\, dy + \tilde{M}\, \Lambda\, dx = d\, y' + \tilde{M}\, d\, x'. \end{aligned} \tag{4.19}$$

Where the 'prime' coordinates are the inertial transformations $d\, x' = \Lambda\, dx$, $d\, y' = \Lambda\, dy$.

Clearly the spacetime $\mathbb{M} \subset \mathbb{P}$ continues to be an invariant submanifold. All observer states, inertial or noninertial, agree on this spacetime submanifold of the phase space. The inertial case of the previous section is $M = 0$. Note that the energy-momentum space $\tilde{\mathbb{M}}$ is no longer an invariant submanifold of these transformations.

The proper time is invariant,

$$d\tau^{\circ 2} = d\tilde{x}^t\, \eta\, d\tilde{x} = dx^t\, \Lambda^t\, \eta\, \Lambda\, dx = dx^t\, \eta\, dx, \tag{4.20}$$

and the energy-momentum line element transforms as

$$d\tilde{\mu}^2 = d\tilde{y}^t\, \eta\, d\tilde{y} = d\mu^2 + dy^t\, \Lambda^t\, \eta\, M\, dx + dx^t\, M^t\, \eta\, \Lambda\, dy + dx^t\, M^t\, \eta\, M\, dx. \tag{4.21}$$

In the factoring coordinate system, the energy-momentum line element transforms as

$$d\tilde{\mu}^2 = d\mu^2 + dy'^t\, \eta\, \tilde{M}\, dx' + dx'^t\, \tilde{M}^t\, \eta\, dy' + dx'^t\, \tilde{M}^t\, \eta\, \tilde{M}\, dx'. \tag{4.22}$$

Defining the 'four' velocity and force, $V, F \in \mathbb{R}^{n+1}$,

$$V = \frac{dx'}{d\tau^\circ}, \quad F = \frac{dy'}{d\tau^\circ}, \tag{4.23}$$

this results in

$$\frac{d\tilde{\mu}^2}{d\tau^\circ} = \frac{d\mu^2}{d\tau^\circ} + (\Lambda F)^t\, \eta\, \tilde{M}\, V + (\tilde{M} V)^t\, \eta\, \Lambda\, F + (\tilde{M} V)^t\, \eta\, \tilde{M}\, V. \tag{4.24}$$

This may be put in component form by defining $M = [\![m^a{}_b]\!]$ are (with units where $c = 1$) and noting from (4.15) that $m_b{}^a = \eta^{a,c}\, \eta_{b,d}\, m^d{}_c$ and therefore $m_{b,a} = m_{a,b}$,

$$d\tilde{\mu}^2 = d\mu^2 + \eta_{a,c}\, m^c{}_b(dx^a\, dy^b + dy^a\, dx^b) + \eta_{a,c}\, m^c{}_d\, m^d{}_b\, dx^a\, dx^b. \tag{4.25}$$

The components of 'four' velocity and 'four' force (for $n = 3$) are

$$V^a = \frac{dx'^a}{d\tau^\circ}, \quad F^a = \frac{dy'^a}{d\tau^\circ}. \tag{4.26}$$

This may be put in the form,

$$\frac{d\tilde{\mu}^2}{d\tau^\circ} = \frac{d\mu^2}{d\tau^\circ} + 2\, m_{a,b}\, V^a\, F^b + \eta^{c,d}\, m_{a,c}\, m_{d,b}\, V^a\, V^b. \tag{4.27}$$

We will see below that $m_{a,b}$ has the physical interpretation as a symmetric (0, 2) power-force-stress tensor that is the proper time derivative of a symmetric traceless (0, 2) energy-momentum-stress tensor.

## 4.3   Inertial Lorentz special relativity transformations

Consider a phase space $\mathbb{P} \simeq \mathbb{R}^{2n+2}$ endowed with a symplectic metric $\omega = ds^t\, \zeta\, ds$, $s \in \mathbb{P}$ that we introduced in Section 1.2 and assumed in the previous section for Hamilton's mechanics. Consider also two degenerate orthogonal metrics that define line elements,

$$d\tau^{\circ 2} = -ds^t\, \tilde{\eta}^x\, ds = -dx^t\, \eta\, dx, \quad d\mu^2 = -ds^t\, \tilde{\eta}^y\, ds = -dy^t\, \eta\, dy, \tag{4.28}$$

where $s = (x, y)$ with $x \in \mathbb{M} \simeq \mathbb{R}^{n+1}$, $dx \in T^*\mathbb{M}$, $y \in \tilde{\mathbb{M}} \simeq \mathbb{R}^{n+1}$, $dy \in T^*\tilde{\mathbb{M}}$ and

$$\tilde{\eta}^x = \begin{pmatrix} \eta & 0 \\ 0 & 0_{n+1} \end{pmatrix}, \quad \tilde{\eta}^y = \begin{pmatrix} 0_{n+1} & 0 \\ 0 & \eta \end{pmatrix}, \quad \eta = \begin{pmatrix} -1 & 0 \\ 0 & 1_n \end{pmatrix}. \tag{4.29}$$

The transformations $d\tilde{s} = \Sigma\, ds$ that leave the symplectic form invariant require $\Sigma^t\, \zeta\, \Sigma = \zeta$. We expand $\Sigma$ in block matrices (2.10)

$$\Sigma = \begin{pmatrix} \Sigma_1 & \Sigma_2 \\ \Sigma_3 & \Sigma_4 \end{pmatrix}, \tag{4.30}$$

where the submatrices must satisfy (2.11),

$$\Sigma_1^t\, \eta\, \Sigma_4 - \Sigma_3^t\, \eta\, \Sigma_2 = \eta, \quad \Sigma_1^t\, \eta\, \Sigma_3 = \left(\Sigma_1^t\, \eta\, \Sigma_3\right)^t, \quad \Sigma_2^t\, \eta\, \Sigma_4 = \left(\Sigma_2^t\, \eta\, \Sigma_4\right)^t. \tag{4.31}$$

Applying the condition that the degenerate orthogonal line elements $d\tau^{\circ 2}$ and $d\mu^2$ are also invariant results in the additional conditions,

$$\begin{aligned}\Sigma_1^t\, \eta\, \Sigma_1 &= \eta, \quad \Sigma_4^t\, \eta\, \Sigma_4 = \eta, \\ \Sigma_2^t\, \eta\, \Sigma_2 &= 0, \quad \Sigma_3^t\, \eta\, \Sigma_3 = 0.\end{aligned} \tag{4.32}$$

These require $\Sigma_2 = \Sigma_3 \equiv 0$ and, together with the above symplectic conditions, require $\Sigma_1 = \Sigma_4 = \Lambda$ with $\Lambda \in O(1, n)$. Therefore, the elements $\Gamma$ defined by

$$\Gamma(\Lambda) = \begin{pmatrix} \Lambda & 0 \\ 0 & \Lambda \end{pmatrix}, \tag{4.33}$$

are matrix realizations of elements of $O(1, n) \subset Sp(2n + 2)$. The resulting transformation equations are just the usual basic special relativistic inertial state transformations,

$$d\tilde{x} = \Lambda\, dx, \quad d\tilde{y} = \Lambda\, dy. \tag{4.34}$$

For these transformations, both the spacetime $\mathbb{M}$ and the energy momentum space $\tilde{\mathbb{M}}$ are invariant. All inertial observers agree on the same spacetime subspace of the phase space $\mathbb{P}$.

## 4.4   The $n = 1$ transformations and the nonrelativistic limit

Certain essential properties of the noninertial transformations of special relativity may
$n = 1$

be most simply presented by considering the case with with $n = 1$ with just one position and momentum dimension.

In this case $n = 1$, and hence $m = 1(1+3)/2 = 2$, the symmetry group is

$$Oa(1, 1) \simeq O(1, 1) \otimes_s \mathcal{A}(2) \simeq \mathbb{Z}_{2,2} \otimes_s \mathcal{L}(1, 1) \otimes_s \mathcal{A}(2) \subset Sp(4). \tag{4.35}$$

$\mathbb{Z}_{2,2}$ is the discrete group of time reversal and parity transformations and the Lorentz group $\mathcal{L}(1, 1)$ is the normal connected component. We further restrict our considerations here to the continuous connected transformations,

$$\mathcal{L}a(1, 1) \simeq \mathcal{L}(1, 1) \otimes_s \mathcal{A}(2) \subset Sp(4). \tag{4.36}$$

The elements $\Gamma(\Lambda, M) \in \mathcal{L}a(1, 1)$ have the form that satisfies (4.6) and hence $\text{tr}(M) = 0$,

$$\Gamma(\Lambda, M) = \begin{pmatrix} \Lambda & 0 \\ M & \Lambda \end{pmatrix}, \quad \Lambda \in \mathcal{L}(1, 1), \quad M = \gamma \begin{pmatrix} m_{0,0} & -m_{0,1} \\ m_{0,1} & -m_{0,0} \end{pmatrix}. \tag{4.37}$$

Consider first the elements $\Gamma(\Lambda, 0)$ that describe the usual inertial transformations of basic special relativity. We can choose to parameterize the transformations with the dimensioned velocity $v$ in the usual manner,

$$\Lambda = \begin{pmatrix} \cosh \beta & \sinh \beta \\ \sinh \beta & \cosh \beta \end{pmatrix} = \gamma^\circ(v) \begin{pmatrix} 1 & \frac{v}{c} \\ \frac{v}{c} & 1 \end{pmatrix}, \quad \tanh \beta = \frac{v}{c}, \quad \gamma^\circ(v) = \sqrt{1 - \frac{v^2}{c^2}}.$$

The velocity is dimensioned and so is scaled by $c$ to provide the dimensionless expression. The condition (4.6) on M requires $\gamma = \gamma^\circ(v)$ and so the group elements of $\mathcal{L}a(1, 1)$ have the form

$$\Gamma(\Lambda, M) = \gamma^\circ(v) \begin{pmatrix} 1 & \frac{v}{c} & 0 & 0 \\ \frac{v}{c} & 1 & 0 & 0 \\ m_{0,0} & -m_{0,1} & 1 & \frac{v}{c} \\ m_{0,1} & -m_{0,0} & \frac{v}{c} & 1 \end{pmatrix}. \tag{4.38}$$

The transformations $d\tilde{s} = \Gamma(\Lambda, M) \, ds$ expand out, with dimensions explicit, as

$$\begin{pmatrix} \frac{1}{\lambda_t} d\tilde{t} \\ \frac{1}{\lambda_q} d\tilde{q} \\ \frac{1}{\lambda_\varepsilon} d\tilde{\varepsilon} \\ \frac{1}{\lambda_p} d\tilde{p} \end{pmatrix} = \gamma^\circ(v) \begin{pmatrix} 1 & \frac{v}{c} & 0 & 0 \\ \frac{v}{c} & 1 & 0 & 0 \\ m_{0,0} & -m_{0,1} & 1 & \frac{v}{c} \\ m_{0,1} & -m_{0,0} & \frac{v}{c} & 1 \end{pmatrix} \begin{pmatrix} \frac{1}{\lambda_t} dt \\ \frac{1}{\lambda_q} dq \\ \frac{1}{\lambda_\varepsilon} d\varepsilon \\ \frac{1}{\lambda_p} dp \end{pmatrix}. \tag{4.39}$$

Collecting the dimensional scales and using (2.14) gives the transformations between the dimensioned bases,

$$\tag{4.40}$$

$$\begin{pmatrix} d\tilde{t} \\ d\tilde{q} \\ d\tilde{\varepsilon} \\ d\tilde{p} \end{pmatrix} = \gamma^\circ(v) \begin{pmatrix} 1 & \frac{\lambda_t}{\lambda_q} \frac{v}{c} & 0 & 0 \\ \frac{\lambda_q}{\lambda_t} \frac{v}{c} & 1 & 0 & 0 \\ \frac{\lambda_\varepsilon}{\lambda_t} m_{0,0} & -\frac{\lambda_\varepsilon}{\lambda_q} m_{1,0} & 1 & \frac{\lambda_\varepsilon}{\lambda_p} \frac{v}{c} \\ \frac{\lambda_p}{\lambda_t} m_{0,1} & -\frac{\lambda_p}{\lambda_q} m_{0,0} & \frac{\lambda_p}{\lambda_\varepsilon} \frac{v}{c} & 1 \end{pmatrix} \begin{pmatrix} dt \\ dq \\ d\varepsilon \\ dp \end{pmatrix}$$

$$= \gamma^\circ(v) \begin{pmatrix} 1 & \frac{v}{c^2} & 0 & 0 \\ v & 1 & 0 & 0 \\ b\,c\,m_{0,0} & -b\,m_{0,1} & 1 & v \\ b\,m_{0,1} & -\frac{b}{c} m_{0,0} & \frac{v}{c^2} & 1 \end{pmatrix} \begin{pmatrix} dt \\ dq \\ d\varepsilon \\ dp \end{pmatrix}.$$

For M = 0, these result in the usual special relativistic inertial transformations,

$$d\tilde{t} = \gamma^\circ(v)\left(dt + \tfrac{v}{c^2} dq\right), \quad d\tilde{q} = \gamma^\circ(v)(dq + v\,dt),$$
$$d\tilde{\varepsilon} = \gamma^\circ(v)(d\varepsilon + v\,dp), \quad d\tilde{p} = \gamma^\circ(v)\left(dp + \tfrac{v}{c^2} d\varepsilon\right). \tag{4.41}$$

For the noninertial case, $M \ne 0$, these must contract with $c \to \infty$ to nonrelativistic limit given in (3.48)

$$\lim_{c \to \infty} \gamma^\circ(v) \begin{pmatrix} 1 & \frac{v}{c^2} & 0 & 0 \\ v & 1 & 0 & 0 \\ b\,c\,m_{0,0} & -b\,m_{0,1} & 1 & v \\ b\,m_{0,1} & -\frac{b}{c} m_{0,0} & \frac{v}{c^2} & 1 \end{pmatrix} = \begin{pmatrix} 1 & 0 & 0 & 0 \\ v & 1 & 0 & 0 \\ r & -f & 1 & v \\ f & 0 & 0 & 1 \end{pmatrix}. \tag{4.42}$$

This leads immediately to the identifications

$$M = \gamma^\circ(v) \begin{pmatrix} m_{0,0} & -m_{0,1} \\ m_{0,1} & -m_{0,0} \end{pmatrix} = \gamma^\circ(v) \begin{pmatrix} \frac{r}{bc} & -\frac{f}{b} \\ \frac{f}{b} & \frac{-r}{bc} \end{pmatrix}. \tag{4.43}$$

$f$ is force and therefore scaled by the universal dimensional scale $b$ for force, $r$ is power and therefore is scaled by $b\,c$. M is the force-power stress tensor that is the time derivative of the momentum-energy stress tensor. Note that $M^t = \eta\,\Lambda^{-1}\,M\,\Lambda^{-1}\,\eta$ as required.

The resulting dimensioned one dimensional noninertial transformations are

$$d\tilde{t} = \gamma^\circ(v)\left(dt + \tfrac{v}{c^2} dq\right), \qquad d\tilde{q} = \gamma^\circ(v)(dq + v\,dt),$$
$$d\tilde{\varepsilon} = \gamma^\circ(v)(d\varepsilon + v\,dp - f\,dq + r\,dt), \quad d\tilde{p} = \gamma^\circ(v)\left(dp + \tfrac{v}{c^2} dt + f\,dt - \tfrac{r}{c^2} dq\right). \tag{4.44}$$

The dimensioned $\mathcal{O}a(1,1) \simeq \mathbb{Z}_{2,2} \otimes_s \mathcal{L}a(1,1)$ transformation equation are

$$\begin{pmatrix} d\tilde{t} \\ d\tilde{q} \\ d\tilde{\varepsilon} \\ d\tilde{p} \end{pmatrix} = \gamma^\circ(v) \begin{pmatrix} 1 & \frac{v}{c^2} & 0 & 0 \\ v & 1 & 0 & 0 \\ r & -f & 1 & v \\ f & \frac{-r}{c^2} & \frac{v}{c^2} & 1 \end{pmatrix} \begin{pmatrix} \epsilon & 0 & 0 & 0 \\ 0 & \tilde{\epsilon} & 0 & 0 \\ 0 & 0 & \epsilon & 0 \\ 0 & 0 & 0 & \tilde{\epsilon} \end{pmatrix} \begin{pmatrix} dt \\ dq \\ d\varepsilon \\ dp \end{pmatrix} \tag{4.45}$$

$$= \gamma^\circ(v) \begin{pmatrix} \epsilon & \frac{\tilde{\epsilon} v}{c^2} & 0 & 0 \\ \epsilon v & \tilde{\epsilon} & 0 & 0 \\ \epsilon r & -\tilde{\epsilon} f & \epsilon & \tilde{\epsilon} v \\ \epsilon f & \frac{-\tilde{\epsilon} r}{c^2} & \frac{\epsilon v}{c^2} & \tilde{\epsilon} \end{pmatrix} \begin{pmatrix} dt \\ dq \\ d\varepsilon \\ dp \end{pmatrix}, \quad \epsilon, \tilde{\epsilon} = \pm 1.$$

## 4.5 Noninertial relativity

Special relativity addresses inertial states that have relative motion on a flat manifold in the absence of gravity. Invariant time is Einstein proper time defined by the the Minkowski metric that is Lorentz invariant. General relativity shows that the apparent force of gravity is described by manifolds with curvature defined by the Einstein gravitational field equations such that apparent noninertial (i.e. accelerating) states due to gravity are locally inertial states on this manifold. Einstein time is made local by defining it relative to the Riemannian metric satisfying the gravitational field equations. For timelike states, locally there always exists at a point locally inertial coordinates where the metric is the Minkowski metric.

This section addresses noninertial states due to forces other than gravity such as, for example, electromagnetism. We restrict to the flat manifold case $\mathbb{P} \simeq \mathbb{R}^{2n+2}$ where gravity is negligible ($\alpha_N = G_N = 0$).

The $\mathcal{O}a(1, n)$ symmetry group with the $\mathcal{L}a(1, n)$ connected subgroup that results defines the relativity of these noninertial states in phase space for which the Lorentz subgroup defines the relativity for the special case of inertial states in special relativity.

The properties of the connected relativity group $\mathcal{L}a(1, n)$ are as follows:

- Invariant time is Einstein proper time $d\tau^\circ$ that is measured in the locally inertial rest frame of the particle state. Observers in timelike states with relative motion, inertial or noninertial, measure the passage of time such that $d\tau^{\circ 2}$ is invariant. These observers measure the rate at which time passes for a clock for a state with a relative velocity $v$ to be $dt = d\tau^\circ / \sqrt{(1 - v^2/c^2)}$. The rate of passage of time does not depend on the relative rate of change of momentum or energy with time. Thus, time is relative and it is not an invariant submanifold of the phase space $\mathbb{P}$. This is a defining property of the relativity groups $\mathcal{L}a(1, n)$ and $\mathcal{O}a(1, n)$.
- Simultaneity is relative in the sense that observers with different rates of change of position may observe different ordering of events. The ordering of events does not depend on the relative rate of change of momentum or energy.
- Rates of change of position with respect to time for timelike states is bounded by the null surface of $d\tau^{\circ 2}$. That is $v$ is bounded by the universal constant $c$ that is the speed of light in a vacuum.
- Rates of change of momentum and energy are unbounded.
- There is an absolute inertial state that all observers in *any* state agree on.
- There is not an absolute rest state that all observers in *any* state agree on.
- Information can propagate at a finite rate in position space. That is, there is a causal 'light cone' constraining the rate of propagation of information in position space. 'Instantaneous action at a distance' in position space is not possible.

- Information can propagate instantaneously in momentum space. That is, there are no 'causal cones' constraining the propagation of information in momentum space. 'Instantaneous action at a distance' in momentum space is possible.
- Position-time space (i.e spacetime) with coordinates $(t, q)$ is an invariant subspace of the phase space $\mathbb{P}$ that all observers in *any* physical states agree on.
- Momentum-energy space with coordinates $(\varepsilon, p)$ is an invariant subspace of the phase space $\mathbb{P}$ only for observers in *inertial* states.

# 5   Noninertial relativity with reciprocally invariant proper time

## 5.1   Introduction

The definition of invariant time and the invariant symplectic metric $\omega$ (2.13) on the phase space $\mathbb{P}$ leads directly to the understanding of the symmetry group for both inertial and noninertial relativity in Newton-Hamilton and Einstein mechanics. Hamilton mechanics has absolute Newton time described by the degenerate orthogonal metric $dt^2$ and Einstein mechanics has invariant proper time where the rate of passage of local time measured by an observer in a different inertial state depends on the velocity relative to the state in question. Einstein invariant proper time is defined in (4.1) as

$$d\tau^{\circ 2} = dt^2 - \frac{1}{c^2} dq^2 = dt^2 \left(1 - \frac{v^2}{c^2}\right). \tag{5.1}$$

This definition of proper time continues to hold for noninertial states with instantaneous velocity $v(t)$ on a flat manifold in the absence of gravity as we discussed in the preceding section. Einstein time $d\tau^{\circ 2}$ contracts to Newton time $dt^2$ in the limit of $c \to \infty$ that models the regime of small velocities relative to $c$. Both of these time line elements are degenerate orthogonal metrics on the phase space $\mathbb{P}$.

We now leave the domain of well established physics to investigate a more general definition of invariant time that is given by a nondegenerate orthogonal metric on the phase space $\mathbb{P}$ in addition to it being endowed with the invariant symplectic metric $\omega$ (2.13). There are several observations that motivate this more general definition of invariant time.

The first is that we are ultimately interested in a quantum theory. We have shown in Chapters 8 and 9 in [24] that the symplectic metric on phase space and the associated symplectic group invariance is fundamental to a theory with a noncommutative structure with invariant Heisenberg commutation relations. This noncommutative structure in a nonrelativistic theory has the property that, while momentum and position cannot be observed simultaneously, we can equally choose to observe either momentum or position. That is, a quantum state propagating in time can be equally diagonalized in position or in momentum; one is not more fundamental than the other.

In the non relativistic theory, the rate of propagation of the quantum state is unbounded and so instantaneous 'action-at-a-distance' is possible.  If position is observed, the rate of change of position (velocity) is diagonalized and is simply additive and is unbounded. If momentum is observed, the rate of change of momentum (force) is diagonalized and is simply additive is unbounded.

We can characterize the 'symmetry' between momentum and position of Newton-Hamilton mechanics mathematically by noting that the symplectic metric $\omega$ and the Newton time line element $dt^2$ are invariant under the discrete symmetry

$$\Delta_B \left(\frac{q}{\lambda_q}, \frac{p}{\lambda_p}\right) \mapsto \left(\frac{p}{\lambda_p}, -\frac{q}{\lambda_q}\right). \tag{5.2}$$

Born[7] noted this discrete symmetry $\Delta_B$ of momentum and position in basic Hamilton

*Principle*

$\mathbb{P}^{\circ}$   $\Delta_B$

mechanics and nonrelativistic quantum mechanics and this led him to postulate it as a *Reciprocity Principle* [4, 5]. This principle conjectured that the laws of physics in the nonrelativistic domain should be invariant under this discrete symmetry. Note that $\Delta_B$ is realized by the symplectic matrix on the phase space $\mathbb{P}^\circ$,

$$\Delta_B = \zeta^\circ = \begin{pmatrix} 0 & -1_n \\ 1_n & 0 \end{pmatrix}.$$

To understand the role of $\Delta_B$ in quantum mechanics, recall the inhomogeneous symplectic quantum symmetry studied in Chapter 9 in [24] where the projective representations of the inhomogeneous symplectic group is computed. This gave rise to the Weyl-Heisenberg group as the algebraic central extension of the normal abelian subgroup of the inhomogeneous symplectic group. The Heisenberg commutation relations are the Hermitian representations of the Lie algebra corresponding to the unitary representations of the Weyl-Heisenberg group. The operator $\Delta_B = \zeta^\circ$ appears in that analysis and is realized in the projective representations as the Fourier transform between momentum and position. This clearly is fundamental to quantum mechanics.

In Einstein relativistic mechanics, we consider the reciprocity between position-time and momentum-energy defined by

$$\Delta_{BB^\circ}(x, y) \mapsto (y, -x), \tag{5.3}$$

where $x = (\frac{t}{\lambda_t}, \frac{q}{\lambda_q})$ and $y = (\frac{\varepsilon}{\lambda_\varepsilon}, \frac{p}{\lambda_p})$. Note that $\Delta_{BB^\circ} = \Delta_B \Delta_{B^\circ}$ where $\Delta_{B^\circ}$ is the reciprocity between time and energy,

$$\Delta_{B^\circ} \left( \frac{t}{\lambda_t}, \frac{\varepsilon}{\lambda_\varepsilon} \right) \mapsto \left( \frac{\varepsilon}{\lambda_\varepsilon}, -\frac{t}{\lambda_t} \right). \tag{5.4}$$

However the Einstein proper time $d\tau^{\circ 2}$ defined by (5.1) is not invariant under either the reciprocity $\Delta_{BB^\circ}$ (nor $\Delta_B$ or $\Delta_{B^\circ}$ separately). In this case, if position is observed, the rate of change of position (velocity) is diagonalized but now it is bounded by the causal cone. This causal cone is defined by the null hypersurface $d\tau^{\circ 2} = 0$ which turns out to be the requirement that $v^2 < c^2$ for timelike states. On the other hand, if momentum is observed, the rate of change of momentum (force) is diagonalized and is simply additive and so is unbounded. It is extremely problematic that our choice of how we measure the quantum state, that is, either diagonalizing momentum or position, should result in such fundamentally different causal properties. In one case the rate of change of the state is bounded by a causal cone, and in the other it is unbounded and so admits 'action-at-a distance'.

Consequently, in order to resolve this, we postulate that the theory must be generalized to be invariant under the Born reciprocity, $\Delta_B$ and $\Delta_{B^\circ}$ (and hence $\Delta_{BB^\circ}$).

Closely related to this is the question of fundamental dimensional constants. As we have discussed in (2.14-17), three universal dimensional constants are required to define the Planck phase space dimensional scales $\{\lambda_t, \lambda_q, \lambda_\varepsilon, \lambda_p\}$. These are usually taken to by $c$, $\hbar$ and $G_N$. $c$ and $\hbar$ are universally applicable to all theories; $c$ for the causal structure of spacetime and $\hbar$ for the noncommutative structure of quantum mechanics. $G_N$ however is a coupling constant specific to gravity, just as $\alpha_e$ is the fine structure constant is a coupling constant specific to electrodynamics; why should $G_N$ be a universal dimensional scale?

This lead us to consider the constant $b$ that has the units of force to be the third fundamental dimensional constant in (2.15). $b$ is a universal constant that must be determined experimentally that is applicable to all physical phenomena in the same manner that $c$ and $\hbar$ are applicable to all physical theories. $b$ plays the same role for rates of change of momentum (force) as $c$ plays for rates of change of position (velocity). $c, b, \hbar$ define the Planck dimensional scales for the phase space $\mathbb{P}$ degrees of freedom, time, position, energy and momentum given in (2.15) that we repeat here,

$$\lambda_t = \sqrt{\frac{\hbar}{bc}}, \quad \lambda_q = \sqrt{\frac{\hbar c}{b}}, \quad \lambda_p = \sqrt{\frac{\hbar b}{c}}, \quad \lambda_\varepsilon = \sqrt{\hbar b c}.$$

The gravitational constant $G_N$ is now an independent constant defined by the gravitational coupling constant $\alpha_N$ that was defined in (2.16), $G_N = \alpha_N c^4/b$. $\alpha_N$ is a dimensionless coupling constant that is specific to gravity that is analogous to other dimensionless coupling constants such as the electromagnetic fine structure constant $\alpha_e \simeq \frac{1}{137}$. As we already know the value of $G_N$, determining $\alpha_N$ fixes $b$ or vice versa.

The third motivating reason is the concept of noninertial relativity itself. Hamilton relativity with invariant Newton time is the limit of the noninertial relativity with invariant Einstein time for small velocities relative to $c$. This suggests that noninertial relativity with invariant Einstein proper time is, in turn, the limit of small forces relative to $b$ for a more general theory. This is precisely the theory that is described in this section. Special relativity has been verified to a very high degree of accuracy for inertial states and so we expect the inertial case of this more general theory theory to continue to reduce directly to special relativity.

We are interested initially in the case where both the phase space $\mathbb{P} \simeq \mathbb{R}^{2n+2}$ and the spacetime submanifold $\mathbb{M} \simeq \mathbb{R}^{n+1}$ are flat and therefore gravity is not present. Mathematically, this is the idealized case where $b$ has some finite value but $\alpha_N$ and hence $G_N$ are set to zero. This enables us to study the effects of this constant $b$ without the full complexity of the curved manifolds and field equations of general relativity or its generalization. In this sense the theory is a 'special theory' where 'special' in this context refers to the underlying manifold being flat rather than inertial states.

This leads us to conjecture that invariant time $\tau$ is defined by a nondegenerate orthogonal metric on $\mathbb{P}$. Using coordinates $s = \left(\frac{t}{\lambda_t}, \frac{q}{\lambda_q}, \frac{\varepsilon}{\lambda_\varepsilon}, \frac{p}{\lambda_p}\right)$,

$$\left(\frac{d\tau}{\lambda_t}\right)^2 = -ds^t \tilde{\eta}\, ds = \left(\frac{dt}{\lambda_t}\right)^2 - \left(\frac{dq}{\lambda_q}\right)^2 + \left(\frac{d\varepsilon}{\lambda_\varepsilon}\right)^2 - \left(\frac{dp}{\lambda_p}\right)^2, \quad \tilde{\eta} = \begin{pmatrix} \eta & 0 \\ 0 & \eta \end{pmatrix}. \quad (5.5)$$

Therefore, using (2.14), this is

$$d\tau^2 = d\tau^{\circ 2} + \frac{1}{b^2} d\mu^2 = dt^2 - \frac{1}{c^2} dq^2 - \frac{1}{b^2} dp^2 + \frac{1}{b^2 c^2} d\varepsilon^2. \quad (5.6)$$

Note immediately that the invariant time line element $d\tau^2$ is invariant under the discrete Born reciprocity symmetries $\Delta_B$, $\Delta_{B^\circ}$ and so we refer to this as *reciprocally invariant proper time*. The introduction of this nondegenerate orthogonal metric and the universal constant $b$ are the only changes to standard physics that we are making. We will see that the consequences are profound.

This metric for invariant time was originally introduced by the author in [14-16]. Born introduced a *metric operator* based on the reciprocity principle that is a similar construct, albeit with a rather different physical interpretation in the papers [4, 5]. In the paper [15], the author named the metric (5.6) the Born metric as a tribute to Born's seminal work.

This metric may be written in the form

$$d\tau^2 = dt^2\left(1 - \frac{v^2}{c^2} - \frac{f^2}{b^2} + \frac{r^2}{c^2 b^2}\right), \tag{5.7}$$

where $v$, $f$ and $r$ are the relative rate of change of position, momentum and energy with respect to time (i.e: velocity, force, power) between the states. This means that the proper invariant time now depends on the noninertial state. Time 'dilation' is

$$dt = \frac{d\tau}{\sqrt{1 - \frac{v^2}{c^2} - \frac{f^2}{b^2} + \frac{r^2}{c^2 b^2}}}. \tag{5.8}$$

In the inertial limit $b \to \infty$, (i.e. $\frac{f}{b} \to 0$, $\frac{r}{cb} \to 0$) this immediately reduces to the usual Einstein proper time $d\tau^{\circ 2}$. We will find that the noninertial relativity with Einstein proper time is the small rate of change of momentum - force - limit $b \to \infty$ of the noninertial relativity with reciprocally invariant time. As we noted above, this limit 'breaks' the reciprocity symmetries $\Delta_B$, $\Delta_{B^\circ}$ and $\Delta_{BB^\circ}$. Furthermore, in the limit $b, c \to \infty$, it contracts to the Hamilton noninertial relativity with Newton absolute time $dt^2$ for which the reciprocity $\Delta_{B^\circ}$ and $\Delta_{BB^\circ}$ are broken but, interestingly, it turns out to be invariant under the Born reciprocity $\Delta_B$.

Note that the symplectic metric $\omega$ in the dimensionally scaled coordinates is

$$\omega = -\frac{dt}{\lambda_t} \wedge \frac{d\varepsilon}{\lambda_\varepsilon} + \delta_{i,j} \frac{dq^i}{\lambda_q} \wedge \frac{dp^j}{\lambda_p} = \frac{1}{\hbar}\left(dt \wedge d\varepsilon + \delta_{i,j} dq^i \wedge dp^j\right). \tag{5.9}$$

Consequently, $\omega$ is unaffected by the limits, $c \to \infty$, $b \to \infty$ and so the symplectic metric remains the same for all of the cases.

The null cones $d\tau^2 = 0$ for this theory defines an elliptical hyperboloid in terms of $v, f, r$. The inertial special case $f = r = 0$ reduces this to the usual null cones of special relativity, $v = \pm c$. Another special case is $v = r = 0$ in which case it reduces to the null cones $f = \pm b$. When we study this further in the sections that follow, we find that this, in fact, ensures that information is be propagated at a finite rate in both position and momentum spaces that was a fundamental motivation for this generalization.

The remainder of this paper is the exploration of the consequences of introducing this metric and the fundamental universal constant $b$. This exposition is based on a set of original papers by the author in the research journals [14-23].

As the symmetries are subgroups of the symplectic group, the quantum generalizations follows directly by studying the projective representations of the inhomogeneous versions of these groups. These representations result in a Weyl-Heisenberg group and its associated Heisenberg commutation relations as we studied in Chapters 8 and 9 in [24].

## 5.2   Reciprocally invariant proper time symmetry group

As in our study of noninertial symmetry in Sections 3.3-4, we start with a flat symplectic manifold $(\mathbb{P}, \omega)$ with $\mathbb{P} \simeq \mathbb{R}^{2n+2}$ and the symplectic metric $\omega = ds^{\mathrm{t}} \zeta\, ds$, $ds \in T^*\mathbb{P}$. In

$$ds^2 = ds^{\mathrm{t}} \tilde{\eta}\, ds$$

addition, we assume that we have an invariant Born metric (5.5) $ds^2 = ds^t \tilde{\eta} \, ds$ with

$$\tilde{\eta} = \begin{pmatrix} \eta & 0 \\ 0 & \eta \end{pmatrix}, \; \zeta = \begin{pmatrix} 0 & \eta \\ -\eta & 0 \end{pmatrix}. \tag{5.10}$$

Under the transformation $d\tilde{s} = \Sigma \, ds$ with $\Sigma \in \mathcal{GL}(2n+2, \mathbb{R})$, invariance of the symplectic and orthogonal metrics requires

$$\Sigma^t \zeta \Sigma = \zeta, \; \Sigma^t \tilde{\eta} \Sigma = \tilde{\eta}. \tag{5.11}$$

The symplectic condition requires $\Sigma \in \mathcal{S}p(2n+2)$ and the orthogonal metric condition requires $\Sigma \in \mathcal{O}(2, 2n)$. The invariance of both these metrics requires $\Sigma$ to be an element of the real unitary group defined by the intersection

$$\mathcal{U}u(1, n) \equiv \mathcal{O}(2, 2n) \cap \mathcal{S}p(2n+2). \tag{5.12}$$

Before continuing, we take a brief aside to explain why this group is referred to as $\mathcal{U}u(1, n)$. We will show that the noncompact symmetry admits an overall $\mathcal{U}(1)$ phase symmetry that is not simply a trivial direct product '$\mathcal{U}(1) \otimes \mathcal{U}(1, n)$'. The reason for this is that we will show that $\mathcal{SU}(1, n)$ is not connected but has two components and may be written as $\mathcal{SU}(1, n) \simeq \mathbb{Z}_2 \otimes_s \mathcal{SU}c(1, n)$ where $\mathcal{SU}c(1, n)$ is the connected normal subgroup. This is analogous to $\mathcal{SO}(1, n) \simeq \mathbb{Z}_2 \otimes_s \mathcal{L}(1, n)$, and in fact, we will show that this is a subgroup of $\mathcal{SU}(1, n)$. We will then show that for this overall $\mathcal{U}(1)$ phase, $\mathcal{U}(1) \cap \mathcal{SU}(1, n) \simeq \mathbb{Z}_2$ and hence $\mathcal{U}(1) \cap \mathcal{U}(1, n) \simeq \mathbb{Z}_2$. Consequently the overall phase is not simply a direct product as it is topologically nontrivially intertwined with the unitary subgroup. However, we will show that $\mathcal{U}u(n) \simeq \mathcal{U}_2(1) \otimes_s \mathcal{SU}c(1, n)$ is well defined where $\mathcal{U}_2(1) = \mathcal{U}(1) \otimes \mathcal{U}(1)$. We will also show that $\mathcal{O}(1, n) \subset \mathcal{U}u(1, n)$ where $\mathcal{O}(1, n) \simeq \mathbb{Z}_{2,2} \otimes_s \mathcal{L}(1, n)$ and $\mathbb{Z}_{2,2} \subset \mathcal{U}_2(1)$ and $\mathcal{L}(1, n) \subset \mathcal{SU}c(1, n)$.

Returning to the main line of the discussion we start by setting

$$\Sigma = \begin{pmatrix} \Sigma_1 & \Sigma_2 \\ \Sigma_3 & \Sigma_4 \end{pmatrix}.$$

and the the symplectic invariance results in conditions expand out as (2.10),

$$\Sigma_1^t \eta \Sigma_4 - \Sigma_3^t \eta \Sigma_2 = \eta, \; \Sigma_1^t \eta \Sigma_3 = (\Sigma_1^t \eta \Sigma_3)^t, \; \Sigma_2^t \eta \Sigma_4 = (\Sigma_2^t \eta \Sigma_4)^t, \tag{5.13}$$

and the orthogonal conditions are

$$\Sigma_1^t \eta \Sigma_1 + \Sigma_3^t \eta \Sigma_3 = \eta, \; \Sigma_1^t \eta \Sigma_2 = -\Sigma_3^t \eta \Sigma_4, \; \Sigma_2^t \eta \Sigma_2 + \Sigma_4^t \eta \Sigma_4 = \eta. \tag{5.14}$$

These may be straightforwardly solved for the solution $\Lambda = \Sigma_1 = \Sigma_4$, $M = \Sigma_3 = -\Sigma_2$. Therefore, the elements of the group $\mathcal{U}u(1, n)$ have the form

$$\Sigma = \Gamma(\Lambda, M) = \begin{pmatrix} \Lambda & -M \\ M & \Lambda \end{pmatrix}, \; \Lambda^t \eta \Lambda + M^t \eta M = \eta, \; \Lambda^t \eta M = (\Lambda^t \eta M)^t \tag{5.15}$$

The group product and inverse follows directly,

$$\Gamma(\Lambda'', M'') = \Gamma(\Lambda', M') \Gamma(\Lambda, M) \\ = \Gamma(\Lambda' \Lambda - M' M, \; \Lambda' M + M' \Lambda), \tag{5.16}$$

$$\Gamma^{-1}(\Lambda, M) = \Gamma(\eta \Lambda^t \eta, -\eta M^t \eta). \tag{5.17}$$

The group elements may also be written in factorized form

$$\Gamma(\Lambda, \tilde{M}\,\Lambda) = \Gamma(1_{n+1}, \tilde{M})\,\Gamma(\Lambda, 0)\,,\ \Lambda^t\left(\eta + \tilde{M}^t\,\eta\,\tilde{M}\right)\Lambda = \eta,\ \tilde{M}^t = \eta\,\tilde{M}\,\eta \tag{5.18}$$

In the special case $\Gamma(\Lambda, 0)$ the group product and inverse simplify to

$$\Gamma(\Lambda'', 0) = \Gamma(\Lambda'\,\Lambda, 0),\ \Gamma^{-1}(\Lambda, 0) = \Gamma(\Lambda^{-1}, 0),\ \Lambda^t\,\eta\,\Lambda = \eta,$$

and so elements of this form define an $O(1, n)$ subgroup. The orthogonal group may be written as

$$O(1, n) \simeq \mathbb{Z}_{2,2} \otimes_s \mathcal{L}(1, n)\ \begin{matrix} \simeq \mathbb{P} \otimes_s SO(1, n),\ SO(1, n) \simeq \mathbb{PT} \otimes \mathcal{L}(1, n),\ n\ \text{odd} \\ \simeq \mathbb{PT} \otimes_s SO(1, n),\ SO(1, n) \simeq \mathbb{P} \otimes \mathcal{L}(1, n),\ n\ \text{even} \end{matrix}.$$

where $\mathbb{P} \simeq \mathbb{Z}_2$ is the parity group with elements $\{1_{n+1}, -\eta\}$, $\mathbb{T} \simeq \mathbb{Z}_2$ is the time-reversal group with elements $\{1_{n+1}, \eta\}$, $\mathbb{PT} \simeq \mathbb{Z}_2$ is the parity-time reversal group with elements $\{1_{n+1}, -1_{n+1}\}$. $\mathbb{Z}_{2,2} \simeq \mathbb{P} \otimes \mathbb{PT} \simeq \mathbb{P} \otimes \mathbb{T} \simeq \mathbb{Z}_2 \otimes \mathbb{Z}_2$ is the 4 element parity, time reversal and parity-time reversal finite group. The orthochronous Lorentz group $\mathcal{L}(1, n)$ is the connected component. Note that elements of the form $\Gamma(1_{n+1}, M)$ do not define a subgroup.

The transformation equations $d\tilde{s} = \Gamma(\Lambda, M)\,ds$ with $s = (x, y)$ are

$$\begin{aligned} d\tilde{x} &= \Lambda\,dx - M\,dy, \\ d\tilde{y} &= \Lambda\,dy + M\,dx. \end{aligned} \tag{5.19}$$

Consequently, there is not an invariant spacetime manifold $\mathbb{M} \subset \mathbb{P}$, $\mathbb{M} \simeq \mathbb{R}^{n+1}$ that all noninertial states agree on. If $M = 0$, these transformation equations reduce to the special relativistic transformations for inertial states. We will return to this in more detail below.

## Complex form of unitary group

The $\mathcal{U}u(1, n)$ group is complex group that is a subgroup of $\mathcal{GL}(n + 1, \mathbb{C})$ when $\mathbb{P}$ is endowed with a complex structure

$$1 \simeq \begin{pmatrix} 1_{n+1} & 0 \\ 0 & 1_{n+1} \end{pmatrix},\ i \simeq \begin{pmatrix} 0 & -1_{n+1} \\ 1_{n+1} & 0 \end{pmatrix}, \tag{5.20}$$

such that $\mathbb{P} \simeq \mathbb{R}^{2(n+1)} \simeq \mathbb{C}^{n+1}$. Define the Hermitian metric metric $dz^\dagger\,\eta\,dz = \eta_{a,b}\,dz^{*a}\,dz^b$. The elements $\Xi \in \mathcal{GL}(n + 1, \mathbb{C})$ map $dz \mapsto d\tilde{z} = \Xi\,dz$ and they leave the Hermitian metric invariant if $dz^\dagger\,\eta\,dz = (\Xi\,dz)^\dagger\,\eta\,(\Xi\,dz)$. This requires the matrices $\Xi$ to satisfy,

$$\Xi^\dagger\,\eta\,\Xi = \eta. \tag{5.21}$$

Define $z = x + i\,y$, $x, y \in \mathbb{R}^{n+1}$ and

$$\Xi = \Lambda + i\,M \simeq \Gamma(\Lambda, M) = \begin{pmatrix} \Lambda & -M \\ M & \Lambda \end{pmatrix}. \tag{5.22}$$

where $\Lambda, M$ are real matrices. The Hermitian metric condition (5.21) expands out to the conditions in real form (5.13-14) that leave the symplectic $\omega$ and orthogonal metrics $\tilde{\zeta}$ invariant.

## Semidirect product form of the unitary group

The unitary group $\mathcal{U}u(1, n)$ has a semidirect product form that is summarized in the following proposition

**Proposition 5.4:** *The group* $\mathcal{U}u(1,n)$ *has the semidirect product form*

$$\mathcal{U}u(1,n) \simeq \mathcal{U}_2(1) \otimes_s \mathcal{S}\mathcal{U}c(1,n) \tag{5.23}$$

*where* $\mathcal{U}_2(1) \simeq \mathcal{U}(1) \otimes \mathcal{U}(1)$ *and* $\mathcal{S}\mathcal{U}c(1,n)$ *with elements* $\Xi$ *is the orthochronous connected subgroup of* $\mathcal{S}\mathcal{U}(1,n)$ *defined by the additional condition* $\text{Re}\,\Xi^0{}_0 \geq 1$.

A corollary of this proposition is

**Corollary 5.5:** $\mathcal{S}\mathcal{U}(1,n)$ *is a subgroup of* $\mathcal{U}u(1,n)$ *that is not connected and has two components and it is given by*

$$\mathcal{S}\mathcal{U}(1,n) \simeq \begin{cases} \mathbb{PT} \otimes \mathcal{S}\mathcal{U}c(1,n), & n \text{ odd}, \; \mathbb{PT} \simeq \mathbb{Z}_2 \\ \mathbb{P} \otimes_s \mathcal{S}\mathcal{U}c(1,n), & n \text{ even}, \; \mathbb{P} \simeq \mathbb{Z}_2 \end{cases}. \tag{5.24}$$

*Furthermore,* $\mathcal{S}\mathcal{U}(1,n) \cap \mathcal{U}_2(1) \simeq \mathbb{Z}_2$.

**Corollary 5.6:** $\mathcal{U}u(1,n)$ *restricted to the reals is the group* $O(1,n) \simeq \mathbb{Z}_{2,2} \otimes_s \mathcal{L}(1,n)$ *where* $\mathbb{Z}_{2,2} \simeq \mathbb{Z}_2 \otimes \mathbb{Z}_2$ *and* $\mathcal{L}(1,n)$ *is the orthochronous subgroup of* $SO(1,n)$. $\mathcal{S}\mathcal{U}c(1,n)$ *restricted to the reals is the group* $\mathcal{L}(1,n)$.

The first step in establishing the proposition is to show that $\mathcal{U}_2(1)$ is a subgroup of $\mathcal{U}u(1,n)$ and then to define $\mathcal{S}\mathcal{U}c(1,n)$ and show that it is a normal subgroup of $\mathcal{U}u(1,n)$. We will then establish that it is a semidirect product by establishing that the intersection of these groups is the identity and that $\mathcal{U}u(1,n)$ is the product of these matrix groups.

### The $\mathcal{U}_2(1)$ group and its $\mathbb{Z}_{2,2}$ subgroup

Complex realizations of the $\mathcal{U}(1)$ are defined by,

$$\Delta^\circ(\vartheta) = \begin{pmatrix} e^{i\vartheta} & 0 \\ 0 & 1_n \end{pmatrix} \in \mathcal{U}(1), \quad \Delta(\theta) = \begin{pmatrix} 1 & 0 \\ 0 & e^{i\theta} 1_n \end{pmatrix} \in \mathcal{U}(1). \tag{5.25}$$

An element of $\mathcal{U}_2(1) \simeq \mathcal{U}(1) \otimes \mathcal{U}(1)$ is given by

$$\Delta(\vartheta,\theta) = \Delta^\circ(\vartheta)\,\Delta(\theta) = \begin{pmatrix} e^{i\vartheta} & 0 \\ 0 & e^{i\theta} 1_n \end{pmatrix}. \tag{5.26}$$

Straightforward matrix multiplication shows that these elements satisfy condition (5.21) to be an element of $\mathcal{U}u(1,n)$,

$$\Delta(\vartheta,\theta)^\dagger \,\eta\, \Delta(\vartheta,\theta) = \eta, \quad \text{Det}\,\Delta(\vartheta,\theta) = e^{i(\vartheta + n\theta)}. \tag{5.27}$$

Note also that elements of the form $\Delta(\theta,\theta) = e^{i\theta} 1_{n+1}$ define another $\mathcal{U}(1)$ subgroup of $\mathcal{U}_2(1)$.

Restricting the complex $\mathcal{U}(1)$ group to the reals results in the $\mathbb{Z}_2$ group. Consequently,

$$\{\Delta_e, \Delta_T, \Delta_P, \Delta_{PT}\} \in \text{Re}\,\mathcal{U}_2(1) \simeq \mathbb{Z}_{2,2} \simeq \mathbb{Z}_2 \otimes \mathbb{Z}_2, \tag{5.28}$$

where

$$\Delta_e = \Delta(0,0) = \begin{pmatrix} 1 & 0 \\ 0 & 1_n \end{pmatrix}, \quad \Delta_T = \Delta(\pi,0) = \begin{pmatrix} -1 & 0 \\ 0 & 1_n \end{pmatrix},$$

$$\Delta_P = \Delta(0,\pi) = \begin{pmatrix} 1 & 0 \\ 0 & -1_n \end{pmatrix}, \quad \Delta_{PT} = \Delta(\pi,\pi) = \begin{pmatrix} -1 & 0 \\ 0 & -1_n \end{pmatrix}. \tag{5.29}$$

This is the parity and time-reversal $\mathbb{Z}_{2,2}$ group. Note that $\mathbb{Z}_{2,2}$ has three $\mathbb{Z}_2$ subgroups $\mathbb{T}, \mathbb{P}, \mathbb{PT}$ with the elements

$$\{\Delta_e, \Delta_T\} \in \mathbb{T}, \ \{\Delta_e, \Delta_P\} \in \mathbb{P}, \ \{\Delta_e, \Delta_{PT}\} \in \mathbb{PT}. \tag{5.30}$$

The real form of the group elements of $\mathcal{U}_2(1)$ follows directly from the complex form given in (5.26).

$$\Delta(\theta, \vartheta) = \begin{pmatrix} \Lambda_\Delta(\theta, \vartheta) & -\mathrm{M}_\Delta(\theta, \vartheta) \\ \mathrm{M}_\Delta(\theta, \vartheta) & \Lambda_\Delta(\theta, \vartheta) \end{pmatrix},$$

$$\Lambda_\Delta(\theta, \vartheta) = \begin{pmatrix} \cos\theta & 0 \\ 0 & \cos\vartheta\, 1_n \end{pmatrix}, \ \mathrm{M}_\Delta(\theta, \vartheta) = \begin{pmatrix} \sin\theta & 0 \\ 0 & \sin\vartheta\, 1_n \end{pmatrix}. \tag{5.31}$$

### The special unitary group $\mathcal{SU}(1, n)$ and its orthochronous subgroup

The determinant of a matrix group defines a homomorphism. Taking the determinant of (5.21) results in $|\mathrm{Det}\,\Xi|^2 = 1$ and therefore the determinate is a phase $\mathrm{Det}\,\Xi = e^{i\phi}$. As $\mathcal{U}_2(1)$ is a subgroup the describes a phase, we have $\mathrm{Det}\,\Xi \in \mathcal{U}_2(1)$.

The group $\mathcal{SU}(1, n)$ is the normal subgroup of $\mathcal{U}(1, n)$ that is the kernel of the determinant homomorphism,

$$\mathrm{Det}: \mathcal{Uu}(1, n) \to \mathcal{U}_2(1), \ \ \mathcal{SU}(1, n) \simeq \ker(\mathrm{Det}).$$

Consequently for $\Xi(\Lambda, \mathrm{M}) \in \mathcal{SU}(1, n)$, $\mathrm{Det}\,\Xi(\Lambda, \mathrm{M}) = 1$.

The intersection $\mathcal{U}_2(1) \cap \mathcal{SU}(1, n)$ is not trivial but rather is $\mathbb{Z}_2$. For $n$ odd, $\Delta_{PT}$ is an element of both $\mathcal{U}_2(1)$ and $\mathcal{SU}(1, n)$ and for $n$ even, $\Delta_P$ is an element of both groups. Consequently, $\mathcal{Uu}(1, n)$ is not a direct product of $\mathcal{U}_2(1)$ and $\mathcal{SU}(1, n)$.

This leads us to define the orthochronous subgroup $\mathcal{SU}c(1, n)$ of $\mathcal{SU}(1, n)$ with the additional condition $\mathrm{Re}\,\Xi(\Lambda, \mathrm{M})^0{}_0 \geq 1$. This follows closely the argument that was used to establish the orthochronous subgroup $\mathcal{L}(1, n)$ of $\mathcal{SO}(1, n)$.

We start by noting that the real elements of $\mathcal{SU}(1, n)$ are of the form $\Xi(\Lambda, 0) \simeq \Lambda$ with the conditions $\Lambda^t \eta \Lambda = \eta$ and $\mathrm{Det}\,\Lambda = 1$. Thus, the special unitary group restricted to the reals is $\mathcal{SO}(1, n) \simeq \mathrm{Re}(\mathcal{SU}(1, n))$. The special orthogonal group has the semidirect product form,

$$\mathcal{SO}(1, n) \simeq \mathbb{Z}_2 \otimes_s \mathcal{L}(1, n), \ \begin{cases} \mathbb{Z}_2 \simeq \mathbb{PT}, & n \text{ odd} \\ \mathbb{Z}_2 \simeq \mathbb{P}, & n \text{ even} \end{cases} \tag{5.32}$$

$\mathcal{L}(1, n)$ is the orthochronous Lorentz group with $\Lambda^0{}_0 \geq 1$. This group is the maximal connected component of both the $\mathcal{SO}(1, n)$ and $O(1, 1)$ groups that preserves orientation and the direction of time for timelike states in the future causal light cone in spacetime $\mathbb{M}$.

We can now prove the proposition that shows that the orthochonus group $\mathcal{SU}c(1, n)$ is a subgroup of $\mathcal{SU}(1, n)$.

**Proposition 5.7:** *The elements $\Xi(\Lambda, \mathrm{M}) \in \mathcal{SU}(1, n)$ with the additional condition $\mathrm{Re}\,\Xi(\Lambda, \mathrm{M})^0{}_0 \geq 1$, that is $\Lambda^0{}_0 \geq 1$, define a matrix group that is path connected to the identity. This group is denoted the orthochronous group $\mathcal{SU}c(1, n)$.*

The condition (5.21)

$$|\Xi^0{}_0| = \sqrt{1 + \sum_{i=1}^{n} |\Xi^0{}_i|^2} \geq 1 \qquad (5.33)$$

and consequently $\text{Re}(\Xi^0{}_0) = \Lambda^0{}_0 \geq 1$. Next, for the group product $\Xi'' = \Xi' \Xi$ and inverse $\Xi^{-1}$, assume $\text{Re}(\Xi^0{}_0) \geq 1$ and $\text{Re}(\Xi'^0{}_0) \geq 1$ and consequently, $\Lambda^0{}_0, \Lambda'^0{}_0 \geq 1$. As we have already established this is the condition for $\mathcal{L}(1, n) \subset SO(1, n)$, this implies $\Lambda''^0{}_0 \geq 1$ and $(\Lambda''^{-1})^0{}_0 \geq 1$. Then as $\text{Re}(\Xi^0{}_0) = \Lambda^0{}_0$, this means $\text{Re}(\Xi^0{}_0) \geq 1$ as required and the proposition is proven.

It follows directly that

$$\mathcal{SU}(1, n) \simeq \mathbb{Z}_2 \otimes_s \mathcal{SU}c(1, n), \begin{cases} \mathbb{Z}_2 \simeq \mathbb{PT}, & n \text{ odd} \\ \mathbb{Z}_2 \simeq \mathbb{P}, & n \text{ even} \end{cases}, \qquad (5.34)$$

and hence our notation, the '*c*' in $\mathcal{SU}c(1, n)$ is for connected subgroup of $\mathcal{SU}(1, n)$. This establishes the Corollary 5.5. It follows that

$$\text{Re}\,\mathcal{SU}c(1, n) \simeq \mathcal{L}(1, n). \qquad (5.35)$$

### The semidirect product form of $\mathcal{U}u(1, n)$

We can now complete the proof of Proposition 5.4. First, $\mathcal{SU}c(1, n)$ is a normal subgroup of $\mathcal{SU}(1, n)$ which, in turn, is a normal subgroup of $\mathcal{U}u(1, n)$, it follows that $\mathcal{SU}c(1, n)$ is a normal subgroup of $\mathcal{U}u(1, n)$.

Next, $\text{Re}\,\Delta(\vartheta, \theta) \geq 1$ if an only if $\vartheta = \theta = 0$ for $\vartheta, \theta \in [0, 2\pi)$. Therefore,

$$\mathcal{U}_2(1) \cap \mathcal{SU}c(1, n) = e. \qquad (5.36)$$

Finally, any element of $\mathcal{U}u(1, n)$ is given by the product $\Xi(\Lambda, M)\,\Delta(\vartheta, \theta)$ and therefore $\mathcal{U}u(1, n) \simeq \mathcal{U}_2(1) \otimes_s \mathcal{SU}c(1, n)$ as claimed in Proposition 5.4. The Corollary 5.5 follows by noting that $\text{Re}\,\mathcal{U}_2(1) \simeq \mathbb{Z}_{2,2}$ and $\text{Re}\,\mathcal{SU}c(1, n) \simeq \mathcal{L}(1, n)$.

## 5.3 Inertial Lorentz special relativity group

Special relativity and the associated orthochronous Lorentz symmetry have been experimentally verified to a high degree of accuracy. The $\mathcal{SU}c(1, n)$ has $\mathcal{L}(1, n)$ as a subgroup for the case $M = 0$,

$$\Gamma(\Lambda, 0) = \begin{pmatrix} \Lambda & 0 \\ 0 & \Lambda \end{pmatrix}, \quad \begin{matrix} d\tilde{x} = \Lambda\,dx \\ d\tilde{y} = \Lambda\,dy \end{matrix}, \quad \Lambda \in \mathcal{L}(1, n). \qquad (5.37)$$

These are the expected Lorentz transformations on spacetime and energy-momentum space. Note that in this case, both the spacetime $\mathbb{M}$ and energy-momentum space $\tilde{\mathbb{M}}$ are invariant subspaces of $\mathbb{P} \simeq \mathbb{M} \otimes \tilde{\mathbb{M}}$.

For the $\mathcal{L}(1, n)$ subgroup, the Born metric 'breaks' into the invariant Einstein proper time line element $d\tau^{\circ 2}$ (1.1) and invariant mass line element $d\mu^2$ (1.5),

$$\begin{aligned} d\tau^{\circ 2} &= -ds^t\,\eta^x\,ds = -dx^t\,\eta\,dx, \quad \eta^x = \begin{pmatrix} \eta & 0 \\ 0 & 0_{n+1} \end{pmatrix}, \\ d\mu^2 &= -ds^t\,\eta^y\,ds = -dy^t\,\eta\,dy, \quad \eta^x = \begin{pmatrix} 0_{n+1} & 0 \\ 0 & \eta \end{pmatrix}. \end{aligned} \qquad (5.38)$$

The subgroup of $\mathcal{U}_2(1)$ that leaves these two degenerate orthogonal metrics invariant is $\mathbb{Z}_{2,2}$. Therefore, the largest group leaving these metrics invariant is the extended Lorentz group,

$$O(1, n) \simeq \mathbb{Z}_{2,2} \otimes_s \mathcal{L}(1, n), \quad O(1, n) \subset \mathcal{U}u(1, n),$$

As we noted in Corollary 5.5, this is equivalent to restricting the complex unitary group to the reals,

$$O(1, n) \simeq \mathcal{U}u(1, n) \cap \mathcal{GL}(n+1, \mathbb{R}). \tag{5.39}$$

## 5.4 The 1 dimensional case: $\mathcal{U}u(1, 1)$ group

The case $n = 1$ again allows us to investigate essential properties of the noninertial transformations in their most simple form. The transformations leaving the symplectic form invariant are $Sp(4)$ and the transformations leaving the orthogonal metric invariant are $O(2, 2)$. The group of transformations leaving both invariant is

$$\mathcal{U}u(1, 1) \simeq \mathcal{U}_2(1) \otimes_s \mathcal{SU}c(1, 1) \simeq Sp(4) \cap O(2, 2). \tag{5.40}$$

### The orthochronous $\mathcal{SU}c(1, 1)$ group

We consider first in this section the $\mathcal{SU}c(1, 1)$ group. A convenient parameterization of the group $\mathcal{SU}c(1, 1)$ that is the connected subgroup of $\mathcal{SU}(1, 1)$ is

$$\Lambda = \begin{pmatrix} \cosh \beta & \sinh \beta \cos \theta \cosh \phi \\ \sinh \beta \cos \theta \cosh \phi & \cosh \beta \end{pmatrix},$$
$$M = \begin{pmatrix} \sinh \beta \sinh \phi & -\sinh \beta \sin \theta \cosh \phi \\ \sinh \beta \sin \theta \cosh \phi & -\sinh \beta \sinh \phi \end{pmatrix}. \tag{5.41}$$

A straightforward calculation shows that for these matrices satisfy conditions (5.13-14) for the symplectic and orthogonal metrics to be invariant and therefore $\Gamma(\Lambda, M) \in \mathcal{U}u(1, 1)$ in its real form.

The complex form, $\Xi(\Lambda, M) = \Lambda + i M \in \mathcal{U}u(1, 1)$ is

$$\Xi(\beta, \theta, \phi) = \cosh \beta \begin{pmatrix} 1 + i \tanh \beta \sinh \phi & \tanh \beta\, e^{-i\theta} \cosh \phi \\ \tanh \beta\, e^{i\theta} \cosh \phi & 1 - i \tanh \beta \sinh \phi \end{pmatrix}. \tag{5.42}$$

A calculation shows that $\text{Det}\,\Xi(\Lambda, M) = 1$ and as $\Lambda^0{}_0 = \cosh \beta \geq 1$, the conditions are satisfied for $\Xi(\Lambda, M) \in \mathcal{SU}c(1, 1)$ as claimed above.

Define $\rho = \tanh \beta$ and these matrices take the form

$$\Xi(\rho, \theta, \phi) = \gamma(\rho) \begin{pmatrix} 1 + i \rho \sinh \phi & \rho\, e^{-i\theta} \cosh \phi \\ \rho\, e^{i\theta} \cosh \phi & 1 - i \rho \sinh \phi \end{pmatrix}, \quad \gamma(\rho) = (1 - \rho^2)^{\frac{-1}{2}}. \tag{5.43}$$

Note that for $\theta = \phi = 0$, then $M = 0$ and if $\rho = \frac{v}{c}$, then $\Lambda$ is the usual Lorentz inertial transformation for the timelike domain. This leads us to consider for the noninertial transformations parameterized by relative rate of change of position (velocity $v$), rate of change of momentum (force $f$) and rate of change of energy (power $r$),

$$\frac{v}{c} = \rho \cos \theta \, \cosh \phi, \quad \frac{f}{b} = \rho \sin \theta \, \cosh \phi, \quad \frac{r}{bc} = \rho \sinh \phi. \tag{5.44}$$

$\{c, b, \hbar\}$ are the fundamental dimensional scales with dimensions of velocity, force and action that we reviewed in Section 2.2. Consequently,

$$\rho(v, f, r) = \frac{v^2}{c^2} + \frac{f^2}{b^2} - \frac{r^2}{c^2 b^2}. \tag{5.45}$$

With this parameterization in terms of $(v, f, r)$, the $\Lambda$ and M matrices have the form,

$$\Lambda = \gamma(v, f, r) \begin{pmatrix} 1 & \frac{v}{c} \\ \frac{v}{c} & 1 \end{pmatrix}, \quad M = \gamma(v, f, r) \begin{pmatrix} \frac{r}{cb} & -\frac{f}{b} \\ \frac{f}{b} & -\frac{r}{cb} \end{pmatrix}, \quad \text{tr}(M) = 0. \tag{5.46}$$

where

$$\gamma(v, f, r) = (1 - \rho^2)^{-\frac{1}{2}} = \left(1 - \frac{v^2}{c^2} - \frac{f^2}{b^2} + \frac{r^2}{c^2 b^2}\right)^{-\frac{1}{2}}. \tag{5.47}$$

As $\Gamma(\Lambda, M) \in \mathcal{SU}c(1, 1)$, $\Lambda_0^0 = \gamma(v, f, r) \geq 1$. Consequently $\rho(v, f, r) \leq 1$.

### The transformation equations for dimensioned coordinates

For $n = 1$, the coordinates of the phase space $\mathbb{P} \simeq \mathbb{R}^4$ are $s = \left\{\frac{t}{\lambda_t}, \frac{q}{\lambda_q}, \frac{\varepsilon}{\lambda_\varepsilon}, \frac{p}{\lambda_p}\right\}$ where, as usual, $t, q, \varepsilon, p$ are the dimensioned coordinates of time, position, energy and momentum. $\{\lambda_t, \lambda_q, \lambda_\varepsilon, \lambda_p\}$ are the Planck scales (2.15) defined in terms of the fundamental dimensional constants $\{c, b, \hbar\}$.

The $\mathcal{SU}c(1, 1)$ transformation equations acting on the cotangent space of the phase space are

$$\Gamma(\Lambda, M) : T_s^* \mathbb{P} \to T_s^* \mathbb{P} : ds \mapsto d\tilde{s} = \Gamma(\Lambda, M)\, ds. \tag{5.48}$$

Expanded this out in the dimensioned coordinates for $n = 1$ results in,

$$\begin{pmatrix} \frac{1}{\lambda_t} d\tilde{t} \\ \frac{1}{\lambda_q} d\tilde{q} \\ \frac{1}{\lambda_\varepsilon} d\tilde{\varepsilon} \\ \frac{1}{\lambda_p} d\tilde{p} \end{pmatrix} = \gamma(v, f, r) \begin{pmatrix} 1 & \frac{v}{c} & -\frac{r}{cb} & \frac{f}{b} \\ \frac{v}{c} & 1 & -\frac{f}{b} & \frac{r}{cb} \\ \frac{r}{cb} & -\frac{f}{b} & 1 & \frac{v}{c} \\ \frac{f}{b} & -\frac{r}{cb} & \frac{v}{c} & 1 \end{pmatrix} \begin{pmatrix} \frac{1}{\lambda_t} dt \\ \frac{1}{\lambda_q} dq \\ \frac{1}{\lambda_\varepsilon} d\varepsilon \\ \frac{1}{\lambda_p} dp \end{pmatrix}. \tag{5.49}$$

Using the relations (2.2) for the dimensional scales, this results in the transformation equations,

$$\begin{pmatrix} d\tilde{t} \\ d\tilde{q} \\ d\tilde{\varepsilon} \\ d\tilde{p} \end{pmatrix} = \gamma(v, f, r) \begin{pmatrix} 1 & \frac{v}{c^2} & -\frac{r}{c^2 b^2} & \frac{f}{b^2} \\ v & 1 & -\frac{f}{b^2} & \frac{r}{b^2} \\ r & -f & 1 & v \\ f & -\frac{r}{c^2} & \frac{v}{c^2} & 1 \end{pmatrix} \begin{pmatrix} dt \\ dq \\ d\varepsilon \\ dp \end{pmatrix}, \tag{5.50}$$

that may be written as

$$\tag{5.51}$$

$$d\tilde{t} = \gamma(v, f, r)\left(dt + \frac{v}{c^2} dq + \frac{f}{b^2} dp - \frac{r}{c^2 b^2} d\varepsilon\right),$$

$$d\tilde{q} = \gamma(v, f, r)\left(dq + v\, dt + \frac{r}{b^2} dp - \frac{f}{b^2} d\varepsilon\right),$$

$$d\tilde{p} = \gamma(v, f, r)\left(dp + f\, dt - \frac{r}{c^2} dq + \frac{v}{c^2} d\varepsilon\right),$$

$$d\tilde{\varepsilon} = \gamma(v, f, r)(d\varepsilon + v\, dp - f\, dq + r\, dt).$$

These equations define how an observers with a relative rate of change of position, momentum and energy (i.e. velocity, force and power) transform how they measure increments of time, position, momentum and energy that are mathematically defined by their respective basis $\{dt, dq, d\varepsilon, dp\}$, $\{d\tilde{t}, d\tilde{q}, d\tilde{\varepsilon}, d\tilde{p}\} \in T_s^* \mathbb{P}$. It is clear that for these transformations, the spacetime $\mathbb{M}$ is no longer an invariant subspace of $\mathbb{P}$. Observers with different noninertial states characterized by nonzero $f, r$ perceive different space-time subspaces of $\mathbb{P}$. All of of the physical degrees of freedom 'mix'.

## Inertial states

For inertial states, $f = r = 0$ and hence $\mathbb{M} = 0$ and $\gamma(v, 0, 0) = \gamma^\circ(v)$. These equation reduce to the usual inertial equations of special relativity.

$$d\tilde{t} = \gamma^\circ(v)\left(dt + \frac{v}{c^2} dq\right),$$
$$d\tilde{q} = \gamma^\circ(v)(dq + v\, dt),$$
$$d\tilde{p} = \gamma^\circ(v)\left(dp + \frac{v}{c^2} d\varepsilon\right),$$
$$d\tilde{\varepsilon} = \gamma^\circ(v)(d\varepsilon + v\, dp).$$
(5.52)

In this case, there is an invariant spacetime $\mathbb{M} \subset \mathbb{P}$ on which all observers in *inertial* states agree. Furthermore, these observers in inertial states also agree on an invariant momentum-energy space $\tilde{\mathbb{M}} \subset \mathbb{P}$.

Returning to the full noninertial equations (5.51), observers in different noninertial states measure the physical degrees of freedom, time, position, momentum, and energy differently. Heuristically, we can say that time, position, momentum and energy can be transformed into each other or *mix*. Note that the terms in the transformations that cause spacetime not to be invariant are always scaled by $\frac{1}{b^2}$. As we do not generally observe these effects, we conclude that $b$ is sufficiently large that forces that we generally observe are much smaller than $b$, $\frac{f}{b} \ll 1$. Thus, only extreme noninertial observers for which $f \sim o(b)$ would observe that the spacetime $\mathbb{M}$ is not an invariant subspace of the phase space $\mathbb{P}$.

## The $b \to \infty$ limit of $\mathcal{SU}_c(1, 1)$ group

The approximate noninertial transformations that are valid in the almost inertial regime $\frac{f}{b} \ll 1$ and $\frac{r}{cb} \ll 1$ are given by the limit $b \to \infty$,

(5.53)

(5.51)

$$\lim_{b \to \infty} \gamma(v, f, r) \begin{pmatrix} 1 & \frac{v}{c^2} & -\frac{r}{c^2 b^2} & \frac{f}{b^2} \\ v & 1 & -\frac{f}{b^2} & \frac{r}{b^2} \\ r & -f & 1 & v \\ f & -\frac{r}{c^2} & \frac{v}{c^2} & 1 \end{pmatrix} = \gamma^\circ(v) \begin{pmatrix} 1 & \frac{v}{c^2} & 0 & 0 \\ v & 1 & 0 & 0 \\ r & -f & 1 & v \\ f & -\frac{r}{c^2} & \frac{v}{c^2} & 1 \end{pmatrix}.$$

This is the matrix realization of the $\mathcal{L}a(1, 1)$ that we reviewed in (1.1) for noninertial transformations.

$$\lim_{b \to \infty} \mathcal{SU}c(1, 1) \to \mathcal{L}a(1, 1). \tag{5.54}$$

Again, the $\mathcal{L}a(1, 1)$ noninertial transformations are the approximation for the near special relativistic inertial regime where relative rates of change of momentum and energy with time are small relative to the fundamental dimensional scale $b$.

### Time dilation, the null hypersurface and causal cones

Returning, to the full noninertial equations, time dilation depends on the noninertial state of the observer,

$$d\tau = \frac{dt}{\gamma(v, f, r)} = dt \sqrt{1 - \frac{v^2}{c^2} - \frac{f^2}{b^2} + \frac{r^2}{c^2 b^2}}. \tag{5.55}$$

Again, as $b \to \infty$, these reduce to the usual time dilation equation of special relativity (1.1). For this general case, simultaneity is different for noninertial states; for clocks to measure the passage at the same rate they must be at inertial rest relative to one another.

The null hypersurface $d\tau^2 = 0$ for general noninertial states is the elliptical hyperboloid

$$1 = \rho(v, f, r) = \frac{v^2}{c^2} + \frac{f^2}{b^2} - \frac{r^2}{c^2 b^2}. \tag{5.56}$$

We can use (5.44) with $\rho = 1$ to define the polar-hyperbolic coordinates of the null hypersurface

$$v = c \cos \theta \cosh \phi, \quad f = b \cos \theta \cosh \phi, \quad r = bc \sinh \phi. \tag{5.57}$$

The timelike cone for causal information propagation is defined by $\rho(v, f, r) < 1$. In the inertial special case, $f = r = 0$, this is the usual special relativistic null cone $v = \pm c$. But note that for observers in states with a small relative change of position and energy with time, $v = r = 0$ that this is a null cone in momentum space, $f = \pm b$. States that have a relative change of energy with time, $r$, experience a 'widening' of the cone enabling $v, f$ to achieve values greater than $c$ and $b$,

$$\frac{v^2}{c^2} + \frac{f^2}{b^2} < 1 + \frac{r^2}{c^2 b^2}. \tag{5.58}$$

For example, for $f = 0$,

$$v = \pm c \sqrt{1 + \frac{r^2}{c^2 b^2}}. \tag{5.59}$$
$$\tag{5.53}$$

This means that states that are dissipating energy with $r \sim o(bc)$ may experience velocities that are larger than $c$ and yet are still constrained by a causal cone. It has the fundamental property of bounding the rate of propagation of information in position space as well as momentum space. This is a fundamental motivation for considering the reciprocal relativity theory of noninertial states with an invariant Born metric.

## Velocity, force, power transformation

The transformation of velocity, force and power may be computed through basic matrix multiplication using the group composition law expanded out for the parameters $f$, $v$, $r$,

$$\Gamma(v'', f'', r'') = \Gamma(v', f', r')\, \Gamma(v, f, r). \tag{5.60}$$

The transformations also follow directly from (5.51). Note that,

$$\frac{dt''}{dt} = \gamma(v, f, r)\left(1 + \frac{v}{c^2}\frac{dq}{dt} + \frac{f}{b^2}\frac{dp}{dt} - \frac{r}{c^2 b^2}\frac{d\varepsilon}{dt}\right),$$

and therefore,

$$\frac{dq''}{d\tilde{t}} = \left(\frac{dq}{dt} + v + \frac{r}{b^2}\frac{dp}{dt} - \frac{f}{b^2}\frac{d\varepsilon}{dt}\right)\bigg/\left(1 + \frac{v}{c^2}\frac{dq}{dt} + \frac{f}{b^2}\frac{dp}{dt} - \frac{r}{b^2 c^2}\frac{d\varepsilon}{dt}\right),$$

$$\frac{dp''}{d\tilde{t}} = \left(\frac{dp}{dt} + f - \frac{r}{c^2}\frac{dq}{dt} + \frac{v}{c^2}\frac{d\varepsilon}{dt}\right)\bigg/\left(1 + \frac{v}{c^2}\frac{dq}{dt} + \frac{f}{b^2}\frac{dp}{dt} - \frac{r}{b^2 c^2}\frac{d\varepsilon}{dt}\right),$$

$$\frac{d\varepsilon''}{d\tilde{t}} = \left(\frac{d\varepsilon}{dt} - f\frac{dq}{dt} + v\frac{dp}{dt} + r\right)\bigg/\left(1 + \frac{v}{c^2}\frac{dq}{dt} + \frac{f}{b^2}\frac{dp}{dt} - \frac{r}{b^2 c^2}\frac{d\varepsilon}{dt}\right). \tag{5.61}$$

With the identification $v' = \frac{dq}{dt}$, $f' = \frac{df}{dt} \simeq f$ and $r' = \frac{d\varepsilon}{dt}$ this leads to the rate of change of position, momentum and energy relativity transformation laws for the $n = 1$ case,

$$v'' = \left(v' + v + \frac{r f'}{b^2} - \frac{f r'}{b^2}\right)\bigg/\left(1 + \frac{v v'}{c^2} + \frac{f f'}{b^2} - \frac{r r'}{b^2 c^2}\right),$$

$$f'' = \left(f' + f - \frac{r v'}{c^2} + \frac{v r'}{c^2}\right)\bigg/\left(1 + \frac{v v'}{c^2} + \frac{f f'}{b^2} - \frac{r r'}{b^2 c^2}\right),$$

$$r'' = (r' - f v' + v f' + r)\bigg/\left(1 + \frac{v v'}{c^2} + \frac{f f'}{b^2} - \frac{r r'}{b^2 c^2}\right). \tag{5.62}$$

The null hypersurface is defined in (5.56) by $\rho(v, f, r) = 1$. For the orthochronous $\mathcal{SU}_c(1, 1)$ group, $\Lambda_0^0 = \rho(v, f, r) \leq 1$ is always the case. Therefore if $\rho(v, f, r) \leq 1$ and $\rho(v', f', r') \leq 1$, then $\rho(v'', f'', r'') \leq 1$ in the above expression and this condition defines the forward causal cone.

## The $\mathcal{Uu}(1, 1)$ group and its limiting form

The full unitary group follows directly from its semidirect product form in (5.40) using the real form of $\mathcal{U}_2(1)$ given in (5.26),

$$\Gamma(\Lambda', M') = \Gamma(\Lambda, M)\, \Gamma(\Lambda_\Delta, M_\Delta) = \Gamma(\Lambda\, \Lambda_\Delta - M\, M_\Delta,\ \Lambda\, M_\Delta + M\, \Lambda_\Delta)\,, \tag{5.63}$$

where

$$\tag{5.64}$$

$$\Lambda' = \gamma(v, f, r) \begin{pmatrix} \cos\theta - \frac{r}{cb}\sin\theta & \frac{v}{c}\cos\vartheta + \frac{f}{b}\sin\vartheta \\ \frac{v}{c}\cos\theta - \frac{f}{b}\sin\theta & \cos\vartheta + \frac{r}{cb}\sin\vartheta \end{pmatrix},$$

$$M' = \gamma(v, f, r) \begin{pmatrix} \sin\theta + \frac{r}{cb}\cos\theta & \frac{v}{c}\sin\vartheta - \frac{f}{b}\cos\vartheta \\ \frac{v}{c}\sin\theta + \frac{f}{b}\cos\theta & \sin\vartheta - \frac{r}{cb}\cos\vartheta \end{pmatrix}.$$

In the limit $b \to \infty$, the reciprocally invariant time line element $d\tau^2$ contracts to the Einstein proper time line element $d\tau^{\circ 2}$. Invariance of $d\tau^{\circ 2}$ requires

$$\Lambda_\Delta(\theta, \vartheta)^t \eta \, \Lambda_\Delta(\theta, \vartheta) = \eta. \tag{5.65}$$

Therefore with $\theta, \vartheta \in [0, 2\pi)$, this requires $\theta, \vartheta \in \{0, \pi\}$. This is precisely the condition to define the discrete subgroup $\mathbb{Z}_{2,2} \subset \mathcal{U}_2(1)$ as given in (5.30). Consequently, the $\mathcal{U}u(1,1)$ group contracts to the the symmetry group defined in (4.45),

$$\lim_{b \to \infty} \mathcal{U}u(1,1) = \mathcal{O}a(1, 1) \simeq \mathbb{Z}_{2,2} \otimes_s \mathcal{L}a(1, 1). \tag{5.66}$$

## 5.5 Noninertial relativity

Sections 3 and 4 do not introduce new physics. Rather, they simply examine noninertial states in the contexts of Newton and proper Einstein time and describe the resulting noninertial relativity groups.

In this section, we introduced a more general concept of time that depends not only on the relative velocity of rate of change of position of states, but also on the relative rate of change of momentum and energy. This is mathematically accomplished by introducing a nondegenerate orthogonal metric to define proper time (5.6) on phase space.

A key physical consequence of this metric is that the rate of change of momentum and energy are relative just as the rate of change of position is relative in special relativity. As a consequence of this, observers in different states do not necessarily agree on the spacetime and energy-momentum subspaces of phase space. That is there is neither an absolute inertial state nor absolute rest state.

For inertial observers, special relativity is recovered and all inertial observers agree on the invariant spacetime and energy-momentum subspaces of phase space.

Furthermore, in the limit of small forces relative to $b$ and power relative to $bc$, the noninertial relativity with Einstein invariant time described in Section 4.4 results. Furthermore, in the limit also of small velocities relative to $c$, the noninertial relativity based on the Hamilton group given in Section 3.3 results.

The most physically consequential result is that the rate of change of momentum is now bounded just as the rate of change of position is bounded. This bound is given by the null hypersurface of $d\tau^2$. This means that that information must be propagated at a finite rate in both position and momentum space. This is of importance in the quantum theory were states can be equally diagonalized measurements of position or momentum. This is embodied in Born's Principle of Reciprocity that defines the discrete Born symmetry. As this reciprocity between position and momentum is foundational, the author named this relativity *Reciprocal Relativity of Noninertial States* in the papers [15,16] that described these concepts.

$$(5.64)$$

The properties of the connected relativity group $\mathcal{SU}c(1,n)$ are as follows:

- Invariant time is proper time $d\tau$ that is measured in the locally inertial rest frame of the particle state. Observers in timelike states with relative motion, inertial or noninertial, measure the passage of time such that reciprocally invariant time $d\tau^2$ is invariant. The local time dilation is $dt = \dfrac{d\tau}{\sqrt{1-\frac{v^2}{c^2}-\frac{f^2}{b^2}+\frac{r^2}{c^2 b^2}}}$ where $v$ is the relative rate of change of position, $f$ is the relative change of momentum and $r$ is the relative change of energy with time. Thus we say that time is relative and it is not an invariant submanifold of either spacetime $\mathbb{M}$ nor the phase space $\mathbb{P}$. This is a defining property of the relativity group $\mathcal{SU}c(1,n)$.
- For inertial states, this reduces to usual special relativity.
- Simultaneity is relative in the sense that observers with different rates of change of position, momentum and energy may observe different ordering of events.
- Rates of change of position and momentum with respect to time are bounded by a 'causal cone' defined by the null surface of $d\tau^2$. The rate of change of energy is not bounded.
- There is not an absolute inertial state that all observers in *any* noninertial states agree on.
- There is not an absolute rest state that all observers in *any* noninertial states agree on.
- Information propagates at a finite rate in position space. That is, there is a 'causal cone' constraining the rate of propagation of information in position space. 'Instantaneous action at a distance' in position space is not possible.
- Information propagates at a finite rate in momentum space. That is, there is a 'causal cone' constraining the rate of propagation of information in momentum space. 'Instantaneous action at a distance' in momentum space is not possible.
- Position-time space (i.e spacetime) $\mathbb{M}$ with coordinates $(t,q)$ is not an invariant subspace of the phase space $\mathbb{P}$ that all observers agree. However inertial observers agree on the spacetime subspace and in this case the usual Lorentz invariance of special relativity applies.
- Momentum-energy space with coordinates $(\varepsilon, p)$ $\tilde{\mathbb{M}}$ is an invariant subspace of the phase space $\mathbb{P}$ only for observers in *inertial* states. In this case the usual Lorentz invariance of special relativity applies.

## 5.6  Born reciprocity

The discussion above on the relativistic symmetry for reciprocally invariant time focussed on the transformation equations that result from the $\mathcal{SU}c(1,n)$. The full $\mathcal{U}u(1,n)$ symmetry includes also a $\mathcal{U}_2(1) = \mathcal{U}(1) \otimes \mathcal{U}(1)$ subgroup. This subgroup has the parity, time-reversal $\mathbb{Z}_{2,2} = \mathbb{Z}_2 \otimes \mathbb{Z}_2$ group as a subgroup, $\{\Delta_e, \Delta_T, \Delta_P, \Delta_{PT}\} \in \mathbb{Z}_{2,2} \subset \mathcal{U}_2(1)$. Restriction to inertial states and also In the limit $b \to \infty$ and also for inertial states, the $\mathcal{U}_2(1)$ symmetry breaks to the discrete $\mathbb{Z}_{2,2}$ symmetry.

The Born reciprocity symmetries $\Delta_B$, $\Delta_{B^\circ}$ and $\Delta_{BB^\circ}$ are also elements of $\mathcal{U}_2(1)$ and together with the parity, time-reversal elements generate a $\mathbb{Z}_{4,4} = \mathbb{Z}_4 \otimes \mathbb{Z}_4 \subset \mathcal{U}_2(1)$ with $\{\Delta_e, \Delta_T, \Delta_{B^\circ}, \Delta_{TB^\circ}\} \in \mathbb{Z}_4$ and $\{\Delta_e, \Delta_P, \Delta_{B^\circ}, \Delta_{PB}\} \in \mathbb{Z}_4$.

In a full reciprocal relativity theory of noninertial states, we have the phenomena

$$\mathbb{Z}_{2,2} \qquad \mathbb{Z}_{4,4}$$
$$\mathcal{U}_2(1)$$

that the discrete parity, time-reversal group $\mathbb{Z}_{2,2}$ and its generalization to $\mathbb{Z}_{4,4}$ with Born reciprocity becomes the continuous $\mathcal{U}_2(1)$ group. A clearer physical understanding of this would provide a key experimental test for the reciprocal relativity.

# 6    Summary

In this paper, we considered a phase space $\mathbb{P} \simeq \mathbb{R}^{2n+2}$ with an invariant symplectic metric

$$\omega = -dt \wedge d\varepsilon + \delta_{i,j} \, dq^i \wedge dp^j,$$

and also a (degenerate) orthogonal metric defining invariant time. The three cases considered are Newton absolute time $t$, Einstein proper $\tau^\circ$ and reciprocally invariant proper time $\tau$,

$$dt^2,$$

$$d\tau^{\circ 2} = dt^2 - \tfrac{1}{c^2} dq^2 = dt^2\left(1 - \tfrac{v^2}{c^2}\right),$$

$$d\tau^2 = dt^2 - \tfrac{1}{c^2} dq^2 - \tfrac{1}{b^2} dp^2 + \tfrac{1}{b^2 c^2} d\varepsilon^2 = dt^2\left(1 - \tfrac{v^2}{c^2} - \tfrac{f^2}{b^2} + \tfrac{r^2}{b^2 c^2}\right).$$

The resulting symmetry groups that leave these definitions of invariant time are

$$\mathbb{Z}_{2,2} \otimes_s \mathcal{HSp}(2n) \subset \mathcal{Sp}(2n+2), \qquad \mathcal{HSp}(2n) \simeq \mathcal{Sp}(2n) \otimes_s \mathcal{H}(n),$$

$$\mathcal{O}a(1,n) \simeq \mathbb{Z}_{2,2} \otimes_s \mathcal{L}a(1,n) \subset \mathcal{Sp}(2n+2), \qquad \mathcal{L}a(1,n) \simeq \mathcal{L}(1,n) \otimes_s \mathcal{A}(m), \; m = \tfrac{n(n+3)}{2},$$

$$\mathcal{U}u(1,n) \simeq \mathcal{U}_2(1) \otimes_s \mathcal{SU}c(1,n) \subset \mathcal{Sp}(2n+2), \quad \mathcal{U}_2(1) = \mathcal{U}(1) \otimes \mathcal{U}(1).$$

The Newton and Einstein cases do not yield any new physics but rather a slightly new perspective on some old physics. For example, in the Newton case with its Jacobi group parameterized by velocity, force and power, the maps $\varrho : \mathbb{P} \to \mathbb{P}$ such that $\varrho^* \omega = \omega$ and $\varrho^*(dt^2) = dt^2$ can be decomposed into the maps $\varrho = \rho \circ \varphi$ where $\rho$ is a solution of Hamilton's equations and $\varphi$ are the usual canonical transformations on the position-momentum subspace $\mathbb{P}^\circ \subset \mathbb{P}$, $\mathbb{P}^\circ \simeq \mathbb{R}^{2n}$ with symplectic metric $\omega^\circ = \delta_{i,j} \, dq^i \wedge dp^j$, $\varphi^* \omega^\circ = \omega^\circ$.

For the Einstein case, the abelian subgroup $\mathcal{A}(m)$, $m = \tfrac{n(n+3)}{2}$ that behaves as a 'power-force-stress' symmetric traceless $(0, 2)$ tensor under Lorentz transformations. It is the proper time derivative of the 'energy-momentum-stress' symmetric traceless $(0, 2)$ tensor. In the Newton time case (i.e 'nonrelativistic'), the relation between noninertial states (for $n = 3$) is parameterized by a 3-vector force $f \in \mathbb{R}^3$ and a scalar power $r \in \mathbb{R}$. In the Einstein proper time case these generalize to this 'power-force-stress' tensor. This is not new result but rather provides a slightly different a slightly different way of looking at it.

The time line elements satisfy the limits

$$d\tau^2 \xrightarrow[b \to \infty]{} d\tau^{\circ 2} \xrightarrow[c \to \infty]{} dt^2,$$

and, for $n = 1$, the symmetry group satisfy the contraction limits,

$$\mathcal{U}u(1, n) \xrightarrow[b \to \infty]{} \mathcal{O}a(1, n) \xrightarrow[c \to \infty]{} \mathcal{HO}(n).$$

The new physics is for the reciprocally invariant time line element with the $\mathcal{U}u(1, n)$ symmetry group. For the inertial case, $\mathcal{U}u(1, n)$ reduces to $O(1, n)$ and standard special relativity applies. It has an orthochronous subgroup $\mathcal{SU}c(1, n) \subset \mathcal{U}u(1, n)$ corresponding to the inertial case $\mathcal{L}(1, n) \subset O(1, n)$. This is the group that defines the physical timelike transformations in the future causal cones in $\mathbb{P}$ defined by the null surfaces $d\tau^2 = 0$,

$$\frac{v^2}{c^2} + \frac{f^2}{b^2} = 1 + \frac{r^2}{b^2 c^2}.$$

Both the rate of change of position, $v$, and the rate of change of momentum, $f$, is bounded by the causal cone. Power dissipation $r$ causes these causal cones to 'flatten' as given by the above expression. The inertial special relativity case $f = r = 0$ is the usual cones $v^2 = c^2$. Note that $\mathcal{SU}c(1, n) \subset \mathcal{SU}(1, n) \subset \mathcal{U}u(1, n)$ similar to the orthogonal case that are the inertial subgroups, $\mathcal{L}(1, n) \subset SO(1, n) \subset O(1, n)$. The parity, time-reversal discrete group is $\mathbb{Z}_{2,2} \subset \mathcal{U}_2(1)$. Also, it can be shown that

$$\mathcal{SU}c(1, n) \xrightarrow[b \to \infty]{} \mathcal{L}a(1, n) \xrightarrow[c \to \infty]{} \mathcal{H}a(n), \quad \mathcal{U}_2(1) \xrightarrow[b \to \infty]{} \mathbb{Z}_{2,2} \xrightarrow[c \to \infty]{} \mathbb{Z}_{2,2}.$$

The discrete parity, time-reversal group $\mathbb{Z}_{2,2}$ becomes the continuous $\mathcal{U}_2(1) = \mathcal{U}(1) \otimes \mathcal{U}(1)$ group in the extreme noninertial regime. It reduces to the finite parity, time-reversal $\mathbb{Z}_{2,2}$ for inertial states. The full physical consequences of this require further investigation.

The connected noninertial relativity groups have the following properties

|  | $\mathcal{H}a(n)$ | $\mathcal{L}a(1, n)$ | $\mathcal{SU}c(1, n)$ |
|---|---|---|---|
| Invariant time | Newton | Einstein | Reciprocally invariant |
| Inertial subgroup | $\mathcal{E}(n)$ | $\mathcal{L}(1, n)$ | $\mathcal{L}(1, n)$ |
| Rest frame | absolute | relative | relative |
| Inertial frame | absolute | absolute | relative |
| Time subspace | absolute | relative | relative |
| Spacetime | absolute | absolute | relative |
| Velocity, $v$ | unbounded | bounded | bounded |
| Force, $f$ | unbounded | unbounded | bounded |

The corresponding quantum symmetries are described by the projective representations of the inhomogeneous groups groups $\mathcal{IHO}(n)$, $\mathcal{IO}a(1, n)$ and $\mathcal{IU}u(1, n)$ with the connected subgroups groups $\mathcal{IH}a(n)$, $\mathcal{IL}a(1, n)$ and $\mathcal{ISU}c(1, n)$. These quantum symmetries are investigated in the papers [16-23].

## Notes

1. The symmetry group on ℙ with Newton invariant time is studied in the papers [17,18,19,21]. The the author named the group $\mathcal{H}a(n) = SO(n)$ the *Hamilton* group in these papers.
2. 'Nonrelativistic' is a misnomer as Newtonian physics is invariant under the Galilean relativity group. However, as nonrelativistic is so embedded in physics terminology, we continue to use it to refer to the effective symmetry for small velocities relative to $c$.
3. The group $\mathcal{H}Sp(2n)$ was named Jacobi group by Eichler and Zagler [8] who studied primarily the $n = 1$ case in a pure mathematics context. This naming was prescient given its role outlined here in Hamilton-Jacobi mechanics. See also [3,2,26,27].
4. Born wrote several papers [4, 5] on the idea of reciprocity and it is discussed in letters 74 and 91 in his correspondence with Einstein [6]. He was clearly motivated by his observation of the 'momentum-position' conjugacy that is manifest in Hamilton's mechanics and quantum mechanics. He stated it as the discrete symmetries $\Delta_B$ and $\Delta_{BB^\circ}$ but does not develop the continuous group theory. There is also no reference to the symplectic structure. His key investigations were to explore reciprocally invariant 'wave operators' and also momentum space with curvature. He did not see it as a relativity principle with the consequent implications on the definition of time.
5. The author originally proposed the reciprocally invariant time line element in together with an invariant symplectic to metric to define the $\mathcal{U}(1, n)$ symmetry in [14]. The author became aware of Born's work and named the metric the *Born*, or *Born-Green* metric in [15,16] and elaborated on its meaning in the context of noninertial relativity. Born viewed the metric in the context of reciprocally wave equations but did not develop it as a definition for a reciprocally invariant time line element nor its relativistic symmetry group. The name *Born geometry* has been adopted by Freidel et al [10,25] for manifolds with an orthogonal and symplectic metric with a compatible para-Hermitian structure that defines a unique compatible connection.
6. A summary of Planck scales with comprehensive references is on the Wikipedia page (Wikipedia.org/wiki/Planck_units)